\documentclass[reprint,twocolumn,floatfix]{revtex4}
\usepackage{amsmath}
\usepackage{graphicx}
\usepackage{dcolumn}
\usepackage{physics}
\usepackage{bm}
\usepackage{hyperref}
\usepackage{epsfig,colordvi}
\usepackage{color}
\usepackage{natbib}
\usepackage{subfiles}

 \def\be{\begin{equation}}
 \def\ee{\end{equation}}
 \def\bea{\begin{eqnarray}}
 \def\eea{\end{eqnarray}}

\newcommand{\beq}[1]{\begin{equation}\label{eq:#1}}
\newcommand{\eeq}{\end{equation}}
\newcommand{\rarr}{\rightarrow}

\newcommand{\lrarr}{\leftrightarrow}

\begin{document}

\title{Dynamical mechanisms for deuteron production at mid-rapidity in relativistic heavy-ion collisions from SIS to RHIC energies}
 
\author{G.~Coci$^{1,2}$, S.~Gl{\"a}{\ss}el$^{3}$, V.~Kireyeu$^{4,1}$, J.~Aichelin$^{5,6}$, C.~Blume$^{3}$, E.~Bratkovskaya$^{7,2,1}$, V.~Kolesnikov$^{4}$,  V.~Voronyuk$^{4,1}$}

\affiliation{$^{1}$ Helmholtz Research Academy Hessen for FAIR (HFHF),GSI Helmholtz Center for Heavy Ion Physics. Campus Frankfurt, 60438 Frankfurt, Germany}
\affiliation{$^{2}$ Institut f\"ur Theoretische Physik, Johann Wolfgang Goethe University,
Max-von-Laue-Str. 1, 60438 Frankfurt, Germany}
\affiliation{$^{3}$ Institute für Kernphysik, Max-von-Laue-Str. 1, 60438 Frankfurt, Germany}
\affiliation{$^{4}$ Joint Institute for Nuclear Research, Joliot-Curie 6, 141980 Dubna, Moscow region, Russia}
\affiliation{$^{5}$ SUBATECH, Universit\'e de Nantes, IMT Atlantique, IN2P3/CNRS
4 rue Alfred Kastler, 44307 Nantes cedex 3, France}
\affiliation{$^{6}$ Frankfurt Institute for Advanced Studies, Ruth Moufang Str. 1, 60438 Frankfurt, Germany} 
\affiliation{$^{7}$ GSI Helmholtzzentrum f\"ur Schwerionenforschung GmbH,
  Planckstr. 1, 64291 Darmstadt, Germany}
  
\date{\today}

\begin{abstract} \noindent
The understanding of the mechanisms for the production of weakly bound clusters, such as a deuteron $d$, in heavy-ion reactions at mid-rapidity is presently one of the challenging problems which is also known as the ``ice in a fire" puzzle.
In this study we investigate the dynamical formation of deuterons within the Parton-Hadron-Quantum-Molecular Dynamics (PHQMD) microscopic transport approach and advance two microscopic production mechanisms to describe deuterons in heavy-ion collisions from SIS  to RHIC energies:
 kinetic production by hadronic reactions and potential production by the attractive potential between nucleons. Differently to other studies, for the ``kinetic" deuterons we employ the full isospin decomposition of the various $\pi NN\leftrightarrow \pi d$, $NNN\leftrightarrow N d$ channels and take into account the finite-size properties of the deuteron by means of an excluded volume condition in coordinate space and by the projection onto the deuteron wave function in momentum space. We find that considering the quantum nature of the deuteron in coordinate and momentum space reduces substantially the kinetic deuteron production in a dense medium as encountered in heavy-ion collisions. If we add the ``potential" deuterons by applying an advanced Minimum Spanning Tree (aMST) procedure, we obtain good agreement with the available experimental data from SIS energies up to the top RHIC energy. 
\end{abstract}


\maketitle

\section{Introduction}\label{sec:sec1}

Quantum Chromodynamics (QCD), the theory describing the strong interaction between quarks and gluons, the elementary components of hadrons, owns important features, which have not yet been understood. 
To study this QCD matter under extreme conditions of temperature and density is the primary purpose of Heavy-Ion Collisions (HICs) at ultra-relativistic energies~\cite{Busza:2018rrf,Jacak:2021imp}, which are performed at the BNL Relativistic Heavy-ion Collider (RHIC) and at the CERN Large Hadron Collider (LHC). 
At very high energies the energy deposited during the initial stage of the collisions creates an almost net-baryon free hot medium consisting of deconfined quarks and gluons, which is called Quark-Gluon Plasma (QGP).
On the theoretical side the knowledge of the QCD phase diagram, describing the pressure as a function of temperature $T$ and baryon chemical potential $\mu_B$, is limited to the region of high $T$ and almost zero $\mu_B$.  There QCD calculations on lattices (lQCD)~\cite{Borsanyi:2013bia,Bazavov:2014pvz} predict a smooth crossover between the QGP phase and a gas of hadrons.

The ongoing Beam Energy Scan (BES) program at RHIC, as well as the future experiments at the Nuclotron based Ion Collider (NICA) and at the Facility for Antiproton and Ion Research (FAIR), under construction in Dubna and Darmstadt, respectively, will extend the study of strongly interacting matter to lower collision energies. The aim is to explore the QCD phase diagram at high net-baryon density and to search for the existence of a Critical End Point (CEP) at the end of a first-order phase boundary at non-zero $\mu_{B}$, predicted by effective theories~\cite{Asakawa:1989bq,Stephanov:2004wx}. 

To explore this transition from QCD matter to hadrons the study of the production of light nuclei, such as $d$, $t$, $^3\!He$, $^4\!He$ and hypernuclei, is an important issue because the production of composite clusters depends on the correlations and fluctuations of the nucleons. The interest in light nuclei comes from both, experiments and theory.\\
From the experimental side, the observation of light nuclei began with the first heavy-ion experiments at the Bevalac accelerator \cite{Westfall:1976fu,Gutbrod:1976zzr,Nagamiya:1981sd} (after some low statistics bubble chamber data~\cite{Grunder:1971ji}). It continued at AGS \cite{E866:1999ktz,E866:2000dog,E878:1998vna,E886:1994ioj,Armstrong:2000gz}, 
the GSI SIS facility~\cite{Reisdorf:2010aa} and the SPS collider~\cite{NA49:2004mrq,NA49:2016qvu}. Nowadays, the measurements of light nuclei and hypernuclei at mid-rapidity represent an important research program for the STAR collaboration~\cite{STAR:2019sjh} at RHIC and for the ALICE collaboration~\cite{ALICE:2015wav,ALICE:2017nuf} at LHC. 
At low (SIS) beam energies between 30\% and 50\% of protons are bound in deuterons, tritons and $^3\!He$, while this fraction decreases with increasing beam energy to values around 1\% at the LHC (see, for instance, Ref.~\cite{Donigus:2022haq}).  
Collective variables, like the directed or the elliptic flow, are different for clusters and nucleons, indicating that clusters test different phase-space regions than nucleons. 
Therefore, bound nucleons represent an interesting probe to study the dynamics of heavy-ion collisions.

From the theoretical side, the reason is even more fundamental since the mechanism of cluster formation in nucleus-nucleus collisions is not well understood. The deuterons with a binding energy of $|E_B| \simeq 2$ MeV appear to be fragile objects compared to the average kinetic energy of hadrons which surround them. At freeze-out, when the QGP is converted into hadrons, the kinetic freeze-out parameters indicate a temperature of $T \simeq 100-150$ MeV. It is surprising that they can survive in such an environment without being destroyed by collisions with the surrounding hadrons. Hence, it is puzzling that light nuclei are observed in central HICs at mid-rapidity at all, and it is even more puzzling that their multiplicity is well described by statistical model calculations~\cite{Andronic:2010qu}. This observation has been portrayed as ``\textit{ice cubes in a fire}"~\cite{Bratkovskaya:2022vqi}, or ``\textit{snowballs in hell}"~\cite{Oliinychenko:2018ugs}. The presence of light clusters one may consider as a hint that they do not come from the same phase-space regions as nucleons, which makes them interesting for the study of the reaction dynamics.
The formation of light nuclei at mid-rapidity at beam energies above 2 AGeV has been modeled by three main approaches:
\begin{itemize}
\item[i)] In the statistical model hadrons at mid-rapidity are assumed to be emitted from a common thermal source, which is characterized by the temperature $T$, the chemical potential $\mu_B$ and a fixed volume $V$~\cite{Cleymans:1992zc,Andronic:2012dm}. All three quantities are determined by fitting the multiplicity of a multitude of hadrons. Surprisingly, the observed cluster multiplicities are also described with the same fit variables $T$, $V$ and $\mu_B$~\cite{Andronic:2010qu,Andronic:2017pug}. The assumption of the statistical model approach is that the hadronic expansion of the system does not change the number of clusters. 

\item[ii)] In the coalescence approach it is assumed that a proton and a neutron form a deuteron if their distance in momentum and coordinate space is smaller than the coalescence parameters $(r_{coal}, p_{coal})$~\cite{Sombun:2018yqh,Butler:1963pp}. This distance is measured when the last nucleon of the pair undergoes its last elastic or inelastic collision. Several variations of the coalescence model are being used. Some of them project the phase-space distribution function of the nucleons to the Wigner density of the relative coordinates of the nucleons in the deuteron. This distribution is usually approximated by a Gaussian form~\cite{Scheibl:1998tk,Zhu:2015voa}.
However, the coalescence approach neglects that a deuteron is a bound object, which cannot be formed by a simple ``fusion'' of two nucleons, since it would violate the energy-momentum conservation. The  formation  of a deuteron is only possible if it interacts with the environment by a potential or via scattering processes. 
Nevertheless, this approach reproduces well the $p_T$-spectrum of deuterons, as well as their multiplicity for a large range of beam energies. For most recent studies on deuteron production with the coalescence approach we refer to Refs.~\cite{Sun:2018jhg,Zhao:2020irc,Kittiratpattana:2022knq}.

\item[iii)] The Minimum Spanning Tree (MST) approach has been originally advanced in Ref.~\cite{Aichelin:1991xy} to study fragments which come from the projectile and target region and later it has also been employed to study mid-rapidity clusters~\cite{Aichelin:2019tnk}. It assumes that at the end of the heavy-ion reaction two nucleons are part of a cluster if their distance is smaller than a radius $r_{clus}$ which is of the order of the range of the nucleon-nucleon interaction. As investigated in a successive study~\cite{Glassel:2021rod}, this model reproduces well the $p_T$ and $dN/dy$ spectra not only for deuterons, but also for all clusters, observed at mid-rapidity, in the energy range from AGS to top RHIC.
\end{itemize}

Recently, a fourth approach has been advanced. In Refs.~\cite{Oliinychenko:2018ugs,Oliinychenko:2020znl,Staudenmaier:2021lrg} it has been claimed that deuterons can also be created by elementary collisions: $p n \pi \leftrightarrow d \pi$, $p n N \leftrightarrow d N$, $N N \leftrightarrow d \pi$. 
Based on~\cite{Danielewicz:1991dh,Kapusta:1980zz,Siemens:1979dz}, where the production (disintegration) of deuterons by $p n N \leftrightarrow d N$ (nucleon catalysis) was studied at low energy HICs, it has been proposed in Ref.~\cite{Oliinychenko:2018ugs} that at relativistic HICs the pion catalyis, i.e. $p n \pi \leftrightarrow d \pi$, becomes more dominant at mid-rapidity due to the large abundance of pions.
To demonstrate this, $d \pi$ inelastic scatterings and the inverse processes have been implemented in the transport approach SMASH~\cite{Weil:2016zrk}, which describes the hadronic stage of HICs. In the study~\cite{Oliinychenko:2018ugs} the catalysis reactions $p n \pi(N) \leftrightarrow d \pi(N)$ have been approximated as simple two-step processes of $pn \leftrightarrow d'$ and $\pi(N) d' \leftrightarrow \pi(N) d'$ where $d'$ is a fictitious dibaryon resonance with mass and width determined by fitting the experimental total inclusive cross section for $d \pi$ inelastic scattering. With this approach the deuteron multiplicity and $p_T$-spectra at mid-rapidity could be reproduced for LHC Pb+Pb collisions $\sqrt{s}=2.76$ TeV and for Au+Au collisions in the energy range of the RHIC BES ($\sqrt{s}=7.7 - 200$ GeV)~\cite{Oliinychenko:2020znl}. Later, in Ref.~\cite{Staudenmaier:2021lrg}, the numerical artifact of employing the intermediate $d'$ state has been replaced by the treatment of multi-particle reactions within the covariant rate formalism, firstly developed in Ref.~\cite{Cassing:2001ds}. In both studies the deuteron was treated as a point-like particle.

In this work we revise and improve two of the above mentioned dynamical processes for deuteron production in HICs, the ``kinetic'' production by hadronic collisions and the  ``potential'' mechanism, where bound nucleons form deuterons and heavier clusters by potential interactions, and combine them to obtain a comprehensive approach for the description of the experimental measurements at mid-rapidity. For this study we use the Parton-Hadron-Quantum-Molecular Dynamics (PHQMD) transport approach \cite{Aichelin:2019tnk}.

Concerning the first approach, we include, in contradistinction to~\cite{Oliinychenko:2018ugs,Oliinychenko:2020znl,Staudenmaier:2021lrg}, all possible isospin  channels for $NN\pi \leftrightarrow d\pi$ reactions which enhances the production rate compared to the $pn\pi \leftrightarrow d\pi$ case. Following Ref.~\cite{Sun:2021dlz}, in this ``kinetic'' mechanism we also take into account the distribution of the relative momentum of the two nucleons inside a deuteron. 
Regarding the second approach, we overcome the problem discussed in Ref.~\cite{Aichelin:2019tnk} that a choice had to be made at which time the cluster analysis with MST takes place. We will show that an asymptotic distribution of stable clusters, which are also ``bound'' in the sense that they have negative binding energies $E_B<0$, can be obtained, independent of the time when the clusters are identified.
In order to do so, we present a novel cluster recognition procedure based on the MST algorithm used in point iii), which is further developed in order to trace the entire dynamical evolution of the baryons which are bound into a stable cluster.
It is the purpose of this paper to show that the combination of such an advanced MST (aMST) approach and the production of deuterons by collisions gives a very good description of the total multiplicity, $p_T$ and the $dN/dy$ spectra of deuterons from SIS ($\sqrt{s}=2.5$ GeV) up to the highest RHIC energy ($\sqrt{s}=200$ GeV).

This paper is organized as follows:
After the introduction given in Sec.~\ref{sec:sec1}, in Sec.~\ref{sec:sec2}.A we recall the basic ideas of the PHQMD transport approach. The identification of deuterons bound by potential interaction by means of the MST clusterization algorithm is the subject of Sec.~\ref{sec:sec2}.B. In particular, after discussing in Sec.~\ref{sec:sec2}.B.1 the basis of the original MST model employed in previous PHQMD studies (see Ref.~\cite{Aichelin:2019tnk}), we present in Sec.~\ref{sec:sec2}.B.2 our new ``advanced'' MST (aMST) approach.
The theoretical formulation of the main hadronic reactions for the production of ``kinetic'' deuterons is the topic of Sec.~\ref{sec:sec3}. In Sec.~\ref{sec:sec4} we test such deuteron reactions in a ``box'' and verify their correct numerical implementation by comparing with analytic rate results.
In Sec.~\ref{sec:sec5} we investigate the main physical effects of production and disintegration of deuterons by hadronic reactions in heavy-ion simulations within the PHQMD approach. The details on how the two ``kinetic'' and ``potential'' mechanisms are combined within the PHQMD framework are reported at the end of this section.
In Sec.~\ref{sec:sec6} we confront our final results with combined kinetic and potential deuterons with the existing experimental data for rapidity and transverse momentum distributions in HICs from invariant center-of-mass collision energies of $\sqrt{s}_{NN}=2.52$ GeV to $\sqrt{s}_{NN}=200$ GeV. Finally, we outline our conclusions in Section \ref{sec:sec7}.

\section{Model description}\label{sec:sec2}

\subsection{PHQMD}
The Parton-Hadron-Quantum Molecular Dynamics (PHQMD) has been recently conceived as a new type of microscopic transport approach which combines the characteristics of baryon propagation from the Quantum Molecular Dynamics (QMD) model~\cite{Aichelin:1991xy,Aichelin:1987ti,Aichelin:1988me,Hartnack:1997ez} and the dynamical properties and interactions in and out of equilibrium of hadronic and partonic degrees of freedom of the Parton-Hadron-String-Dynamics (PHSD) approach~\cite{Cassing:2008sv,Cassing:2008nn,Cassing:2009vt,Bratkovskaya:2011wp,Linnyk:2015rco}.
 
In this section we provide a short summary of these two building blocks. For more details of the PHQMD model we refer to Ref.~\cite{Aichelin:2019tnk}.\\

{\bf I.} 
In QMD the baryons are described by single-particle wave functions of Gaussian form with a time independent width. The Wigner density of each particle is obtained by a Fourier transformation of the density matrix. Then, the n-body Wigner density is constructed by the direct product of the single-particle Wigner densities and its propagation is determined by a generalized Ritz variational principle~\cite{Feldmeier:1989st}.
Contrary to mean-field approaches, where the n-body phase-space correlations are integrated out and the dynamics is reduced to a single-particle propagation in a mean-field potential, in QMD these correlations are preserved and the fluctuations not suppressed. This allows to investigate the dynamical formation of clusters, which are correlations between nucleons.

In PHQMD a baryon $i$ is represented by the single-particle Wigner density,
which is given by  
\begin{equation}\label{fdefinition}
 f (\mathbf{r}_i, \mathbf{p}_i,\mathbf{r}_{i0},\mathbf{p}_{i0},t) = \frac{1}{\pi^3 \hbar^3}
{\rm e}^{-\frac{2}{L} [\mathbf{r}_i - \mathbf{r}_{i0} (t) ]^2 }
{\rm e}^{-\frac{L}{2\hbar^2} [\mathbf{p}_i - \mathbf{p}_{i0} (t) ]^2},
\end{equation}
the Gaussian width $L$ is taken as $L=8.66$~fm$^2$.\\

The QMD equations of motion (EoMs) for the centroids $(\mathbf{r}_{i0},\, \mathbf{p}_{i0})$ are similar to those of the Hamilton equations for a classical particle~\cite{Aichelin:1991xy}
\begin{equation}\label{eq:QMDeom}
\dot{\mathbf{r}}_{i0} = \frac{\partial \langle H \rangle}{\partial \mathbf{p_{i0}}} \quad , \quad \dot{\mathbf{p}}_{i0} = - \frac{\partial \langle H \rangle}{\partial \mathbf{r_{i0}}} \, ,
\end{equation}   
where the difference lies in the fact that on the right hand side of both equations the expectation value of the quantum Hamiltonian of the many-body system appears. We note in passing that for a non-Gaussian form of the wave function the time evolution equations are different.
The Hamiltonian is the sum of the kinetic energies of the particles and of their (density dependent) two-body interaction. The expectation value can be written as
\begin{equation}\label{eq:QMDexpH}
 \langle H \rangle = \sum_i \langle H_i \rangle = \sum_i ( \langle T_i \rangle + \sum_{j \ne i} \langle V_{i,j} \rangle ). 
\end{equation}
The potential interaction consists of two parts: the Coulomb interaction and a local density dependent Skyrme potential $V_{i,j}=V_{Coul}+V_{Skyrme}$. The expectation value of the Coulomb interaction can be calculated analytically. The expectation value of the Skyrme part contains terms $\propto \rho^2$ and $\propto \rho^\gamma$, where $\rho$ is the local density.  
Their weights, as well as the exponent $\gamma$, are tuned to the Equation of State (EoS) of infinite nuclear matter $E(T=0,\rho/\rho_0=1)=-16$ MeV, where $\rho_0=0.16$~fm$^{-3}$ is the saturation density at zero temperature. This fixes two of the three parameters.
In PHQMD two parameter sets have been introduced, which yield a ``soft'' and a ``hard'' EoS, respectively. 
For details on the realization of the QMD dynamics and the impact of different EoS on bulk and cluster observables we refer to Ref.~\cite{Aichelin:2019tnk}. For bulk and strangeness particle production in PHQMD with a ``hard'' and a ``soft'' EoS at low energy HICs and the comparison with other transport models see also Ref.~\cite{Reichert:2021ljd}.      
In our present study of the deuteron production mechanisms, we employ the PHQMD in its ``hard'' EoS setup like in Ref.~\cite{Glassel:2021rod}.

Finally, we want to stress two aspects: \textit{i)} in this study, as in the previous PHQMD publications, we employ a ``static'', i.e. momentum independent, Skyrme potential. A form of the Skyrme interaction which contains also a momentum dependent part will be reserved for future work.  
\textit{ii)} The PHQMD approach aims to provide a dynamical description of cluster formation in HICs at low-energy, as well as at relativistic energies. The discussed QMD model uses non-relativistic quantum wave functions.
The relativistic formulation of QMD dynamics as a n-body theory has been developed in Refs.~\cite{Sorge:1989dy,Marty:2012vs} by introducing extra constraints in order to reduce the 8N-dimensional phase-space to the (6N + 1)-dimensional phase-space in which particle trajectories can be defined.
However, this method is computationally extremely expensive, requiring the inversion of a matrix of size $N\times N$ in each time step and thus, it is presently not applicable for high statistics simulations.\\
Therefore, in PHQMD the original framework of QMD is extended to relativistic energies by introducing a modified single-particle Wigner density for each nucleon $i$:
\bea
&& \tilde f (\mathbf{r}_i, \mathbf{p}_i,\mathbf{r}_{i0},\mathbf{p}_{i0},t) =  \label{fGam} \\
&& =\frac{1}{\pi^3} {\rm e}^{-\frac{2}{L} [ \mathbf{r}_{i}^T(t) - \mathbf{r}_{i0}^T (t) ]^2} 
  {\rm e}^{-\frac{2\gamma_{cm}^2}{L} [ \mathbf{r}_{i}^L(t) - \mathbf{r}_{i0}^L (t) ]^2}  \nonumber \\
&& \times {\rm e}^{-\frac{L}{2} [ \mathbf{p}_{i}^T(t) - \mathbf{p}_{i0}^T (t) ]^2} 
  {\rm e}^{-\frac{L}{2\gamma_{cm}^2} [ \mathbf{p}_{i}^L(t) - \mathbf{p}_{i0}^L (t) ]^2},  \nonumber 
\eea
which accounts for the Lorentz contraction of the nucleus in the beam $z$-direction
in coordinate and momentum space by  including $\gamma_{cm} =1/\sqrt{1-v_{cm}^2}$, where $v_{cm}$ is the velocity of projectile and target in the computational frame, which is the center-of-mass system of the heavy-ion collision.  
Accordingly, the interaction density modifies as 
\bea
 \tilde \rho_{int} (\mathbf{r}_{i0},t) 
 &\to & C  \sum_j \Big(\frac{1}{\pi L}\Big)^{3/2} \gamma_{cm} 
 {\rm e}^{-\frac{1}{L} [ \mathbf{r}_{i0}^T(t) - \mathbf{r}_{j0}^T (t) ]^2} \nonumber \\ 
 &&\times {\rm e}^{-\frac{\gamma_{cm}^2}{L} [ \mathbf{r}_{i0}^L(t) - \mathbf{r}_{j0}^L (t) ]^2}.  
 \label{densGam}
\eea
We refer again to Ref.~\cite{Aichelin:2019tnk} and Ref.~\cite{Aichelin:1991xy} for a more detailed discussion and the explicit formulas.
We note that in PHQMD the nuclei are initialized in their rest frame with the Gaussian distributions Eq.~\eqref{fdefinition}. The Lorentz squeezing of nuclei by the gamma factor $\gamma_{cm}$ is done after the initialization of the nuclei in their rest frame, so it does not affect the initialization. \\

{\bf II.}
As in the PHSD (Parton-Hadron-String Dynamics) approach  \cite{Cassing:2008sv,Cassing:2008nn,Cassing:2009vt,Bratkovskaya:2011wp,Linnyk:2015rco},
in PHQMD the strongly interacting medium is described by off-shell hadrons and off-shell massive quasi-particles representing the deconfined quarks and gluons of the QGP phase, which is created if the local energy density is larger than a critical value of $\epsilon_c \approx 0.5 \!$ GeV/fm$^3$.
The propagation of these off-shell degrees of freedom, including their spectral functions, is based on the Kadanoff-Baym transport theory~\cite{KadanoffBaym} in first-order gradient expansion from which the Cassing-Juchem generalized off-shell transport equations~\cite{Cassing:1999wx,Cassing:1999mh,Cassing:2008nn} in test-particle representation are derived (see Ref.~\cite{Cassing:2021fkc}). 
The hadronic part is taken from the early development of the Hadron-String-Dynamics (HSD) approach (see Ref.~\cite{Cassing:1999es} for a detailed description of the baryon, meson and resonance species implemented). 

The elementary baryon-baryon ($BB$), meson-baryon ($mb$) and meson-meson ($mm$) reactions for multi-particle production are realized according to the Lund string model~\cite{Nilsson-Almqvist:1986ast} via the event generators FRITIOF 7.02~\cite{Nilsson-Almqvist:1986ast,Andersson:1992iq} and PYTHIA 7.4~\cite{Sjostrand:2006za}. Both generators are ``tuned'' for a better description of the experimental data for elementary $pp$ collisions at intermediate energies~\cite{Kireyeu:2020wou}. 

The partonic part, which describes the QGP phase, follows the description of the so-called Dynamical Quasi-Particle Model (DQPM)~\cite{Cassing:2007yg,Cassing:2007nb,Peshier:2005pp}. In the DQPM quarks and gluons are represented by massive, strongly interacting quasi-particles. They have spectral functions whose pole positions and widths are defined by the real and imaginary terms of parton self-energies. The parton masses and widths are functions of the temperature $T$ (and in the most recent extension~\cite{Soloveva:2019xph} also of the baryon-chemical potential $\mu_B$) through an effective coupling constant, which is fixed by fitting lQCD results from Refs.~\cite{Aoki:2009sc,Borsanyi:2013bia,Bazavov:2014pvz,Cheng:2007jq,Borsanyi:2015waa}. These DQPM partons are evolved with their self-energies according to the same off-shell transport equations and scatter by microscopically computed cross sections.

We recall that in PHQMD only the propagation of mesons and partons relies on the PHSD approach, while the baryons evolve according to the QMD dynamics. However, it is always possible to run PHQMD in the ``(P)HSD-mode'' by switching the baryon propagation back to the mean-field dynamics of HSD. Again we refer to Ref.~\cite{Aichelin:2019tnk} for a description and detailed studies.
 
As already stated, in PHQMD the collision integral, which encodes the main scattering/dissipative processes of hadrons and partons, is adopted from the PHSD model. 
The main hadronic reactions have been implemented for many observables, like strangeness, dileptons, photons, heavy quarks, etc. (cf. examples in the reviews~\cite{Linnyk:2015rco,Bleicher:2022kcu}).
It contains also in-medium effects, such as a dynamical modification of vector meson spectral functions by  collisional broadening~\cite{Bratkovskaya:2007jk}, and the modification of strange degrees of freedom in line with many-body G-matrix calculations~\cite{Cassing:2003vz,Song:2020clw}, as well as chiral symmetry restoration via the Schwinger mechanism for the string decay~\cite{Cassing:2015owa,Palmese:2016rtq} in a dense medium.
The important and pioneering development in the PHSD is related to the formulation and development of the theoretical formalism in order to realize the detailed balance for $m\leftrightarrow n$ reactions based on covariant rates~\cite{Cassing:2001ds}.  This  formalism has been implemented in PHSD in Ref.~\cite{Cassing:2001ds} for the description of baryon-antibaryon $B\bar{B}$ annihilation of $B=p, \Lambda$ and the inverse reaction of multi-meson fusion to $B +\bar B$  pairs; an extension of this first study accounting for all baryon-antibaryon combinations in PHSD has been presented in Ref.~\cite{Seifert:2017oyb}. We also mention  that the implementation of detailed balance on the level of  $2\leftrightarrow 3$ reactions is realized for the main channels of strangeness production/absorption by baryons ($B=N, \Delta, Y$) and pions \cite{Song:2020clw}.
 
One of the main goals of this work is to extend this formalism to the $2 \leftrightarrow 2$ and $3 \leftrightarrow 2$ processes, which  are relevant for the production and disintegration of deuterons. Therefore, we dedicate a separate section to the detailed description of this formalism and its application to deuteron reactions.

\subsection{The {``}advanced{''} MST approach (aMST)}

\subsubsection{The original MST approach}
The Minimum Spanning Tree (MST) method has been employed in the PHQMD transport approach in Ref.~\cite{Aichelin:2019tnk} to identify clusters at different stages of the dynamical evolution of the system. We stress that MST is a cluster recognition procedure, not a ``cluster building" mechanism, since PHQMD propagates baryons, not clusters. 
The possibility to trace back in time the baryons, which combine to clusters due to their potential interaction, allows to investigate more quantitatively the nature of cluster formation, and to answer some fundamental questions, for example how clusters can survive the dense medium~\cite{Glassel:2021rod}.

The principle of MST in its original version described in Ref.~\cite{Aichelin:1991xy} is to collect nucleons which are close in coordinate space. At a given time $t$ a snapshot of the positions and momenta of all nucleons is recorded and the MST clusterization algorithm is applied: two nucleons $i$ and $j$ are considered as ``bound" to a deuteron or to any larger cluster $A>2$ if they fulfill the condition
\begin{equation}\label{eq:MSTcond}
| \mathbf{r}_i^* - \mathbf{r}_j^* | < r_{clus} \, ,
\end{equation}
where on the left hand side the positions are boosted in the center-of-mass of the $ij$ pair. The maximum distance between cluster nucleons $r_{clus}=4$ fm corresponds roughly to the range of the (attractive) $NN$ potential.
Additional momentum cuts do not change the result because the trajectories of baryons, which are not bound, diverge. Therefore the formation of baryons in MST is a consequence of the attractive \textit{potential interaction}.
A nucleon $i$ belongs to a cluster $A \ge 2$ if it is ``bound'' with at least one nucleon of that cluster, i.e. if there exists a nucleon $j$ for which the condition Eq.~\eqref{eq:MSTcond} is fulfilled.
Recently, MST has been developed to an independent tool, which can be coupled to any theoretical transport approach or to any theoretical framework for detector calibration~\cite{Kireyeu:2021igi}.

We want to stress the fact that MST does not lead to the \textit{formation} of real deuterons. It rather marks only whether a given nucleon is a part of a deuteron at times $t_i = t_0+i\cdot\Delta t$. 
During the time step $\Delta t$ each nucleon continues to propagate according to the QMD EoMs Eq.~\eqref{eq:QMDeom}. One can also identify whether in the subsequent time steps the same nucleons form a deuteron and at each time step the binding energy of the deuteron can be calculated.
In particular, the binding energy of the produced cluster of size $A$ is calculated in its center-of-mass (rest) frame as 
$E_{B} = \sum_i^A E_i-\sum_i^A {M_N}_i +\sum_{i\neq j} V_{ij}$; where $E_i$ ($M_{Ni}$) is the energy (mass) of the $i$th nucleon of the cluster in the rest frame of the cluster.  For its calculation the energy and momentum of nucleons are boosted into this frame from the calculational $NN$ frame.
Even if there are no elastic collisions between one of the cluster nucleons and a third hadron the cluster binding energy can change its sign. This is due to the fact that for the propagation the forces between the nucleons are calculated at the same time in the computational frame. On the other hand, to calculate the binding energy in MST one has to take the positions of the baryons after Lorentz boost into the cluster center-of-mass. However, in this frame the baryons have different times, in principle should be corrected but can hardly be done in practice. The larger the $\gamma$ factor between the computational frame and the cluster center-of-mass frame, the more these time differences in the cluster center-of-mass system become important. 

In order to overcome this problem of the semi-classical approach, we recall that in our previous study~\cite{Glassel:2021rod} we calculated cluster observables at the ``physical time" $t$,  which accounts for the time dilatation between the cluster rest frame and the center-of mass system of heavy-ion reaction: $t=t_0\cdot \cosh{y_{cm}}$, where $t_0 $ is the cluster ``freeze-out" time at mid-rapidity and $y_{cm}$ is the rapidity of the center-of-mass of the cluster in the calculational frame, the center-of-mass system of the heavy-ion reaction.
We called $t$ the ``physical time" because it marks the identical times in the rest systems.
The time $t_0 $ was determined such that we could reproduce the total experimental multiplicities of the clusters at  mid-rapidity. Also we verified that the choice of $t$ affects only the multiplicity and neither the form of the rapidity distribution nor that of the transverse momentum distribution.

\subsubsection{{``}Advanced{''} MST}
\begin{figure}[ht]
\centering
\includegraphics[width=0.5\textwidth]{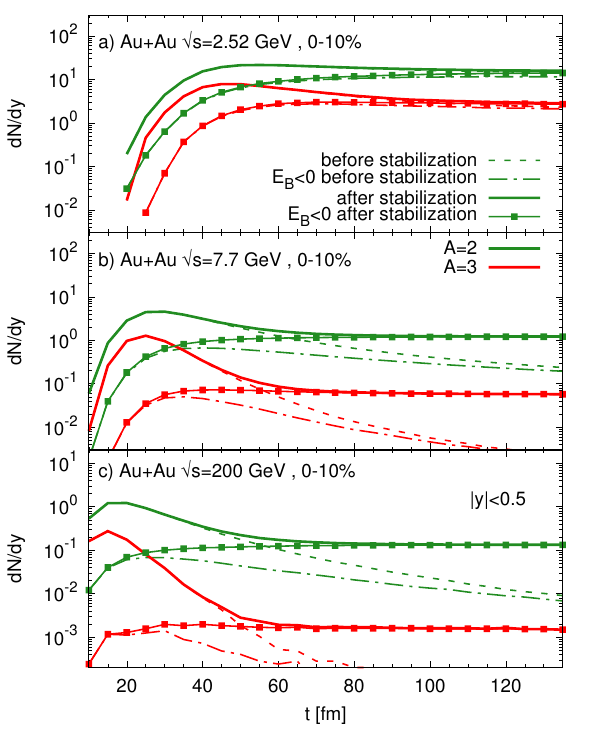}
\caption{\label{fig:AS} (color online) Multiplicity of $A=2$ and $A=3$ clusters at mid-rapidity, $|y|<0.5$, in PHQMD simulations of Au+Au central collisions at three different energies: a) $\sqrt{s}_{NN}=2.52$ GeV (up), b) $\sqrt{s}_{NN}=7.7$ GeV (middle), c) $\sqrt{s}_{NN}=200$ GeV (low). The dashed lines (green for $A=2$, red for $A=3$) are the results obtained with the standard MST, while the dash-dotted lines indicate the MST identified clusters which are bound, i.e. with $E_B<0$. The solid lines with same color coding are the results of the advanced MST (aMST), whose difference to MST is explained in the text. The solid lines with filled squares show the aMST bound clusters, i.e. with $E_B<0$.}  
\end{figure}

This numerical artifact can be surmounted by freezing the internal degrees of freedom of the cluster when it is not anymore in contact (neither by collisions nor by potential interactions) with fellow hadrons, which are not part of the cluster. This freezing can be applied to the ``collision history" file which contains the positions and momenta of the baryons as a function of time. Therefore, it does not influence the dynamics of the reaction. This so-called \textit{stabilization} procedure works as follows and the results are presented in Fig.~\ref{fig:AS}:
\begin{itemize}
\item[1)] Nucleons can be part of a cluster only after they have had their last elastic or inelastic collision. At each time step $t_i=t_0+i\cdot \Delta t$ the positions and momenta of all nucleons are recorded and clusters are identified by means of the MST algorithm. This is the standard MST method shown as dashed lines in Fig.~\ref{fig:AS}, for $A=2$ (green line) and $A=3$ (red).
    
\item [2)] Clusters have to have a negative binding energy $E_B<0$. Applying this selection on the clusters identified by MST after point 1), the result is shown by dash-dotted lines in Fig.~\ref{fig:AS}. Shortly after the collision starts until kinetic freeze-out, unbound nucleons with quite different momenta can be found at the same position in coordinate space. If time proceeds their trajectories diverge and they do not form a cluster anymore. Indeed, only if the clusters are bound, the nucleons are forced to stay together. Therefore, at late times each dashed line joins the corresponding dash-dotted line. 
    
\item[3)] PHQMD conserves energy strictly and the cluster nucleons are maximally separated from the other nucleons with a MST radius of 4 fm. Due to the non-relativistic Skyrme potential, the time shift between the nucleons in the cluster center-of-mass system (where the binding energy is calculated) and an approximation used to extrapolate the interaction density to the relativistic case \cite{Aichelin:1991xy}, 
it may happen that the sign of the binding energy $E_B$ (which is tiny as compared to the total energy of the cluster) changes from negative to positive between the time $t_i$ and the next time step $t_i+\Delta t$ (although the cluster is composed of the same nucleons) and the cluster starts to disintegrate. This disintegration is artificial and, therefore, we preserve the cluster by freezing the internal degrees of freedom.

\item[4)] Due to the semi-classical nature of our approach, it may happen
that a ``bound'' cluster $A>2$ with $E_B<0$ at time step $t_i$ spontaneously disintegrates between $t_i$ and $t_i+\Delta t$, because the available kinetic energy is given to one nucleon which then can leave the cluster. In a quantum system, where the energy of the ground state is larger than in a semi-classical system (because the wave function cannot have zero momentum), this process is much less probable. Therefore, we consider this evaporation as artificial and restore the cluster of the previous time step $t_i$. The result, if we include 3) and 4), is shown by the full lines in Fig.~\ref{fig:AS}. 
We see that at large times the fragment yield becomes stable. Due to the larger $\gamma$ factor the freeze-out of the internal cluster degrees of freedom is important at high beam energies. At SIS energies it is almost negligible.
\end{itemize}

In Fig.~\ref{fig:AS} the multiplicity of $A=2$ (green lines) and $A=3$ (red lines) clusters at mid-rapidity, $|y|<0.5$, from PHQMD simulations of central Au+Au collisions are shown as a function of time for three different collision energies; from upper to lower panel: a) $\sqrt{s}_{NN}=2.52$ GeV, b) $\sqrt{s}_{NN}=7.7$ GeV, c) $\sqrt{s}_{NN}=200$ GeV. The dashed lines correspond to the clusters identified by the original MST as in Ref.~\cite{Aichelin:2019tnk} according to the description ``1)" from above. The dash-dotted lines denote those clusters, which are effectively bound having a binding energy $E_B<0$, corresponding to case ``2)" from above. The solid lines show the cluster yield obtained with the advanced MST (aMST), i.e. employing after the MST identification the stabilization procedure according to the points ``3)" and ``4)". It is clearly visible that MST and aMST give the same cluster multiplicity at low beam energies (top panel) while at higher energies (center and bottom panel) - where the relativistic effects discussed above play a major role - aMST stabilizes the cluster multiplicity. Therefore, it is not anymore necessary to define a time at which the cluster analysis is performed. This represents a remarkable improvement of our previous study in Ref.~\cite{Glassel:2021rod} where we still had to select such a time. This procedure can be applied to any type of cluster of any size, including light nuclei and hypernuclei. In this study we present only the results for deuterons. The study of hypernuclei and heavy clusters we reserve for a future publication.

\section{Kinetic approach for deuteron reactions}\label{sec:sec3}

\subsection*{Collision Integral}
As described, the collision processes involving the formation and the breakup of a deuteron are implemented in PHQMD by means of a covariant rate formalism which has been firstly developed in Ref.~\cite{Cassing:2001ds} and applied within the PHSD microscopic approach in order to study the baryon-antibaryon production by multi-meson fusion~\cite{Seifert:2017oyb,Seifert:2018sbg}. Following the steps of Ref.~\cite{Cassing:2001ds}, we start by writing the covariant Boltzmann transport equation for the single phase-space distribution function of an on-shell hadron $f_i(x,p)$  
\begin{equation}
\label{vlasov}
p_\mu \partial^\mu_x  f_i(x,p)  = I_{coll}^i \, ,
\end{equation}
using the notation $x=(t,\vec{x})$ and $p=(E,\vec{p})$ with the on-shell condition $p^2=m_i^2$ ($m_i$ is the rest mass).
The left hand side of Eq.~\eqref{vlasov} contains only the free streaming term because we neglect for simplicity any mean-field interaction. On the right hand side the collision integral $I_{coll}^i$ encodes all the multi-particle transitions which involve the hadron \textit{i} either in the initial or in the final state. Hence, $I_{coll}^i$ can be written as a sum over all scattering processes with increasing number of participant particles
\begin{equation}
\label{icoll}
I_{coll}^i = \sum_n \sum_m \ I^i_{coll} [n \leftrightarrow m] .
\end{equation}
Each collision term $I^i_{coll} [n \leftrightarrow m]$ accounts for a particular forward scattering process ($\rightarrow$) with $n$-particles in the initial state and $m$-particles in the final state, as well as for the corresponding backward reaction ($\leftarrow$). The forward and backward reactions can be grouped together and in collision theory one usually distinguishes a gain and loss term. Therefore, the on-shell collision term $I^i_{coll} [n \leftrightarrow m]$ becomes
\begin{eqnarray}
\label{icollnm}
&& I^i_{coll} [n \leftrightarrow m]  
= \frac{1}{2} \frac{1}{N_{id}!} \sum_\nu \sum_{\lambda} \left( \frac{1}{(2 \pi)^{3}} \right)^{n+m-1} \times \nonumber \\
&& \left ( \prod_{j=2}^n \! \int \! \frac{d^{3} \vec{p}_j}{2E_j} \right ) 
\left ( \prod_{k=n+1}^{n+m} \! \int \! \frac{d^{3} \vec{p}_k}{2E_k} \right ) \times   \nonumber \\
&& (2\pi)^4  \! \delta^{4}(p_1^{\mu} + \sum_{j=2}^n p^{\mu}_j - \sum_{k=1}^{n+m} p_k^{\mu}) 
 W_{n,m}(p_1, p_j; i,\nu \mid p_k; \lambda) \nonumber \\
&& \times \left[  \prod_{k=n+1}^{n+m} f_k(x,p_k)  -  f_i(x,p_1) \prod_{j=2}^n f_j(x,p_j) \right] .
\end{eqnarray}
In Eq.~\eqref{icollnm} there are $n+m-1$ integrals over the initial $\vec{p}_j$ and final $\vec{p}_k$ momenta of all particles, excluding the tagged hadron \textit{i} (the deuteron) whose momentum is denoted as $p_1$.

The quantity $W_{n,m}(p_1, p_j; i,\nu \mid p_k; \lambda)$ is called \textit{transition amplitude} and is related to the square of the scattering matrix for the transition $p_1+\sum_{j=2}^n p_j \rightarrow \sum_{k=n+1}^{n+m} p_k $ where $\nu$ and $\lambda$ denote a particular set of allowed discrete quantum numbers for the particles (except the hadron \textit{i}) in the initial and final states. The $\delta$ function guarantees the energy and momentum conservation in each individual reaction. Finally, the single-particle distribution functions of the hadrons appear, in particular the functions $f_1$ and $f_j$ for the forward/loss term $n \rightarrow m$ and the function $f_k$ for the backward/gain term $n \leftarrow m$.  
We assign the arbitrary $\pm$ sign to distinguish between the gain and the loss term such that in our study the reaction which leads to the production of a deuteron (playing the role of the tagged hadron \textit{i}) is associated to the backward/gain term, while the inverse reaction where the deuteron is destroyed, is associated to the forward/loss term. 
This choice agrees with the original formulation in Ref.~\cite{Cassing:2001ds}.
In the collision integral of Eq.~\eqref{icollnm} we have also neglected the quantum statistical corrections, i.e. the Pauli-blocking or Bose-enhancement factors, which multiply the product of the phase-space distribution functions, as well as any anti-symmetrization procedure in the transition amplitude for the fermions involved in the reactions. Only the statistical factor $1/N_{id}!$, which counts the number of identical particles (either bosons or fermions) in the initial and final states, survives. 
The expression Eq.~\eqref{icollnm} can be straightforwardly generalized to the off-shell case by including an additional integration over the energy of each single hadron weighted by the associated spectral function (cf. Ref.~\cite{Cassing:2001ds}). In PHQMD/PHSD such an off-shell version of the collision integral is adopted for the dynamical interaction of other baryons and mesons which are propagated with self-consistent off-shell transport equations.

The covariant collision number for the $n(i) \rightarrow m$ scattering process is the number of forward reaction events in the covariant 4-volume $d^4 x$ = $dV dt$ and, therefore, the \textit{covariant collision rate} is obtained by dividing the loss term in Eq.\eqref{icollnm} by the energy $E_1$, followed by the integration over the momentum $d^3\vec{p}_1/(2\pi)^3$ and and a summation over the quantum numbers $\tau(i)$ of the tagged hadron \textit{i} in the initial state of the transition,
\begin{eqnarray}
\label{icollmc} 
\frac{d N_{coll} [n(i) \rightarrow m]}{dt dV}  = \frac{1}{N_{id}!}
\sum_{\tau(i),\nu} \sum_{\lambda} \left( \frac{1}{(2 \pi)^{3}} \right)^{n} \times \nonumber \\
\int \! \left( \prod_{j=1}^n \!  \frac{d^{3} p_j}{2 E_j}  f_j(x,p_j) \right) \int \! \left( \prod_{k=n+1}^{n+m} \!  \frac{1}{(2\pi)^3} \frac{d^{3} p_{k}}{2E_k} \right)  \times \nonumber  \\
(2\pi)^4  \! \delta^{4}( \sum_{j=1}^n p^{\mu}_j - \sum_{k=n+1}^{n+m} p_k^{\mu}) W_{n,m}( p_j; \tau(i),\nu \mid p_k; \lambda) .  
\end{eqnarray}
Similarly, the covariant collision rate of the backward reaction $n(i) \leftarrow m$ is obtained from the gain term of Eq.~\eqref{icollnm}
\begin{eqnarray}
\label{icollmd} 
\frac{d N_{coll} [m \rightarrow n(i)]}{dt dV}  = \frac{1}{N_{id}!}
\sum_{\tau(i),\nu} \sum_{\lambda} \left( \frac{1}{(2 \pi)^{3}} \right)^{m}  \times \nonumber \\
\int \! \left( \prod_{k=n+1}^{n+m} \! \frac{d^{3} p_k}{2 E_k}  f_k(x,p_k) \right) \int \! \left( \prod_{j=1}^n \! \frac{1}{(2\pi)^3} \frac{d^{3} p_{j}}{2E_j} \right) \times \nonumber  \\
(2\pi)^4 \delta^{4} \! ( \sum_{j=1}^n p^{\mu}_j - \sum_{k=n+1}^{n+m} p_k^{\mu}) W_{n,m}( p_j; \tau(i),\nu \mid p_k; \lambda) . 
\end{eqnarray}
The transition amplitude $W_{n,m}$ in Eq.~\eqref{icollmc} and Eq.~\eqref{icollmd} is the same because of the equivalence of the scattering matrix under the detailed balance condition for forward and backward processes. 
Therefore, both expressions can be analytically or numerically solved knowing the dependence of the transition amplitude $W_{n,m}$ on the kinematic variables. This has been suggested in Ref.~\cite{Cassing:2001ds} where, in particular, it has been shown that the collision probabilities of forward and backward reactions can be written in terms of the corresponding many-body phase-space integrals (cf. Appendix C) if one assumes that the transition amplitude $W_{n,m}$ is a function only of the invariant energy of the collision $\sqrt{s}=(\sum_{j=1}^n p_j)^2=(\sum_{k=n+1}^{n+m} p_k)^2$. We apply this procedure for deuteron reactions.

The goal of this work is to implement in the PHQMD transport approach the following deuteron reactions:
\textit{i)} the elastic $d \pi \rightarrow d \pi$ and $d N \rightarrow d N$ reactions, as well as  $2\leftrightarrow 2$ inelastic $d \pi \leftrightarrow NN$ reactions;
\textit{ii)} the $2\leftrightarrow 3$ inelastic reactions $d \pi \leftrightarrow N N \pi$ and $d N \leftrightarrow N N N$ with all pion species $\pi=\pi^+,\pi^0,\pi^-$ and $N=p,n$.

Employing the covariant expressions in Eq.~\eqref{icollmc} and \eqref{icollmd} with $n=m=2$ and $i=d$, the collision rate for the elastic and inelastic $d\pi \leftrightarrow NN$ reactions can be written as follows,
\begin{equation}
 \label{icoll22}
\frac{d N_{coll} [ 2(d) \leftrightarrow 2 ]}{dt dV}  = \int \!
 \left( \prod_{j=1}^2 \! \frac{d^3 p_j}{(2\pi)^3} f_j(x,p_j) \right) v_{rel}  \sigma^{2,2}_{tot} \, , 
\end{equation} 
where $\sigma^{2,2}_{tot} $ is the total cross section for a two-to-two scattering process which is defined from the $W_{2,2}$ transition amplitude by the well known definition 
\begin{eqnarray}
\label{icross22} 
\sigma^{2,2}_{tot}(\sqrt{s}) = \frac{1}{4 I_{flux}} \sum \!\! \int \! \frac{d^3 p_3}{(2\pi)^3 2 E_3}  \int \! \frac{d^3 p_4}{(2\pi)^3 2 E_4} \nonumber \\ 
W_{2,2}(\sqrt{s})(2\pi)^4 \! \delta^4(p_1+p_2 - p_3 - p_4) \, ,
\end{eqnarray}
with the flux factor $I_{flux}$ related the (relativistic) relative velocity $v_{rel}$ of the incident on-shell particles of masses $m_1$ and $m_2$
\begin{equation}
I_{flux} = \sqrt{(p_1 \cdot p_2)^2 - m_1^2 m_2^2} = E_1 E_2 v_{rel} .
\end{equation}
In Eq.~\eqref{icross22} the sum is performed over the quantum numbers involved in the reaction and it includes also the statistical factor $1/N_{id}!$ which we absorb in the cross section.

The PHQMD/PHSD collision integral for the deuteron reactions is solved numerically by dividing the coordinate space in a grid of cells of volume $\Delta V_{cell}$ and sampling the on-shell single-particle distribution function $f(x,p)$ at each time step $\Delta t$ by means of the test-particle ansatz~\cite{Cassing:2008nn} 
\begin{equation}
\label{testp} 
f(x,p) = \frac{(2\pi)^3}{\Delta V_{cell}} \sum_{j=1}^{N_{test}} \delta(\vec{r}_j(t)-\vec{x}) \delta(\vec{p}_j(t)-\vec{p}) \, ,
\end{equation}
where $\vec{r}_j(t)$ and $\vec{p}_j(t)$ are, respectively, the position and the momentum of the particle $j$ at time $t$.
By inserting Eq.~\eqref{testp} in Eq.~\eqref{icoll22} we obtain the collision probability for the $2\leftrightarrow 2$ reactions in the unit volume $\Delta V_{cell}$ and unit time $\Delta t$
\begin{equation}
\label{Prob22}
 P_{2,2} \left(\sqrt{s} \right) = \sigma^{2,2}_{tot} v_{rel} \frac{\Delta t}{\Delta V_{cell}} \, ,
\end{equation}
which depends on $\sqrt{s}$ through $v_{rel}$ and $\sigma^{2,2}_{tot}$. 
Employing sufficiently small values of $\Delta V_{cell}$ and $\Delta t$ we solve numerically the $2 \leftrightarrow 2$ collisions for the deuterons by the stochastic method, i.e. by calculating the invariant energy $\sqrt{s}$ of each possible reaction and then the associated probability $P_{2,2}$ which is confronted with a random number between 0 and 1. To calculate $P_{2,2}$ for the inelastic $d\pi \leftrightarrow NN$ process we use parametrized expressions for the total cross section $\sigma^{2,2}_{tot}$ which are reported in the Appendix A.

Now we describe the inelastic reactions $d \pi \leftrightarrow N N \pi$ and $d N \leftrightarrow N N N$ and, in particular, how the backward reaction $2 \leftarrow 3$ can be fully implemented within the covariant rate formalism adopted in our PHQMD approach. 
On the one hand, this is physically motivated by the fact that these are the dominant reactions for the production of deuterons in HICs due to their large cross sections, $\sigma_{tot} \simeq 200 \, mb$, compared to the sub-dominant channel $NN \rightarrow d \pi$ with $\sigma_{tot} \simeq 10 \, mb$.
On the other hand, it provides an effective method to describe reactions with more than 2 particles in the entrance channel, which cannot be formulated in terms of cross sections as in Eq.~\eqref{icross22}.

For the forward reaction, the breakup of deuterons by an incident $N$ or $\pi$, the definition of the covariant collision rate follows straightforwardly and is given by Eq.~\eqref{icollmc} with $n=2$, $m=3$ and $i=d$,
\begin{equation}
\label{icoll23} 
\frac{d N_{coll} [ 2(d) \rightarrow 3]}{dt dV}  = \int \!
 \left( \prod_{j=1}^2 \! \frac{d^3 p_j}{(2\pi)^3} f_j(x,p_j) \right) \sigma^{2,3}_{tot} v_{rel} \, ,
\end{equation}
where $\sigma^{2,3}_{tot}$ is the total inelastic cross section for either the $d \pi \rightarrow N N \pi$ or the $d N \rightarrow N N N$ scattering process, which is defined similarly to Eq.~\eqref{icross22} with an extra integration over the momentum of the third particle in the final state. The sum over the internal quantum numbers appearing in Eq.~\eqref{icoll23} is also absorbed in $\sigma^{2,3}_{tot}$. In Appendix A we provide the parametrized expressions of such inelastic cross sections as a function of $\sqrt{s}$ and we describe in detail how they are obtained from the experimental measurements of the total inclusive cross section for $d\pi$ and $dN$ collisions. Employing again the test-particle ansatz in Eq.~\eqref{icoll23}, we derive the collision probability for the forward reaction
\begin{equation}
 \label{Prob23}
 P_{2,3} = \sigma_{tot}^{2,3} v_{rel} \frac{\Delta t}{\Delta V_{cell}} \, ,
 \end{equation}
which is a function of $\sqrt{s}$, and we sample stochastically the collisions in the unit volume $\Delta V_{cell}$ and the unit time $\Delta t$ for each PHQMD/PHSD parallel ensemble event. When a collision occurs, we construct the final state of three particles in the center-of-mass of the incident pair by means of standard kinematic routines~\cite{Kajantie:1971rj,Byckling:1971vca}.

The covariant rate for the backward $N N \pi \rightarrow d \pi$ and $N N N \rightarrow d N$ reactions follows from Eq.~\eqref{icollmd}, but we cannot write it in terms of a cross section. However, what is important for us is to obtain a collision probability $P_{3,2}$ in order to apply the stochastic method.  
With the assumption $W_{3,2}=W_{2,3}=W(\sqrt{s})$ \cite{Cassing:2001ds} the transition amplitude can be moved outside the integration over the momenta of the two particles in the final state. As a result, these integrations can be combined with the $\delta$ function into the two-body phase-space $R_2(\sqrt{s},m_1,m_2)$ (cf. Appendix C), so that we can write as intermediate step,  
\begin{eqnarray}
\label{icoll32a} 
\frac{d N_{coll} [ 3 \rightarrow 2(d)]}{dt dV}  =  \int \!
 \left( \prod_{k=3}^5 \! \frac{d^3 p_k}{(2\pi)^3 2 E_k} f_j(x,p_k) \right) \, ,  \nonumber \\
 \sum W(\sqrt{s}) R_2(\sqrt{s},m_1,m_2)
\end{eqnarray}
with the sum running over the quantum numbers and taking into account also the statistical factor for identical particles. Next, we introduce the $\sigma^{2,3}_{tot}$ of the forward process by inverting its definition from the transition amplitude and use again the condition $W(\sqrt{s})$ to isolate the three-body phase-space $R_3(\sqrt{s},m_3,m_4,m_5)$ of the initial particles. Hence,
\begin{equation}
\label{icoll32b} 
\sum W(\sqrt{s}) = F_{spin} F_{iso} \frac{4 E_1^f E_2^f \sigma^{2,3}_{tot}}{R_3(\sqrt{s},m_3,m_4,m_5)} \, ,
\end{equation}

where $F_{spin}$ and $F_{iso}$ denote the factors coming from the sum over the spin and isospin quantum numbers in the transition matrix. For the spin contribution we have
\begin{equation}
F_{spin} = \left( \frac{g_1^f g_2^f}{g_3 g_4 g_5}  \right) \, ,
\end{equation}
where in $NN\pi(N) \rightarrow d\pi(N)$ reactions the particles are ordered as follows,
\begin{equation}
g_1^f = g_d \, , \, g_2^f=g_5=g_{\pi(N)} \, , \, g_3=g_4=g_N \, .      
\end{equation}
For the isospin part a separate discussion is given at the end of the section.

Combining Eq.~\eqref{icoll32a} and Eq.~\eqref{icoll32b} with the test-particle ansatz, we finally obtain the collision probability for the backward $3 \rightarrow 2$ process in the unit volume $\Delta V_{cell}$ and the unit time $\Delta t$,
\begin{eqnarray} 
\label{Prob32}
 && P_{3,2} = F_{spin} F_{iso} \frac{E_1^f E_2^f}{2 E_3 E_4 E_5} \frac{P^{2,3}}{\Delta V_{cell}}\frac{R_2(\sqrt{s},m_1,m_2)}{R_3(\sqrt{s},m_3,m_4,m_5)} \nonumber \\
  &&= F_{spin} F_{iso} \frac{E_1^f E_2^f}{2 E_3 E_4 E_5}  \frac{ \sigma_{tot}^{2,3} v_{rel} \Delta t}{\Delta V_{cell}^2} \frac{R_2(\sqrt{s},m_1,m_2)}{R_3(\sqrt{s},m_3,m_4,m_5)} \, , \nonumber \\ 
\end{eqnarray}
where in the second line we employ the collision probability for the forward $2 \rightarrow 3$ reaction from Eq.~\eqref{Prob32}. We notice that on the right hand side of Eq.~\eqref{icoll32b} and ~\eqref{Prob32} the energy of the produced particles $E_1^f$ and $E_2^f$ appear. That means that in our numerical implementation we have to sample the possible kinematics of the final state before the collisions take place. If the collision occurs, we reconstruct the kinematics of the emitted particles in the center-of-mass of the three interacting initial particles according to our previous sampling.
In this sense, we can implement the forward and the backward reactions consistently within the same stochastic model. 

In Ref.~\cite{Oliinychenko:2018ugs} the deuteron reactions $\pi pn \rightarrow \pi d$ and $N pn \rightarrow N d$ have been implemented in the SMASH transport approach for relativistic HICs. To do this numerically a fictitious $d'$ resonance was introduced and the $3 \rightarrow 2$ process was divided into a two 2-to-2 steps. 

Later on, in Ref.~\cite{Staudenmaier:2021lrg} the same multi-particle production mechanisms have been described according to the covariant rate formalism of Ref.~\cite{Cassing:2001ds}. In particular, Eq.~(6) of Ref.~\cite{Staudenmaier:2021lrg} shows the same probability for the stochastic treatment of the 3-to-2 process as the one we have just derived in Eq.~\eqref{Prob32}. Comparing both expression we can make some comments:\\
\textit{i)} The two- and three-body phase-spaces $R_2$ and $R_3$ appear in both equations as a function of $\sqrt{s}$ and particle masses. For $R_2$ we employ the well known analytic expression, while for $R_3$ we adopt the parametrization of Ref.~\cite{Seifert:2018sbg}. We collect all formula in Appendix D. In Ref.~\cite{Staudenmaier:2021lrg} it is done similarly, so we do not expect any discrepancy due to this part.\\
\textit{ii)} The probability is proportional to the total cross section for the inverse 2-to-3 process. For deuteron disintegration into 3 particles by $\pi$ and $N$ scattering  we employ a parametrization of the cross section as a function of $\sqrt{s}$, as reported in Appendix A, which is fitted to the available experimental data in the peak region. At high $\sqrt{s}$ we let our cross section to tend to zero because the 2-to-3 phase-space closes and other inelastic processes with final particles $m>3$ open. In Ref.~\cite{Staudenmaier:2021lrg} cross sections from Ref.~\cite{Oliinychenko:2018ugs} are used which differ only for a constant behavior at high $\sqrt{s}$. We investigated the possible difference arising from the different asymptotic behavior of the cross sections and we did not find any impact on deuteron yields and the $p_T$-spectra.\\
\textit{iii)} Our spin factor $F_{spin}$ is the same as the one in Ref.~\cite{Staudenmaier:2021lrg}, while the isospin coefficient $F_{iso}$ does not appear there. This represents the novelty of our work.

In Ref.~\cite{Staudenmaier:2021lrg} as from Ref.~\cite{Oliinychenko:2018ugs} the $\pi$-catalysis is considered only for the channel where there is no charge difference between the initial and the final pion, i.e. $\pi d \leftrightarrow \pi p n$. We extend the deuteron production to all possible $\pi  NN$ channels which fulfill the conservation of total isospin. We list all implemented channels in the table \ref{Tab:deut3to2}.

  \begin{table}[h]
\centering
\begin{tabular}{|c|c|c|}
\hline
$ (i) \quad \pi+N+N$  & $ (f) \quad d+\pi$ & $Q_{tot}$   \\
\hline \hline
$n+n+\pi^-$ & $\emptyset$ & -1   \\
\hline
$n+n+\pi^0$ & $d+\pi^-$ & 0  \\
$p+n+\pi^-$ & $d+\pi^-$  & 0  \\
\hline
$n+n+\pi^+$ & $d+\pi^0$ & 1   \\
$p+n+\pi^0$ & $d+\pi^0$ & 1   \\
$p+p+\pi^-$ & $d+\pi^0$ & 1   \\
\hline
$p+p+\pi^0$ & $d+\pi^+$ & 2   \\
$p+n+\pi^+$ & $d+\pi^+$ & 2   \\
\hline
$p+p+\pi^+$ & $\emptyset$ & 3  \\
\hline
\end{tabular}
\caption{Reactions for deuteron production by $\pi$-catalysis implemented in the PHQMD collision integral. In the first column the initial $\pi NN$ states, which are allowed to form the final $d \pi$ state in the second column, are collected by increasing total electric charge $Q_{tot}$  written in the third column. When the final state is a $\emptyset$ it means that deuteron production is not possible for the specific $\pi NN$ state. The probability for each transition depends on the isospin factors $F_{iso}$ which are calculated in Appendix C.}
\label{Tab:deut3to2}
\end{table}

Since the deuteron has isospin zero, the state $\pi d$ is a state with defined isospin 1 provided by the pion. In general, a three particle state $\pi NN$ has not a definite value of total isospin (i.e. it is not an eigenstate of this quantum number), rather it is formed by a superposition of eigenstates.
Therefore, for each channel of the table the $F_{iso}$ represents the projection of the state $\pi NN$ on the isospin 1 state.
We perform the calculation in detail in Appendix C. For the inverse reaction $\pi d \rightarrow \pi NN$ the initial state has total isospin 1. The total cross section $\sigma_{tot}^{2,3}$ describes the reaction of $d \pi$ to any of the possible $\pi NN$ channels. In order to correctly evaluate the disintegration reaction, we weight the transition to one specific channel with the corresponding isospin factor calculated in Appendix C. 
Similarly we calculate the isospin factors $F_{iso}$ for the $N$-catalysis, where in this case there are no multiple channels available, i.e. $NNN \leftrightarrow dN$ does not account for other channels with respect to $Npn \leftrightarrow dN$.

\section{Box validation and analytic results}\label{sec:sec4} 

We first study deuteron reactions in the static ``box" framework where we can compare the results from our stochastic multi-particle approach with expectations from so-called rate equations.  Rate equations differ from the transport approach because they involve the solution of chemical rates of a kinetically equilibrated gas. They have been used for example in Ref.~\cite{Pan:2014caa} to study the dynamical evolution of baryon-antibaryon annihilation and regeneration by solving fugacity equations or in Ref.~\cite{Neidig:2021bal} where the time evolution of light cluster abundancies has been investigated in an expanding medium. In this sense they represent an alternative approach to the covariant rate formalism of Refs. \cite{Cassing:2001ds,Seifert:2017oyb}.

In a static medium at equilibrium with temperature $T$ the rate equations can be taken as analytic reference to verify the correct implementation of the numerical collision criteria.  
Here we follow a one-by-one comparison with Section B of the work done by the SMASH group in Ref.~\cite{Staudenmaier:2021lrg} and check the agreement of our results. 

As a model we consider the $\pi$-catalysis reaction with no isospin factors. Using the same notation of Ref.~\cite{Staudenmaier:2021lrg} we introduce the fugacities $\lambda_i(t)$ for the particle species involved in $\pi d \leftrightarrow \pi pn$ reactions. Without isospin factor the initial and final pion have the same charge. Therefore, we can see immediately that the number of pions remains constant. Hence, we can write the system of rate equations for $d$ and $N=p,n$ as follows

\begin{equation}\label{eq:RateEq0}
\begin{cases}
 \dot{\lambda}_d = \sum \! <v_{rel} \sigma_{\pi d}> \!\left( \frac{g_d g_{\pi}}{g_N^2 g_{\pi}}\lambda_N^2 - \lambda_d \right) n_{\pi}^{eq}\lambda_{\pi} \\
   \dot{\lambda}_N = - \sum \! <v_{rel} \sigma_{\pi d}> \!\left( \frac{g_d g_{\pi}}{g_N^2 g_{\pi}}\lambda_N^2 - \lambda_d \right) n_{\pi}^{eq}\lambda_{\pi} \\
\dot{\lambda_{\pi}} = 0
\end{cases}
\end{equation}
in units of fm$^{-1}$ and denoting the time derivative $\frac{d\lambda_i}{dt}$ as $\dot{\lambda}_i$. In Eq.~\eqref{eq:RateEq1} the sum runs over all pions which are initialized in the system according to an equilibrium density at given temperature $T$ times a constant fugacity 
\begin{equation}
\rho_{\pi} = \lambda_{\pi} n_{\pi}^{eq}(T) = \lambda_{\pi} g_{\pi} \frac{m_{\pi}^2 T}{2\pi^2} K_2 \left( \frac{m_{\pi}}{T} \right) .
\end{equation}
The factors $g_i$ are the spin degenerancies with values \begin{equation}
g_d/3 = g_N/2 = g_{\pi} = 1 \, .
\end{equation} 
Finally, $\sigma_{\pi d}$ is the cross section for $2 \rightarrow 3$ deuteron breakup by an incident pion reported in Appendix A and the thermal average $<v_{rel} \sigma>$ is calculated using the formula 
\begin{align}\label{eq:RateEq1}
&\langle v_{rel}  \sigma_{ij} \rangle = \frac{1}{4 m_i^2 m_j^2 T K_2(\frac{m_i}{T}) K_2(\frac{m_j}{T})} \times \\
& \int_{m_i+m_j}^{\infty} \!\!\! d \sqrt{s} \left[ (s-m_i^2-m_j^2)^2 - 4m_i^2m_j^2 \right] K_1\left(\frac{\sqrt{s}}{T}\right) \sigma_{ij} \, , \nonumber
\end{align}
which generalizes the expression in Ref.~\cite{Cannoni:2013bza} for different particle masses.
The time evolution of the nucleon and deuteron density can be directly calculated from the fugacities by 
$$\rho_i(t) = n_i^{eq}(T) \lambda_i(t),$$ 
where $n_i^{eq}(T)$ are the densities at equilibrium at temperature $T$. We set as initial values $\rho_N(0) = 2\rho_p(0) = 2\rho_n(0) =$~0.12 fm$^{-3}$ and $\rho_d(0)=0$. Provided with these initial conditions Eq.~\eqref{eq:RateEq0} is a system of coupled first order ordinary differential equations (ODEs) which can be solved applying Runge-Kutta methods.

\begin{figure}[ht]
\centering
\includegraphics[width=0.45\textwidth]{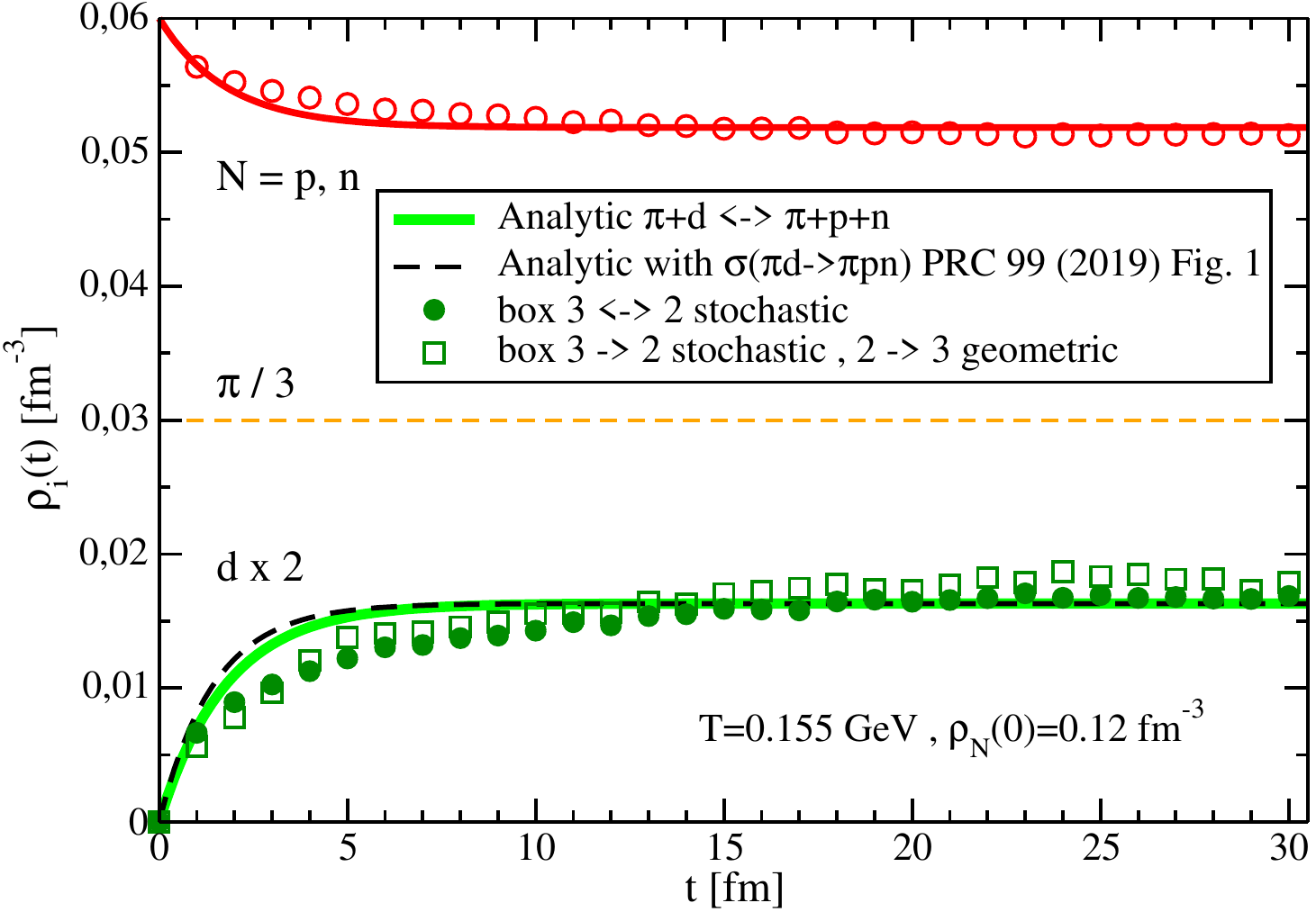}
\caption{\label{fig:B0} (color online) Time evolution of particle densities from box simulations for $d\pi \leftrightarrow pn\pi$ reactions compared to solutions of rate equations (see text). 
The box is initialised with temperature $T=0.155$ GeV, equal densities of protons, neutrons $\rho_N(0) = 2\rho_p(0) = 2\rho_n(0) =$~0.12 fm$^{-3}$ and pion density $\rho_\pi(0)=$~0.09 fm$^{-3}$ for the 3 isospin states. The initial density of deuterons is set to zero, i.e. $\rho_d(0)=0$.} 
\end{figure} 

We prepare a cubic box of volume $V=$~(10 fm)$^3$ in which particles are initially distributed uniformly in coordinate space with a density $\rho_N(0)=0.12$~fm$^{-3}$ for nucleons and a density $\rho_{\pi}(0)=0.09$~fm$^{-3}$ for pions and in momentum space according to a Boltzmann distribution with temperature $T$. Then, we divide the box volume into unit cells $\Delta V_{cell}=$~(2.5 fm)$^3$ where deuteron reactions are sampled numerically at each time step $\Delta t=0.2$~fm. We set the parameters $\Delta V_{cell}$ and $\Delta t$ in order to fulfill the main conditions of the stochastic method~\cite{LANG1993391,Xu:2004mz}. In particular, in each unit cell there are sufficient particles to perform $2 \leftrightarrow 3$ collisions with probabilities Eq.~\eqref{Prob23} and Eq.~\eqref{Prob32} which must be always smaller than unity.

In Fig.~\ref{fig:B0} we show the evolution of particles densities $\rho_i(t)=N_i(t)/V$ as function of time in a static medium at temperature $T=0.155$ GeV due to the reactions $d\pi \leftrightarrow pn\pi$. The labels and the colors in the plot identify the different particles species: red for nucleons $N=p,n$, green for deuterons and orange for pions.
The solid lines are the solutions from the system of Eq.~\eqref{eq:RateEq0} for nucleons and deuterons using our parametrized form for the $\sigma_{\pi d}^{2,3}$ cross section plotted in Appendix A. The dashed black line is the expectation value for the deuterons derived with the same rate equations, but employing the parametrized cross section taken from the SMASH study~\cite{Oliinychenko:2018ugs}. 
The symbols represent the results obtained from box simulations. In particular, for deuterons we show two cases: \textit{i)} the case where the $2 \leftrightarrow 3$ reactions are solved numerically by means of the multi-particle stochastic approach in both directions (filled circles); \textit{ii)} the second case where the forward $2 \rightarrow 3$ channel is perfomed by means of the geometric criterium where the deuteron collides with a pion and it is disintegrated into final $\pi p n$ system if it fulfills the condition
\begin{equation}
d_T < b_{max} = \sqrt{\frac{\sigma_{\pi d}}{\pi}} .
\end{equation}
Here $d_T$ is the distance of closest approach as defined in Ref.~\cite{Kodama:1983yk} (see also Ref.~\cite{Hirano:2012kj}). The geometric criterium is used to describe many reactions in the original PHSD collision integral, which is also employed within PHQMD. 
As follows from Fig.~\ref{fig:B0}, both methods - stochastic and geometric criterium - give the same equilibrium values for $2\to 3$ reactions.
We note that more details of the box simulations for the other deuteron reactions, i.e. $Nd \leftrightarrow pnN$ and $\pi d \leftrightarrow NN$, are reported in Appendix B.

\begin{figure}[th]
\centering
\includegraphics[width=0.48\textwidth]{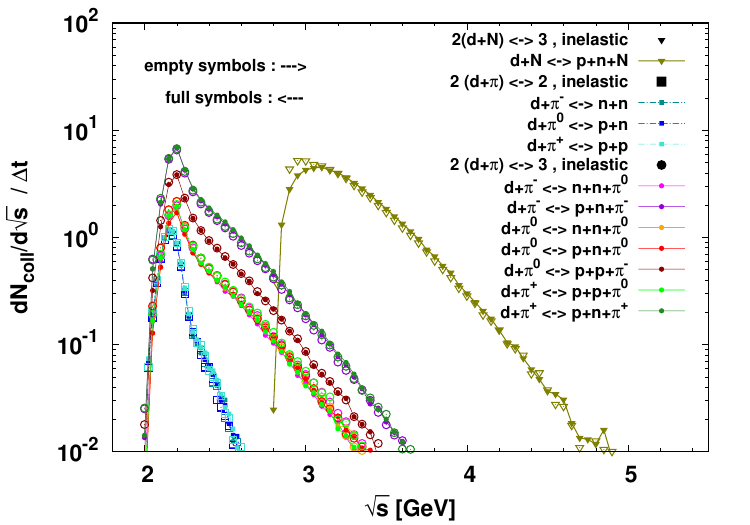}
\caption{\label{fig:B3} (color online) The differential collision rate as a function of the invariant center-of-mass energy $\sqrt{s}$ is shown for all deuteron reactions implemented in PHQMD. The forward direction, i.e. deuteron disintegration, for each channel is represented by an empty symbol, while the backward direction, i.e. deuteron production, is represented by the corresponding full symbol. The various processes are shown using different symbols and line styles: inelastic $N d \leftrightarrow pnN$  (upside down triangles and solid lines); inelastic $\pi d \leftrightarrow NN$ (squares with dash-dotted lines); inelastic $\pi d \leftrightarrow NN\pi$ (circles with solid lines). The different isospin channels are displayed using different colors.}   
\end{figure}

In Fig.~\ref{fig:B3} the detailed balance condition is verified by checking the differential collision rate
as a function of the invariant energy $\sqrt{s}$ for each implemented scattering process:  inelastic $N d \leftrightarrow pnN$ (upside down triangles and solid lines); inelastic $\pi d \leftrightarrow NN$ (squares with dash-dotted lines); inelastic $\pi d \leftrightarrow NN\pi$. The symbols, lines and colors for each channel are described in the figure legend.
As follows from Fig.~\ref{fig:B3}, the reaction rate for the forward direction is equal to the reaction rate of the backward direction for all channels. Thus, detailed balance is fulfilled in our calculations for all isospin channels.
The static box calculations show that:\\
\textit{i)} the numerically computed densities of protons and deuterons are in a good agreement with analytical results for the equilibrium values;
\textit{ii)} the stochastic and geometrical methods for $2 \to 3$ reactions agree;
\textit{iii)} at equilibrium  the detailed balance is verified for $2 \leftrightarrow 3$ and $2 \leftrightarrow 2$ reactions.
 This ensures the validity of our implementation of the $2 \leftrightarrow 3$ and $2 \leftrightarrow 2$ reactions for deuteron production and absorption within the static box study. We note also that our box results agree with the SMASH calculations~\cite{Staudenmaier:2021lrg} when considering the same isospin reaction channel with the same cross section. 
After the box tests all deuteron reactions are implemented in the PHQMD framework. In particular, the deuteron production by $\pi N N \rightarrow \pi d$ and $N N N \rightarrow N d$ reactions are sampled stochastically within each PHQMD/PHSD parallel ensemble, while the inverse processes $\pi d \rightarrow \pi N N$, $N d \rightarrow N N N$ and the sub-dominant $NN \leftrightarrow d \pi$ and elastic reactions are performed by means of the geometric criterium described above, in order to speed up the computations.
In contrast to the ``box" model, for the stochastic method in realistic HICs with PHQMD we simulate the phase-space evolution of the fireball on an expanding 3D-grid which we divide into cells of volume $\Delta V_{cell}=\Delta x \, \Delta y \, \Delta z $ where $\Delta x=\Delta y=2.5$ fm and $\Delta z= 2.5/ \gamma_{cm}$ fm and the longitudinal expansion of the fireball is taken into account through the factor $\gamma_{cm}^{-1}=\sqrt{1-v_{cm}^2}$, where $v_{cm}$ is the velocity of a projectile or target in the cm frame. Inside each cell there are sufficient particles to sample stochastically the deuteron reactions at each time-step $\Delta t$.
Moreover, the time-step $\Delta t$ is initially increasing with time as $\Delta t \sim 1/\gamma_{cm}$ in order to let particles in each cell evolve smoothly at the beginning of the nucleus-nucleus collision. However, we employ the condition $\Delta t \le 0.1$ fm/c at later times in order to keep the collision probability below unity.

\section{Kinetic deuterons in HICs}\label{sec:sec5}

\begin{figure*}[t]
\centering
\includegraphics[width=\textwidth]{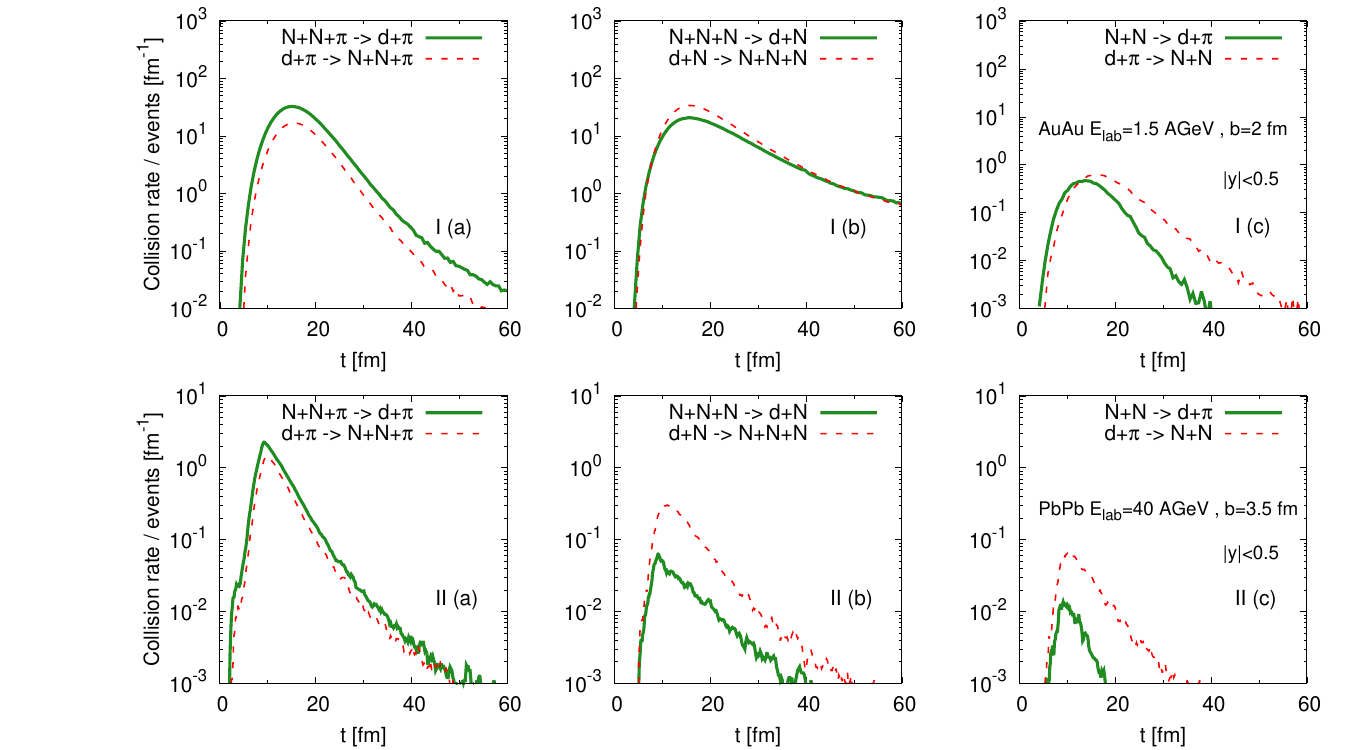}
\caption{\label{fig:C0} (color online) Collision rates for deuteron production (solid green lines) and breakup (dashed red lines) reactions at mid-rapidity $|y|<0.5$ as a function of evolution time are shown for two different HICs setups: (I) top panels show, respectively, the reaction rates for $N+N+\pi \leftrightarrow d+\pi$ (a), $N+N+N \leftrightarrow d+N$ (b) and $N+N\leftrightarrow d+\pi$ (c) in Au+Au collisions at $E_{lab}=1.5$ AGeV at fixed impact parameter $b=2$ fm; (II) bottom panels show the reaction rates for the same channels in Pb+Pb collisions at $E_{lab}=40$ AGeV at fixed impact parameter $b=3.5$ fm.}  
\end{figure*}

We start this section by showing in Fig.~\ref{fig:C0} the collision rates for all inelastic processes for deuteron production (solid green lines) and disintegration (dashed red lines) implemented in PHQMD for two different HIC systems. The top panels (I) show the reaction rates for Au+Au collisions at $E_{lab}=1.5$ AGeV (impact parameter $b=2$ fm), while the bottom panels (II) show the reaction rates for Pb+Pb collisions at $E_{lab}=40$ AGeV (impact parameter $b=3.5$ fm).
 Confronting (I) and (II) we clearly see that at the lower collision energy the formation and breakup of deuterons is mainly driven by the $N N N \leftrightarrow d N$ channel, involving only nucleons, while at higher energies the reaction $\pi N N \leftrightarrow d \pi$ becomes dominant because pions are more abundant. The two-body inelastic reaction $N N \leftrightarrow d \pi$ (right panels) has a much lower rate compared to three-body inelastic $N N \pi \leftrightarrow d \pi$ (left panel) and $N N N \leftrightarrow d N$ (middle panel) channels because the cross section is smaller.

\subsection{Effect of charge exchange reactions}

Before showing our final results, we study the impact of the different isospin channels on deuteron production in relativistic HICs. The reaction $d\pi \leftrightarrow NN\pi$ is important only in the case where the pion catalysis is dominant compared to the nucleon catalysis. Therefore, we select Au+Au central collisions at the energy $\sqrt{s}_{NN}=7.7$ GeV to study the production of deuterons through all the implemented reactions. 

\begin{figure}[ht]
\centering
\includegraphics[width=0.45\textwidth]{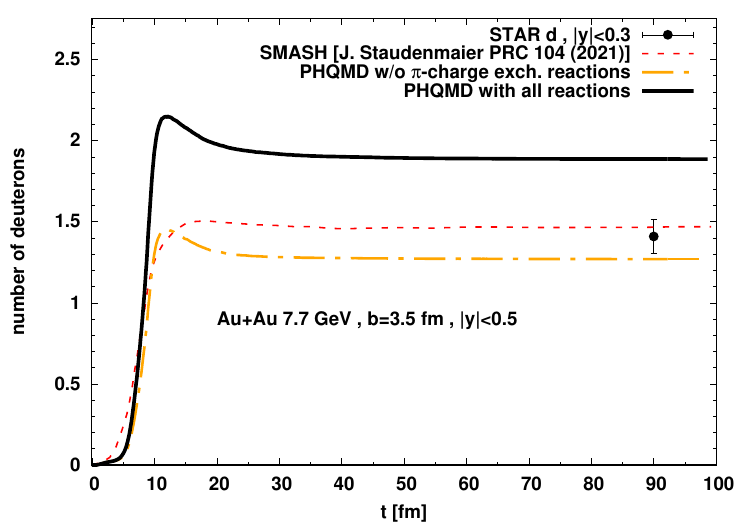}
\caption{\label{fig:C1} (color online) Mid-rapidity number of deuterons as function of evolution time from PHQMD simulations for Au+Au $\sqrt{s_{NN}}=7.7$ GeV central collisions. The full circle represents the STAR data point at mid-rapidity. The different lines are described in the text.}  
\end{figure}

Moreover, for this system we can also compare our results with those obtained recently by the SMASH collaboration~\cite{Staudenmaier:2021lrg} where for deuteron production by multi-particle reactions the same covariant rate formalism is applied. This substitutes the previous SMASH work, where the $3 \rightarrow 2$ channel was simulated numerically by a sequence of two $2 \rightarrow 2$ processes passing through the formation of a fictitious $d'$ resonance (Ref.~\cite{Oliinychenko:2018ugs} for the LHC energy,~\cite{Oliinychenko:2020znl} for RHIC BES).\\
In SMASH studies only the kind of reactions $\pi d \leftrightarrow \pi pn$, where the isospin degrees of freedom are not taken into account, were considered within the pion catalysis.
However, isospin conservation allows for two types of pion catalyzed reactions:
$\pi^+ d \leftrightarrow \pi^+ pn$, in which the $\pi$ charge is conserved and the charge exchange reaction $\pi^+ d \leftrightarrow \pi^0 pp$.
As discussed in the previous section, a goal of this work is to study the impact of including all the charge exchange reaction channels on the production of ``kinetic'' deuterons in HICs.\\
In Fig.~\ref{fig:C1} the mid-rapidity ($|y|<0.5$) multiplicity of deuterons in Au+Au collisions at $\sqrt{s_{NN}}=7.7$ GeV for a fixed impact parameter $b=3.5 \, fm$ is shown as a function of time. The full black circle is the STAR measurement~\cite{STAR:2019sjh} and the full black line is the PHQMD result if we include both types of $\pi$ catalyzed channels.\\ 
It is useful to compare our results with those of SMASH in which only the $\pi^+ d \leftrightarrow \pi^+ pn$ channel (and similar for $\pi^-$ and $\pi^0$) is included and displayed as red dashed line (taken from Fig.~(4)a in Ref.~\cite{Staudenmaier:2021lrg}). If we omit the $\pi$ charge exchange reaction and retain only what is also employed by SMASH, we find the dash-dotted orange line.
Our result gives a slightly smaller number of deuterons and shows also a different time behavior. The small difference of the order of 10\% is not surprising because SMASH and PHQMD are completely different transport approaches. In SMASH, a hydrodynamical evolution of the fireball is followed by a particlization at the hypersurface of an energy density of $\epsilon \approx 0.26 \!$ GeV/fm$^3$. Then particles propagate without potential interaction (cascade) and collide in the hadronic phase until chemical and kinetic freeze-out is achieved. Therefore, in SMASH the kinetic production of deuterons is limited to the latest stages and, more important, it is not affected by the potential interactions among nucleons. On the contrary, in PHQMD the deuteron reactions are embedded in a transport environment in which the baryons are propagated after their creation by the QMD equations, which include potential interactions. If the local energy density exceeds the critical value $\epsilon_c \approx 0.5 \!$ GeV/fm$^3$, the hadrons dissolve into the partons, which follow the description of the Dynamical Quasi-Particle Model (DQPM) implemented within the PHSD framework (see Sec.~\ref{sec:sec2} and references therein). Deuteron formation is therefore only possible in regions in which the energy density is smaller than $\epsilon_c$, where hadrons are the degrees of freedom of the system.

Such a different description of the expanding system makes it difficult to disentangle the differences of these two approaches. We mention that, just for test purposes, we have implemented in PHQMD an additional energy density cut for deuteron production, $\epsilon < 0.26$ GeV/fm$^3$, mimicking the transition from the hydro to the hadronic phase as in the SMASH approach, however, the results are similar within statistical uncertainties.
 
Comparing the full black and the orange dot-dashed curve we observe that the $\pi$ charge exchange reaction increases the deuteron yield by 50\% (for the isospin factors we refer to Appendix C) at this beam energy and brings the complete calculation outside of the experimental error bars. 
To complete our study, we made a similar check for collisions at lower energies. In particular, we confirm that at the energy of the  GSI-SIS accelerator $E_{lab}=1.5$ AGeV, i.e. $\sqrt{s_{NN}} = 2.52$ GeV, where the production of deuterons occurs mainly by $NNN \rightarrow dN$ (see the collision rate in Fig.~\ref{fig:C0}) the contribution of $\pi$ catalyzed deuteron production is negligible, while at $\sqrt{s_{NN}}=3$ GeV, the energy of the STAR FiXed Target (FXT) experiment, the $\pi$ charge exchange channel increases the deuteron production by 20\%.

\subsection{Modelling of Finite-size of deuteron}

In dense nuclear matter the binding energy of a deuteron is reduced and becomes eventually positive because (for a deuteron at rest and a zero temperature environment) the quantum states with the lowest energy are occupied by protons and neutrons up to the Fermi momentum which is related to the density $\rho_N$ of nuclear matter by 
\begin{equation}
p_F = (3\pi^2 \rho_N )^{1/3} .
\end{equation}
Therefore, only the momentum components above the Fermi surface can contribute to the deuteron binding energy and the expectation value of the deuteron hamiltonian with respect to the $pn$ pair wavefunction $\Phi(p_1,p_2)$ is given by
\begin{equation}
\int_{p_F}^{\infty} \!\! d^3p_1 \!\int_{p_F}^{\infty} \!\! d^3p_2 \bra{\Phi(p_1,p_2)} \hat{H}_d \ket{\Phi(p_1,p_2)} = E_d(p_F) \, ,
\end{equation}
where $E_d(p_F)$ is the binding energy of the deuteron in nuclear matter. If $\rho_N$ increases $E_d(p_F)$ becomes positive and the deuteron becomes unbound. The value of the nuclear density $\rho$ at which the deuteron binding energy vanishes is known as \textit{Mott density} \cite{Ropke:1982ino,Ropke:1988ymk}. However, only the case of low-density cold ($T \simeq 0$) infinite nuclear matter $E_d(p_F)$ can be calculated analytically. 
In the hot fireball ($T \simeq 100$ MeV), created in relativistic HICs at mid-rapidity, deuterons are in addition destroyed by collisions with fellow particles (mostly pions), which scatter with a thermal transverse momentum $p \simeq T \gg E_d$, i.e. which is much larger than the deuteron binding energy.

\subsection*{Excluded-volume}

In collision integrals the final-state particles are considered as point-like particles. In vacuum this is the proper description but in matter, where the final-state hadrons are surrounded by other hadrons, modifications are necessary if the produced particles have a finite extension.

A deuteron with a rms radius of about $\sqrt{<r^2_d>} \simeq 2.1$ fm cannot be formed if between the $p$ and the $n$ other hadrons are located. One possibility to take this into account is to include in our covariant rate formalism an \textit{excluded volume condition}. 
As discussed in Sec.~\ref{sec:sec3}, for the dominant production channels $NN\pi \rightarrow d \pi$ and $NNN \rightarrow d N$ the probability $P_{3,2}$ that the collision occurs and the deuteron is formed is given by Eq.~\eqref{Prob32}.
We include the excluded volume condition in our calculation in the following way. When, according to the collision rate, a deuteron should be produced at time $t$, we compute the position and momentum of the center-of-mass of the ``candidate'' deuteron $d$. Subsequently, we loop over all hadrons, which exist  at that time $t$, and for each particle $i$ we check the following condition
\begin{equation}\label{eq:EXCLcond}
| \mathbf{r}_i - \mathbf{r}_d | > R_d \, ,
\end{equation}
where the parameter $R_d$ is the radius of the excluded volume. The particle positions, $\mathbf{r}_i$ and $\mathbf{r}_d$, are calculated in the center-of-mass frame of the candidate deuteron. 
In order to produce a deuteron at the final state of the reaction the condition Eq.~\eqref{eq:EXCLcond} must be fulfilled by all the surveyed particles. Otherwise, the candidate deuteron is considered not formed and the system is restored to the initial state, as if the participant hadrons had never scattered. 
Thus, like for the MST condition for cluster formation Eq.~\eqref{eq:MSTcond}, we use also here a geometrical criterium to take into account the finite extension of the deuteron. However, both criteria work differently: \textit{i)} in MST $i$ and $j$ can only be baryons, while for the excluded-volume we account for all spectator hadrons in the fireball, i.e. those particles which do not participate in the initial stage of the deuteron reaction; \textit{ii)} the excluded-volume condition, which excludes deuteron formation if a third hadron is too close, works oppositely to the MST clustering where two baryons form a cluster if they are sufficiently close.
The excluded radius $R_d$ in Eq.~\eqref{eq:EXCLcond} is related to the root-mean-square (rms) radius $r_m$ of the deuteron by
\begin{equation}\label{eq:EXCLRd}
R_d^2 \simeq <r_m^2> =  \int_0^{\infty} \! dr \, r^2 |\phi_d(r)|^2 \, ,
\end{equation}
where $\phi_d(r)$ is the Deuteron Wave-Function (DWF) in coordinate space, which is obtained by solving the radial Schr\"odinger equation using a phenomenological parametrization of the nucleon-nucleon potential $V_{NN}$, which correctly reproduces its ground state properties. In particular, the function $\phi_d(r)$ takes into account the fact that the deuteron ground state is a mixture of a $S$- and a $D$-states, with assigned real functions $u(r)/r$ and $v(r)/r$ respectively, and it is normalized so that the total probability to find the deuteron in one of the two states is one,
\begin{equation}\label{eq:DWFrnorm}
\int_{0}^{\infty} \! dr |\phi_d(r)|^2 = \int_{0}^{\infty} \! dr \left[ u^2(r) + v^2(r) \right] = 1 \, .
\end{equation}
More specifically, we employ the DFW parametrization from Ref.~\cite{PhysRev.158.1500}, where the functions $u(r)$ and $v(r)$ can be expressed as discrete superposition of Hankel functions. Similar calculations were performed by the Paris group~\cite{Lacombe:1980dr,Lacombe:1981eg} using their own parametrization of the $V_{NN}$ potential.

\begin{figure}[ht]
\centering
\includegraphics[width=0.45\textwidth]{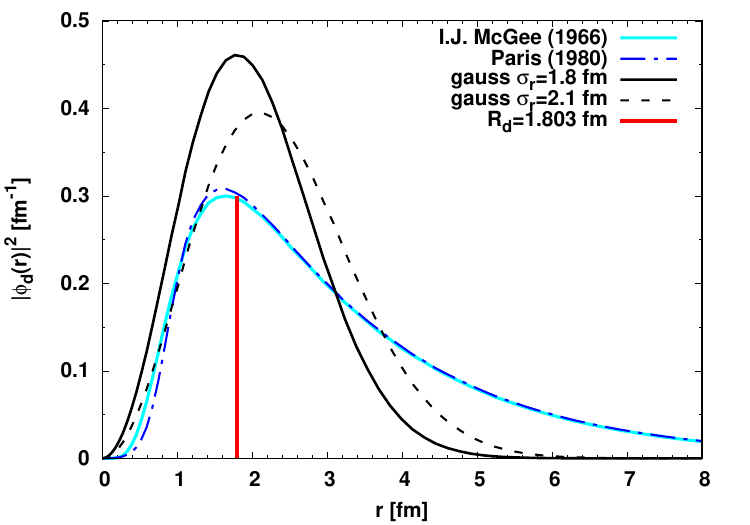}
\caption{\label{fig:C2} (color online) The square modulus of the Deuteron Wave Function (DWF) in coordinate space, normalized to unity, is used as the probability distribution function to evaluate the physical radius parameter $R_d$. The colored solid lines are the parametrizations from Ref.~\cite{PhysRev.158.1500} (solid light blue) and Ref.~\cite{Lacombe:1980dr,Lacombe:1981eg} (dash-dotted blue) employed in  Eq.~\eqref{eq:EXCLRd}. The black solid lines is a Gaussian function with width $\sigma_r = 1.8$ fm which is equivalent to the $R_d$ from the solution of Eq.~\eqref{eq:EXCLRd}. The estimated value of the excluded radius $R_d=1.803$ fm is represented by the vertical red line. The dashed black curve shows another Gaussian shape with larger $\sigma_r=2.1$ fm.}

\end{figure}
\begin{figure}[ht]
\centering
\includegraphics[width=0.45\textwidth]{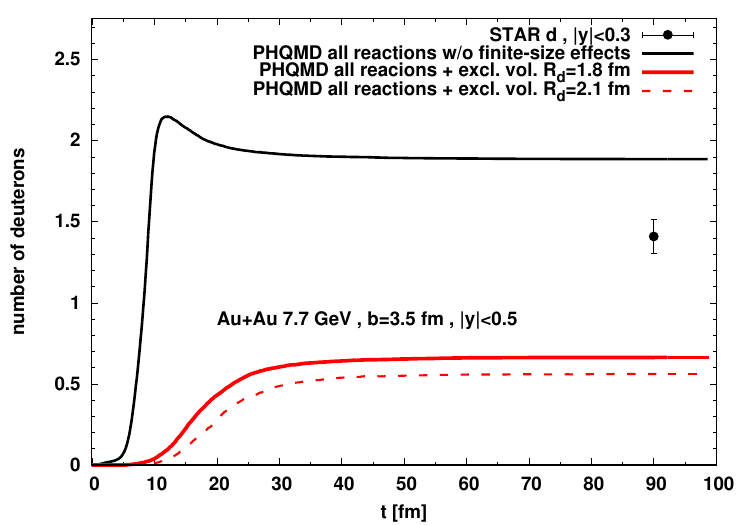}
\caption{\label{fig:C3} (color online) Time evolution of the number of deuterons at mid-rapidity from PHQMD simulations for central Au+Au at $\sqrt{s_{NN}}=7.7$ GeV. The full circle is the STAR data at mid-rapidity. The black solid line is the PHQMD result with full isospin decomposition (same as Fig.~\ref{fig:C1}). The red lines correspond to the result where the excluded-volume procedure is applied to all deuteron production channels. The excluded radius parameter is indicated in the legend: $R_d=1.8$ fm (red solid), $R_d=2.1$ fm (red dashed).}    
\end{figure}

In Fig.~\ref{fig:C2} the DWF from Ref.~\cite{PhysRev.158.1500} is represented by the solid blue line. 
The dashed blue line is the result employing the Paris potential from  Ref.~\cite{Lacombe:1980dr,Lacombe:1981eg}. Both are showing the same behavior. Inserting this function in Eq.~\eqref{eq:EXCLRd} and solving the integral numerically we find $R_d=1.803$ fm (red vertical line). \\
In Fig.~\ref{fig:C3} we study the impact of excluded-volume condition on deuteron production for central ($b=3.5$  fm) Au+Au collisions at $\sqrt{s_{NN}}=7.7$  GeV. The full circle is the measured value of the STAR experiment~\cite{STAR:2019sjh} at mid-rapidity. The lines represent the time evolution of the deuteron yield in PHQMD for the rapidity range $|y|<0.5$. The black solid line is the result if all production channels are included, which has already been shown in Fig.~\ref{fig:C1}. The red lines are the results if we include the excluded-volume condition. Here we present the results for two excluded volume radius parameter, $R_d=1.8 \, fm$ (red thick solid line) and $R_d=2.1 \, fm$ (red dashed line).  
As seen from Fig.~\ref{fig:C3}, the inclusion of the excluded volume condition has a large impact on the formation of deuterons - at the considered energy it reduces their abundance at mid-rapidity by a factor of about 3. This is due to the high density of hadrons at mid-rapidity in the initial phase of the reaction. 
The final abundance of deuterons depends on $R_d$. Two choices of the excluded radius, $R_d$ = 1.8 fm and $R_d$ = 2.1 fm - two values around the rms radius of the deuteron - give a difference of the final number of deuterons of 15\%.

\subsection*{Momentum projection}

\begin{figure}[ht]
\centering
\includegraphics[width=0.45\textwidth]{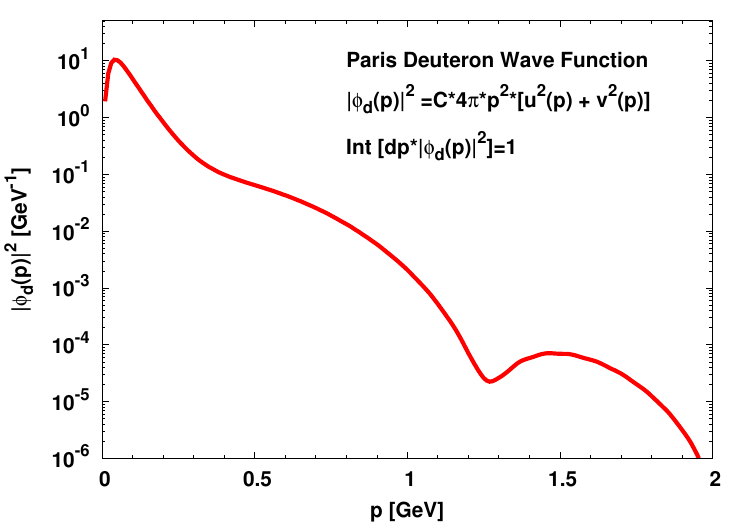}
\caption{\label{fig:C4} (color online) The squared modulus of the Deuteron Wave Function (DWF) in momentum space normalized to unity (red curve) from the Paris parametrization Ref.~\cite{Lacombe:1981eg} as function of $p$, which is the relative momentum of the bound proton-neutron pair in the deuteron center-of-mass frame.}
\end{figure}

As we have seen, the excluded-volume condition models the fact that the deuteron is an extended object in coordinate space with a root-mean-square radius determined from Eq.~\eqref{eq:EXCLRd}, where $|\phi_d(r)|^2$ is the square of its ground state wave function represented in Fig.~\ref{fig:C2} with the colored lines.
The square of the relative momentum $<p^2>$ of the bound $pn$ pair can be obtained from the deuteron wave function represented in momentum space, which can be derived by taking the Fourier transforms of the $S-$ and $D-$ state components. The square of the DWF in momentum space $|\phi_d(p)|^2 \propto 4\pi p^2 [u^2(p) + v^2(p)]$,  calculated using the Paris potential, Ref.~\cite{Lacombe:1981eg}, is presented in Fig.~\ref{fig:C4} as a solid red line and its integral is normalized to unity.
Using the uncertainty principle, we can calculate the expected relative momentum of the  bound $pn$ pair 
\begin{equation}\label{eq:PROJd}
\sqrt{<p^2>}\simeq \frac{1}{\sqrt{<r_m^2>}}=\frac{1}{R_d}     
\end{equation}
For $R_d=1.8$ fm, we obtain $\sqrt{<p^2>} \simeq 0.1$ GeV a value very close to the value calculated using the DWF in Fig.~\ref{fig:C4} which is about $0.13$ GeV. 

The probability amplitude that a proton and a neutron collide and form a deuteron is given by $<\phi_d(p)|\phi(p)>$ where $\phi_d(p)$ is the DWF, whose square is shown in Fig.~\ref{fig:C4}, while $\phi(p)$ is the wave function of the relative momentum $p$ of a proton and a neutron just after the collision occurred. In both cases, the relative momentum $p$ is calculated in the center-of-mass frame of the deuteron. This indicates that a proton and a neutron have the highest chance to form a deuteron in the region where their wave function $\phi$ overlaps most with $\phi_d$, which happens if their relative momentum is of the order of 0.1 GeV.

The covariant collision rate for $3 \rightarrow 2$ reactions derived in Eq.~\eqref{icoll32a} assumes that the transition amplitude depends only on the center-of-mass energy, $\sqrt{s}$. For the deuteron case, this is an oversimplified assumption because nucleons with a very large relative momentum have a small probability to form a deuteron. In order to relax this assumption, we assume that the momentum transfer of the third body is small and that, therefore, in the $\pi NN \rightarrow \pi d$ and $NNN \rightarrow Nd $ reactions, 
the initial relative momentum of the nucleons is close to that of the two nucleons bound in a deuteron. 
This allows to determine the probability that a deuteron is produced in these reactions by weighting the initial relative momentum with  $|\phi_d(p)|^2$ (which is normalized to unity). Consequently, a pair with a smaller relative momentum in its center-of-mass system has a higher chance to produce a deuteron than a pair with a larger relative momentum. 

\begin{figure}[th]
\centering
\includegraphics[width=0.45\textwidth]{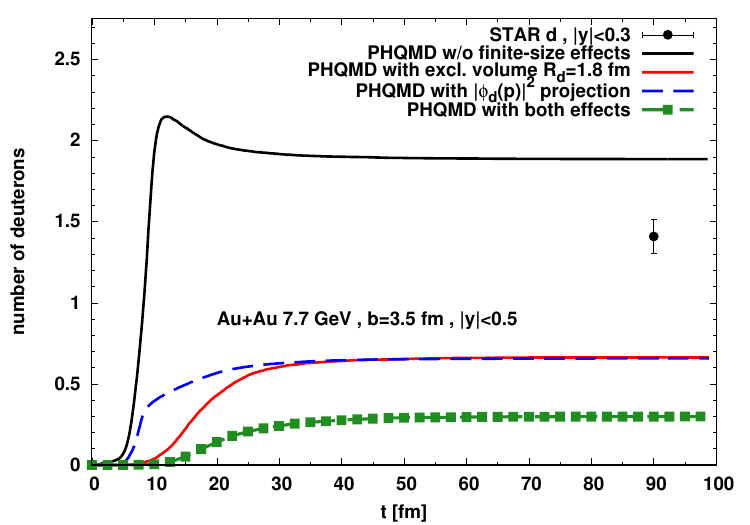}
\caption{\label{fig:C5} (color online) Number of deuterons at mid-rapidity as function of time from PHQMD simulations for central Au+Au collisions at $\sqrt{s_{NN}}=7.7$ GeV. The different lines correspond to different models of finite-size effects: \textit{i)} with excluded-volume from Fig.~\ref{fig:C3} (red solid), \textit{ii)} with momentum projection only (dashed blue), \textit{iii)} including both effects (thick dashed green line with full squares). The case of all production channels without finite-size effects is taken again from Fig.~\ref{fig:C1} (black solid).}
\end{figure}

In the transport calculations we employ a Monte Carlo procedure to decide whether a deuteron is produced or not. If three nucleons are in the entrance channel we randomly determine which of the possible pairs is considered as a possible deuteron candidate.  
Here we study the effect of this \textit{momentum projection} on the deuteron production in HICs. In Fig.~\ref{fig:C5} we display the time evolution of the number of deuterons at mid-rapidity in PHQMD simulations for central Au+Au collisions at $\sqrt{s_{NN}}=7.7$ GeV, as compared to STAR data at mid-rapidity (black point). The black solid line and red solid line are those from Fig.~\ref{fig:C3} and represent the deuteron yield including all possible channels, respectively without and with the excluded-volume condition with a parameter $R_d=1.8$ fm. The dashed blue line is the result applying the momentum projection only.
One can see that momentum projection strongly suppresses the deuteron production at the initial stage of Au+Au collisions where dominantly the collisions of energetic nucleons take place. Moreover, we find that
the momentum projection and the excluded-volume condition give for large times the same suppression at mid-rapidity. 
They both lead to a strong suppression of deuteron formation at mid-rapidity at the time of  10-20 fm, due to the presence of the dense medium populated by many particles (especially pions) which can exist in the volume occupied by the deuteron. At later times deuteron production becomes important but asymptotically the production is only 30\% of that without projection.   

\begin{figure}[ht!]
\centering
\includegraphics[width=0.36\textwidth]{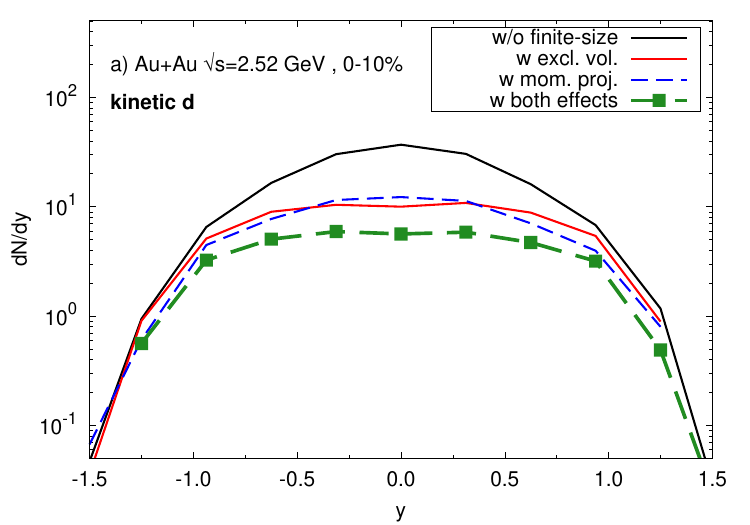}
\includegraphics[width=0.36\textwidth]{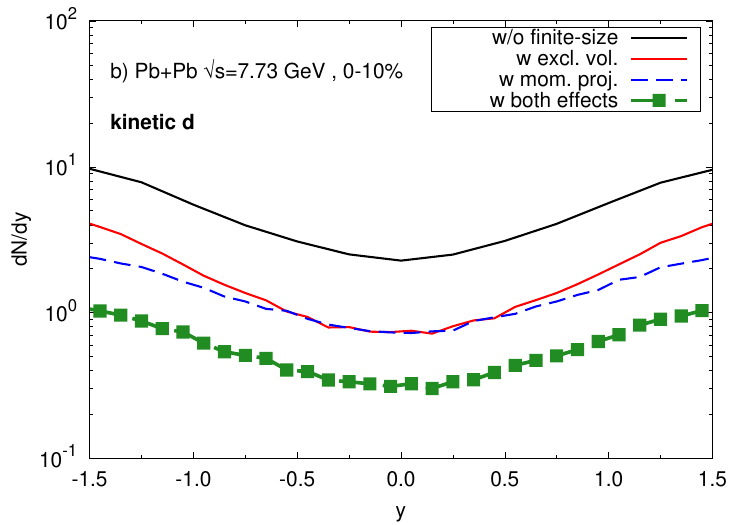}
\includegraphics[width=0.36\textwidth]{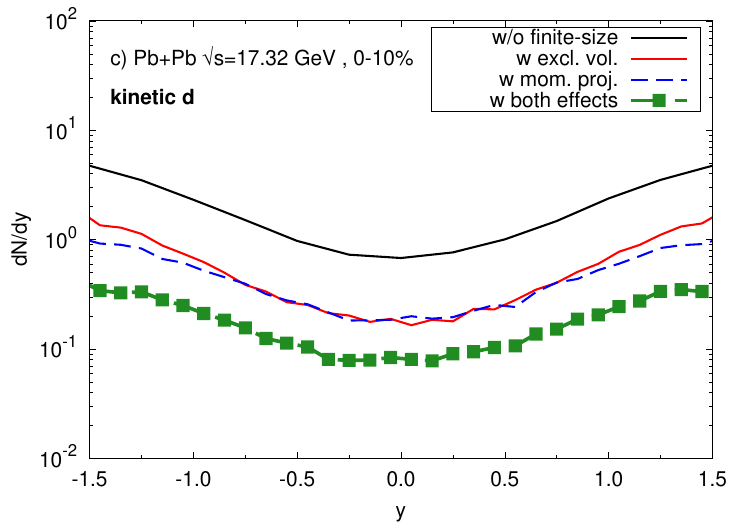}
\includegraphics[width=0.36\textwidth]{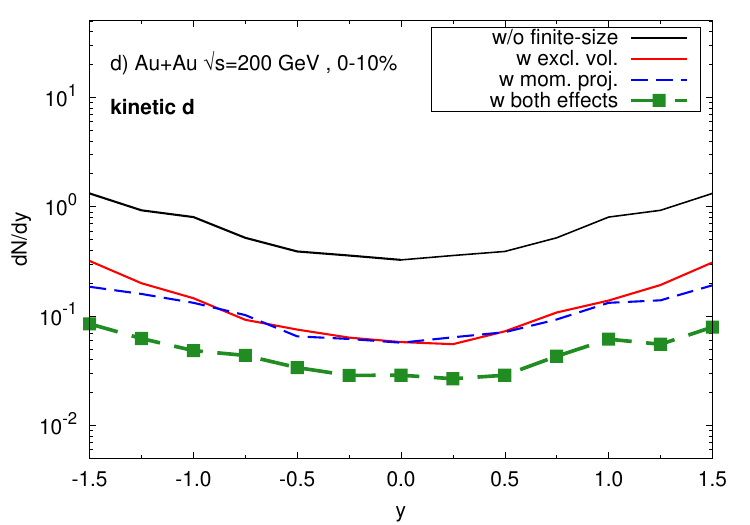}
\caption{\label{fig:C6}(color online) PHQMD rapidity distributions $dN/dy$ of kinetic deuterons in central nucleus-nucleus collisions for four different colliding systems: a) Au+Au at $\sqrt{s_{NN}}=2.52$ GeV (top panel), b) Pb+Pb at $\sqrt{s_{NN}}=7.73$ and c) $\sqrt{s_{NN}}=17.32$ GeV (middle panels), d) Au+Au at $\sqrt{s_{NN}}=200$ GeV (bottom panel). The different models for finite-size effects implemented in kinetic deuteron reactions are denoted by various lines with the same color coding of Fig.~\eqref{fig:C5}: excluded volume condition (solid red), momentum projection only (dashed blue), both conditions (thick dashed line with full squares), without any effect (solid black).}
\end{figure}

As a next step, we investigate the case where the two finite-size effects, namely the excluded volume in coordinate space and the momentum projection in momentum space, are simultaneously applied to the production of kinetic deuterons. This scenario is shown in Fig.~\eqref{fig:C5} as the thick dashed green line with full squares. One can see that the inclusion of both conditions produces a suppression which is about a factor 2 stronger than the case where the excluded volume or the momentum projection are applied individually - which means an overall suppression factor of 6 with respect to the case where no finite-size effects are considered (Fig.~\eqref{fig:C5} solid black line). 

We studied the impact of finite-size effects on kinetic deuterons at mid-rapidity for different collision systems and found that the amount of suppression is quite similar at all collision energies. Only for top RHIC energy we have noticed a larger factor - about an order of magnitude between the case without and with both effects - which we could explain by the higher density of particles (pions) at the initial stages which makes the excluded volume condition more effective. It is also interesting to study this effect in different rapidity intervals.

In Fig.~\eqref{fig:C6} we present the rapidity distributions of kinetic deuterons from PHQMD simulations in central nucleus-nucleus collisions for four different collision systems, which are reported in the legend. The color coding is the same as in Fig.~\eqref{fig:C5}. At the lowest energy,  $\sqrt{s}=2.52$ GeV, corresponding to $E_{Lab}=1.5$ AGeV, where projectile and target decelerate almost completely and form a mid-rapidity source, the
maximum of proton as well as the maximum of the deuteron distribution is peaked at mid-rapidity. At the higher energies projectile and target pass each other and the proton as well as the deuteron distributions have a minimum at mid-rapidity. 

It is remarkable that the excluded volume and the momentum projection approach leads at all beam energies to an almost identical rapidity distribution around mid-rapidity. Only towards the edge of the rapidity interval, which we investigated here, momentum projection leads to a larger suppression of the deuteron yield because at this rapidity the relative momentum of the nucleons is larger.
The suppression of the deuterons is always of the order of 3 at mid-rapidity. At finite rapidities the momentum projection gives always a larger suppression than the excluded volume approach. If we apply the excluded volume and momentum projection simultaneously we obtain an additional suppression, which is, however, at mid-rapidity small compared to the suppression due to the individual application of one of these finite-size corrections. This is a sign that the relative distance and momentum of the proton-neutron pair are correlated. The form of $dN/dy$ is close to the one obtained for momentum projection only and is shallower than the distribution without finite-size effects.
We studied also the slope of  the $p_T$-spectra of the kinetic deuterons and found that it is not changed by finite-size effects.
Before moving to the results, we want to mention that:\\ 
\textit{i)} currently, the kinetic deuterons, which are created in collisions, are treated in PHQMD as point-like particles which stream freely until they eventually disintegrate due to collisions with pions or nucleons. This means that they have no potential interaction with other nucleons;\\
\textit{ii)} the nucleons of the kinetic deuterons do not enter into the MST algorithm, otherwise they would be double counted. One could argue that nothing prevents a deuteron to interact with a surrounding nucleon and gets bound into a larger cluster. However, this cannot happen when applying the excluded volume condition, as the presence of another nucleon close by would not allow the formation of the kinetic deuteron itself.

\section{Results}\label{sec:sec6}

\begin{figure*}[ht!]
\centering
\includegraphics[width=0.9\textwidth]{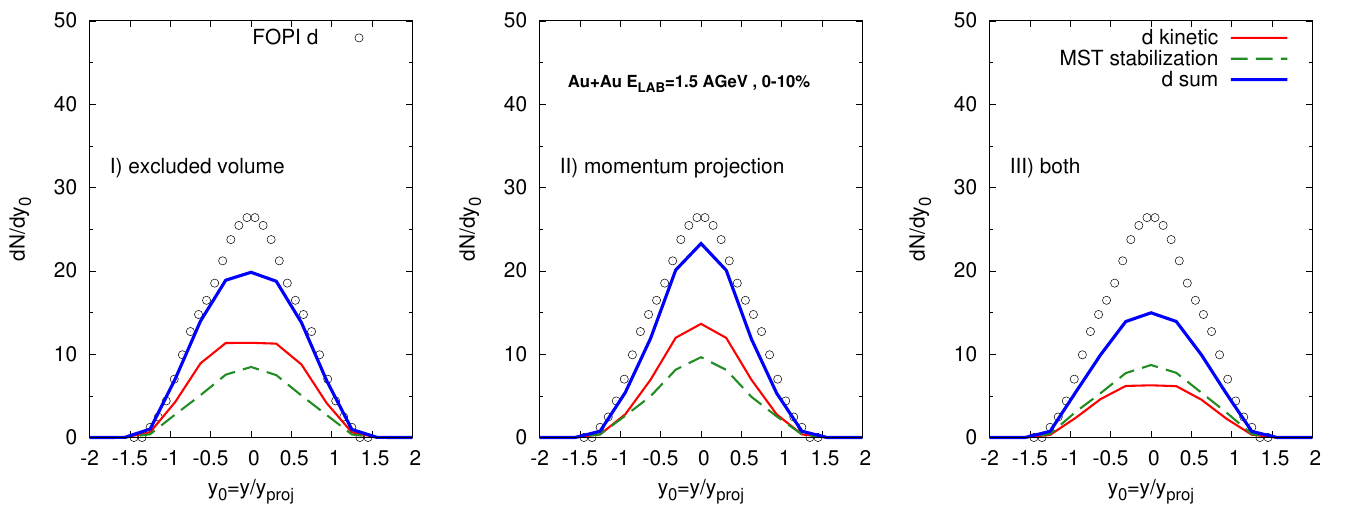}
\caption{\label{fig:yFOPI} (color online) Scaled rapidity distributions, $dN/dy_0$, with $y_0=y/y_{proj}$, of deuterons in central Au+Au collisions at $E_{lab}=1.5 \, AGeV$ measured by the FOPI collaboration \cite{Reisdorf:2010aa}. Experimental data (open circles) are compared with the rapidity distributions from PHQMD simulations. The kinetic deuterons (solid red line) and potential deuterons identified by aMST (dashed green line) are added together to give the solid blue line. The three plots correspond to three different models of finite-size effects in the kinetic production. From left to right: I) excluded-volume only, II) momentum-projection only, III) sum of both effects.}  
\end{figure*}

In this section we compare, from SIS to top RHIC energies,  the  rapidity and transverse momentum distribution of deuterons, calculated with PHQMD, with the experimental data.  
As mentioned in the previous sections we consider two sources of deuteron production:
\begin{itemize}
 \item {\bf Deuterons produced by collisions (kinetic deuterons)} \\
 Kinetic deuterons can be produced by the inelastic reactions $\pi NN\leftrightarrow \pi d$, $NNN\leftrightarrow N d$ and $NN \leftrightarrow d \pi$. We include all possible charge exchange channels in the $\pi-$catalysis. The quantum properties of deuterons are modeled 
 through the three finite-size corrections, discussed in the last section:
   \begin{itemize}
    \item[I)] by the excluded-volume condition choosing the radius $R_d=1.8$ fm;
    \item[II)] by the momentum projection on the DWF $|\phi_d(p)|^2$;
    \item[III)] by taking into account simultaneously I+II.
   \end{itemize}
 \item {\bf Deuterons produced by potential interaction (potential deuterons)}\\
The deuterons, which are produced due to the potential interactions between baryons are identified by the "advanced" Minimum Spanning Tree (aMST) algorithm during the fireball evolution and reconstructed as ``bound'' clusters ($E_B<0$) according to the stabilization procedure described in Sec.~\ref{sec:sec2}~B.
\end{itemize}

\subsection{Au+Au at SIS $E_{Lab}=1.5$ AGeV}

We start by presenting in Fig.~\ref{fig:yFOPI}  the PHQMD results for central Au+Au collisions at $E_{Lab}=1.5$ AGeV ($\sqrt{s}=2.52$ GeV) the scaled rapidity distribution $dN/dy_0$ as function of $y_0=y/y_{proj}$, where $y_{proj}$ is the projectile rapidity in the center-of-mass frame of the colliding nuclei.  Kinetic deuterons  are presented by a thin red line, potential deuterons  by a dashed green line and the sum of both as a blue line.   The three panels display three different approaches to finite-size effects in the kinetic production via $N N N \rightarrow N d$, $\pi N N \rightarrow \pi d$ and $N N \rightarrow d \pi$. From left to right we display: I) deuterons obtained when applying the excluded-volume condition, II) deuterons obtained when the momentum-projection is employed and  III) deuterons if both finite-size corrections are simultaneously considered.   
The results are compared with the FOPI experimental data~\cite{Reisdorf:2010aa}, displayed as open circles. 

We see in Fig.~\ref{fig:yFOPI} that the aMST deuterons alone give less than 40\% of the measured yield for all scenarios, while the contribution of the kinetic deuterons varies strongly: the scenario II with "momentum projection" only gives the best description of the experimental data while an additional application of the excluded volume leads to a strong suppression of the deuteron production.
At this energy a baryon rich, almost equilibrated fireball is created at mid-rapidity which makes both finite-size corrections very effective.  Even together with the aMST deuterons, the deuteron yield is underpredicted.

We note that the underprediction of the cluster multiplicity at mid-rapidity for low beam energies has been already observed in non-relativistic IQMD calculations \cite{LeFevre:2019wuj}. At this low energy  the mid-rapidity region is very complex because it contains decelerated projectile and target nucleons as well as fireball nucleons and is characterized by a high baryon density. Indeed, it seems that in our approach some correlations, which contribute to deuteron formation, are absent.
We think that further improvement of cluster recognition algorithm as well as improvement of the QMD dynamics (e.g. by using a momentum-dependent potential instead of simple Skyrme potential used in this study) might improve the situation.

\subsection{Au+Au at $E_{Lab}=11$ AGeV}

\begin{figure*}[ht!]
    \centering
    \begin{minipage}{0.32\textwidth}
        \centering
        \includegraphics[width=\textwidth]{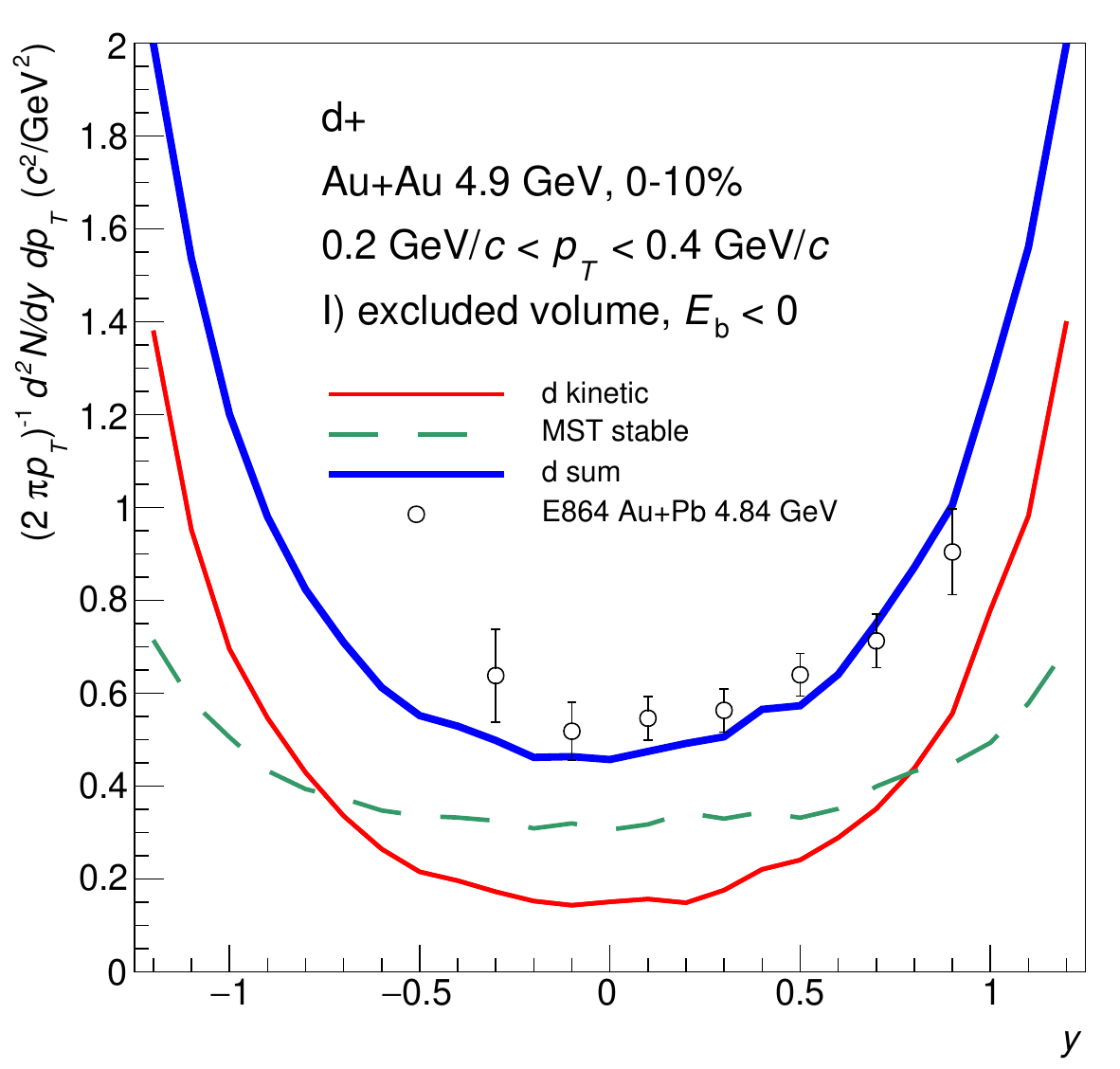}
    \end{minipage}
    \begin{minipage}{0.32\textwidth}
        \centering
        \includegraphics[width=\textwidth]{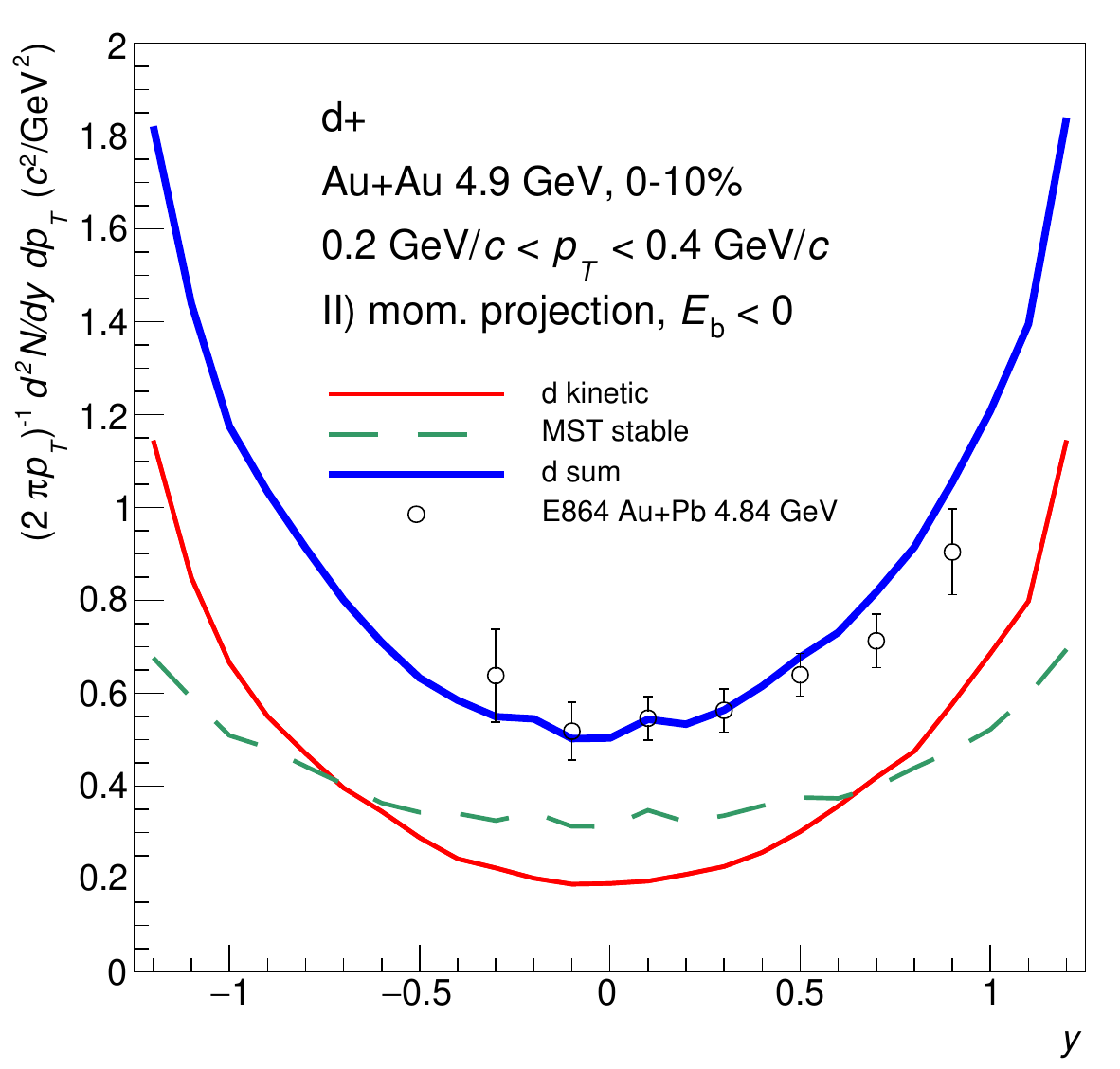}
    \end{minipage}
    \begin{minipage}{0.32\textwidth}
        \centering
        \includegraphics[width=\textwidth]{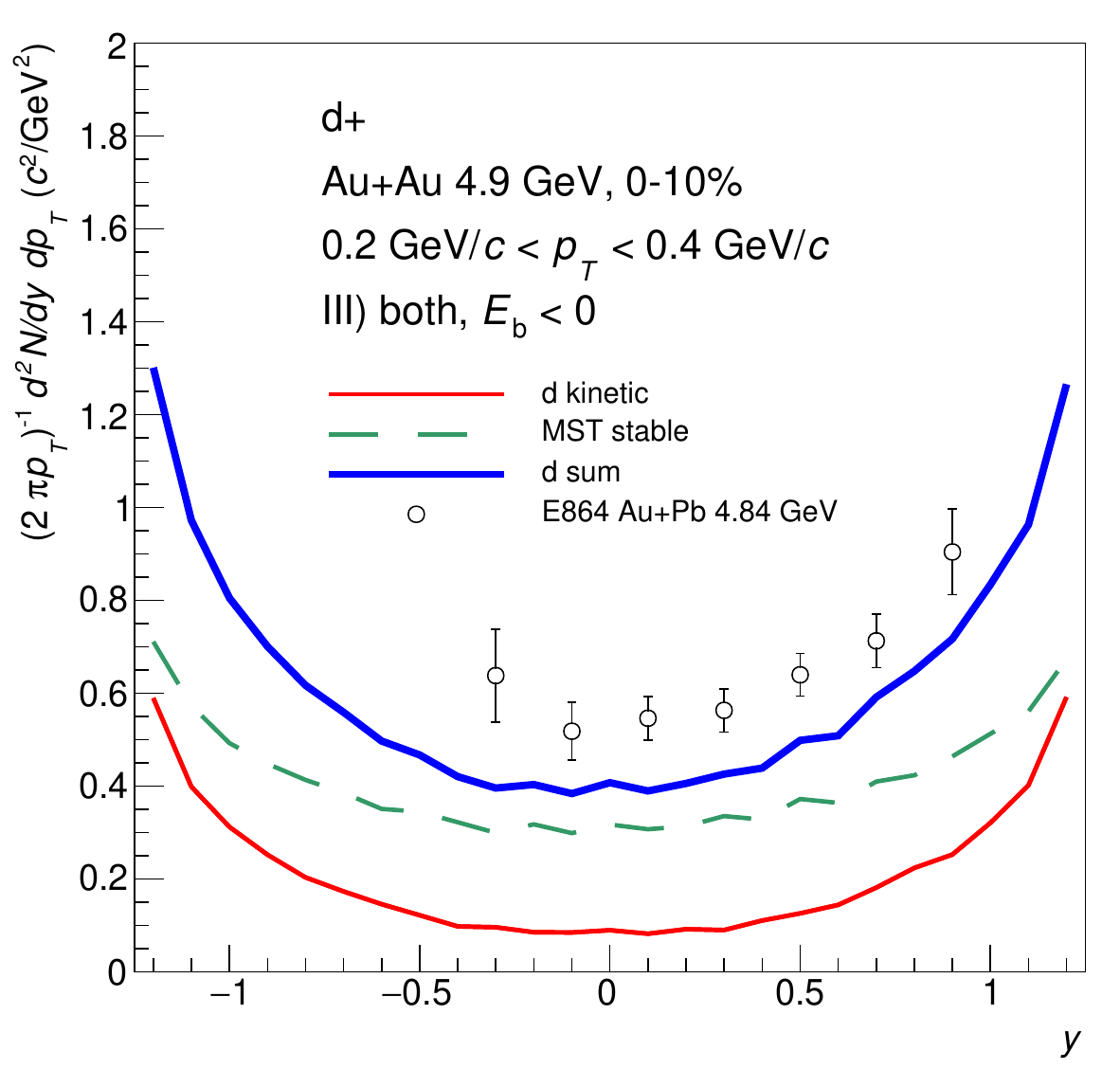}
    \end{minipage}
\caption{\label{fig:yCBM} (color online) The invariant rapidity distributions $(2\pi p_T)^{-1} d^2N/dp_Tdy$ of deuterons in  10\% most central Au+Au collisions at $E_{Lab}=11$ AGeV ($\sqrt{s}=4.9$ GeV). The lines correspond to the PHQMD calculations for deuterons coming from the two production processes: kinetic by hadronic reactions (solid red line) and potential (dashed green line), where in the latter mechanism the stable and bound ($E_B<0$) deuterons are identified via the aMST procedure. The three panels display the three different approaches to finite-size effects in the kinetic production. From left to right: I) deuterons obtained applying the excluded-volume condition, II) deuterons obtained when the momentum-projection is introduced, III) deuterons obtained when both effects are taken into account. The experimental measurements in Au+Pb central collisions at AGS taken from the E864 collaboration~\cite{Armstrong:2000gz} are shown with open circles. To compare the PHQMD results with these data the same cut $0.2<p_T<0.4$ GeV, is applied, as reported in each plot.}
\end{figure*}

\begin{figure*}[ht!]
    \centering
    \begin{minipage}{0.32\textwidth}
        \centering
        \includegraphics[width=\textwidth]{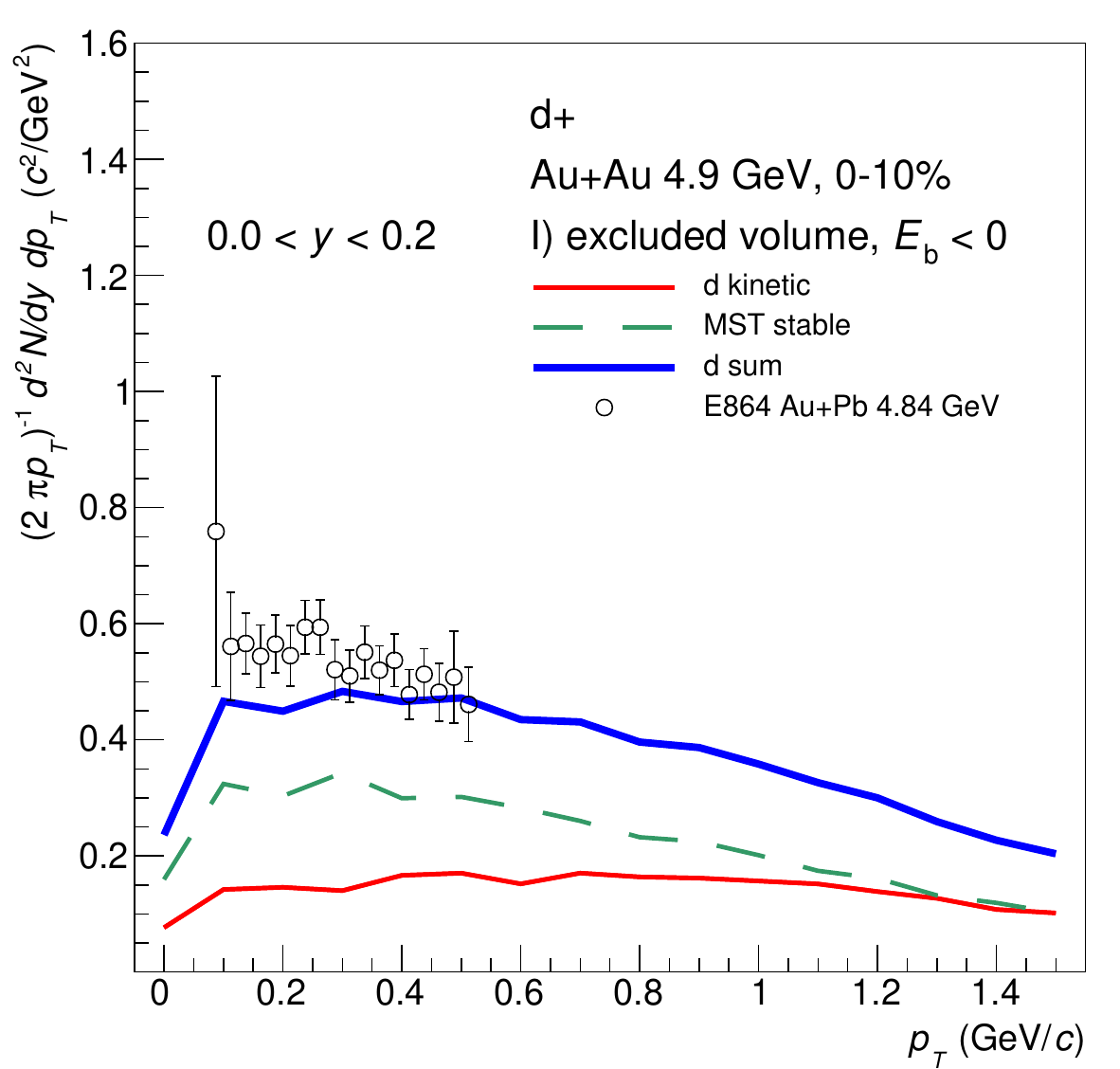}
    \end{minipage}
    \begin{minipage}{0.32\textwidth}
        \centering
        \includegraphics[width=\textwidth]{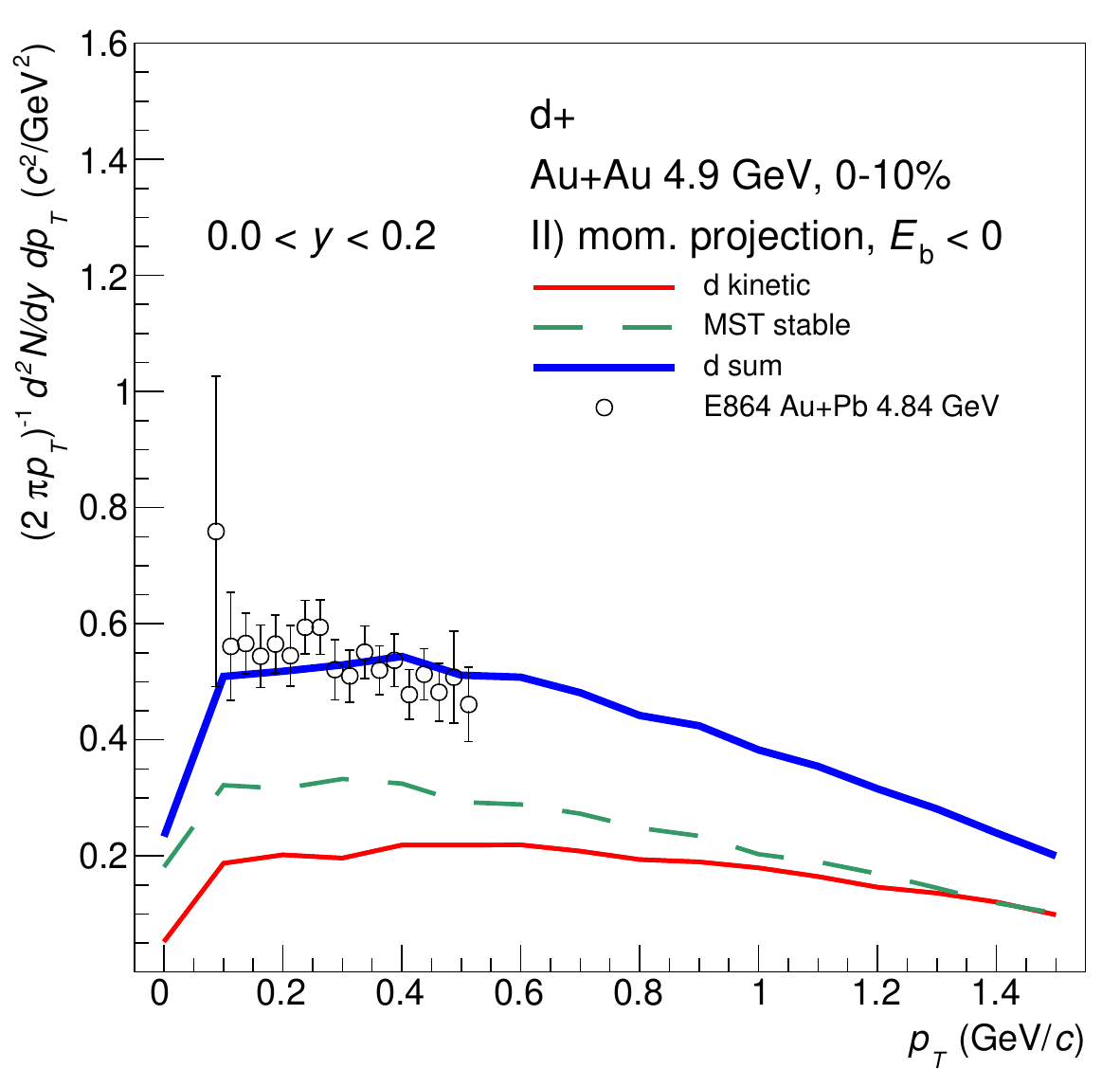}
    \end{minipage}
    \begin{minipage}{0.32\textwidth}
        \centering
        \includegraphics[width=\textwidth]{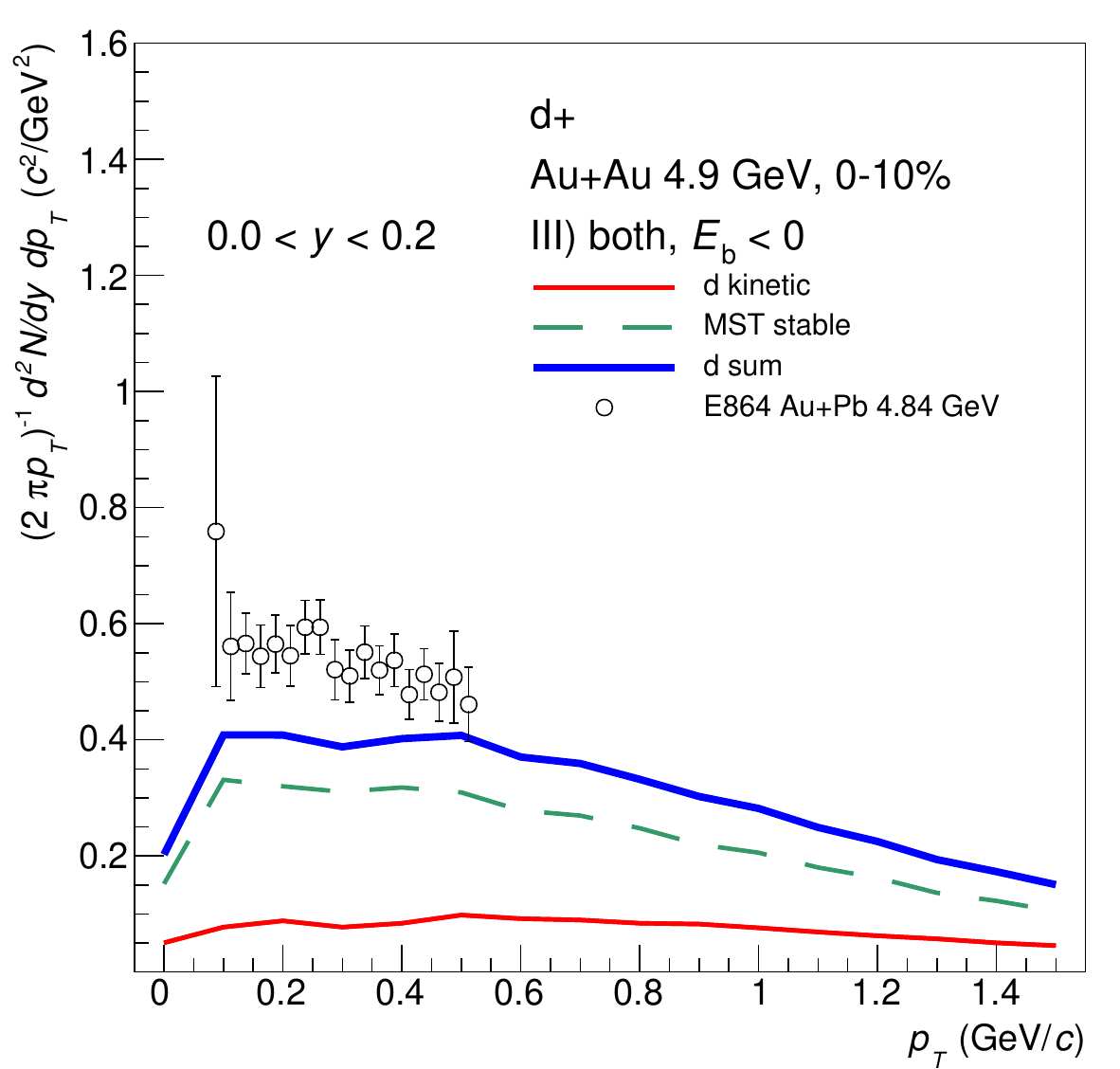}
    \end{minipage} \\
    \begin{minipage}{0.32\textwidth}
        \centering
        \includegraphics[width=\textwidth]{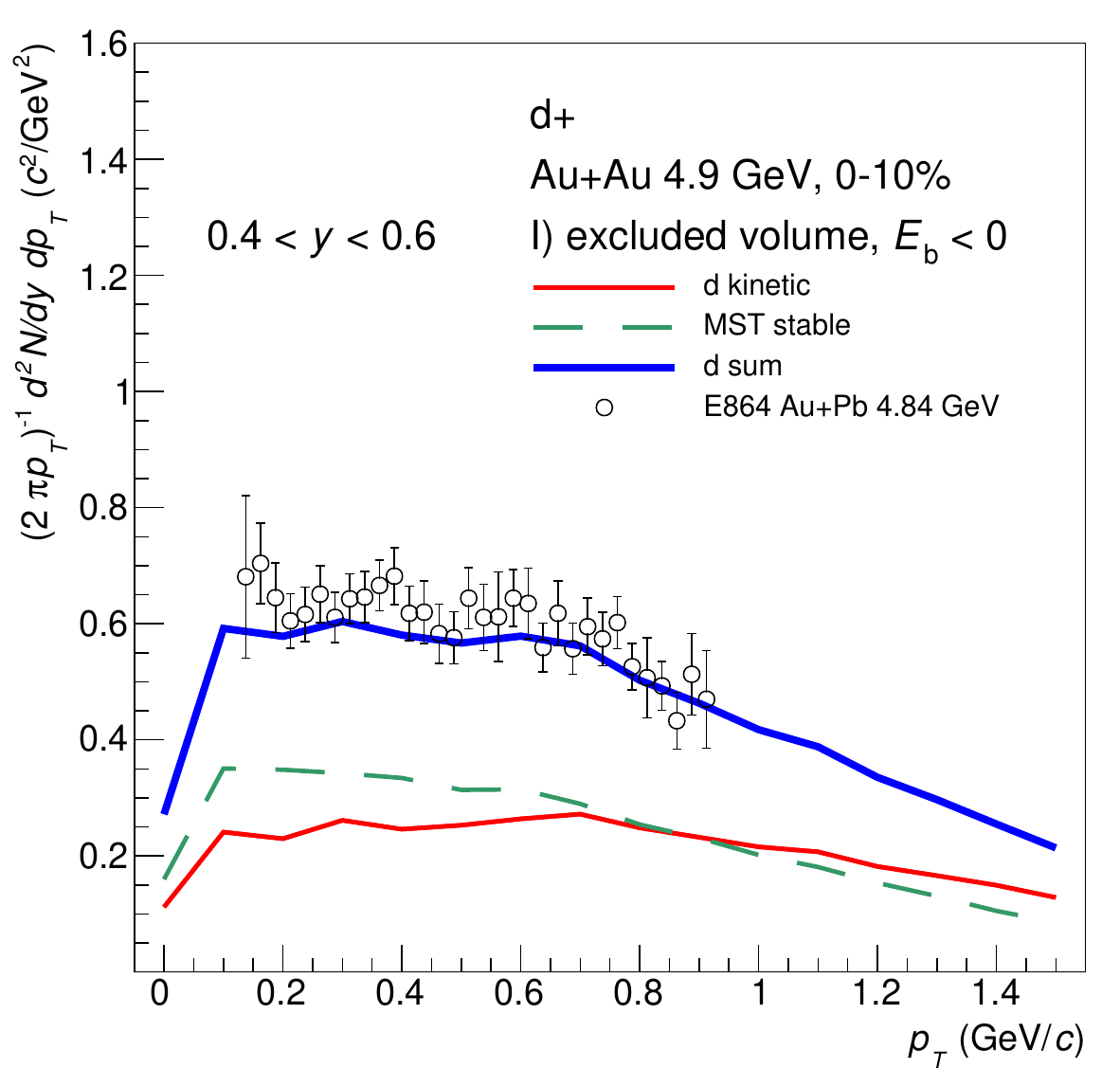}
    \end{minipage}
    \begin{minipage}{0.32\textwidth}
        \centering
        \includegraphics[width=\textwidth]{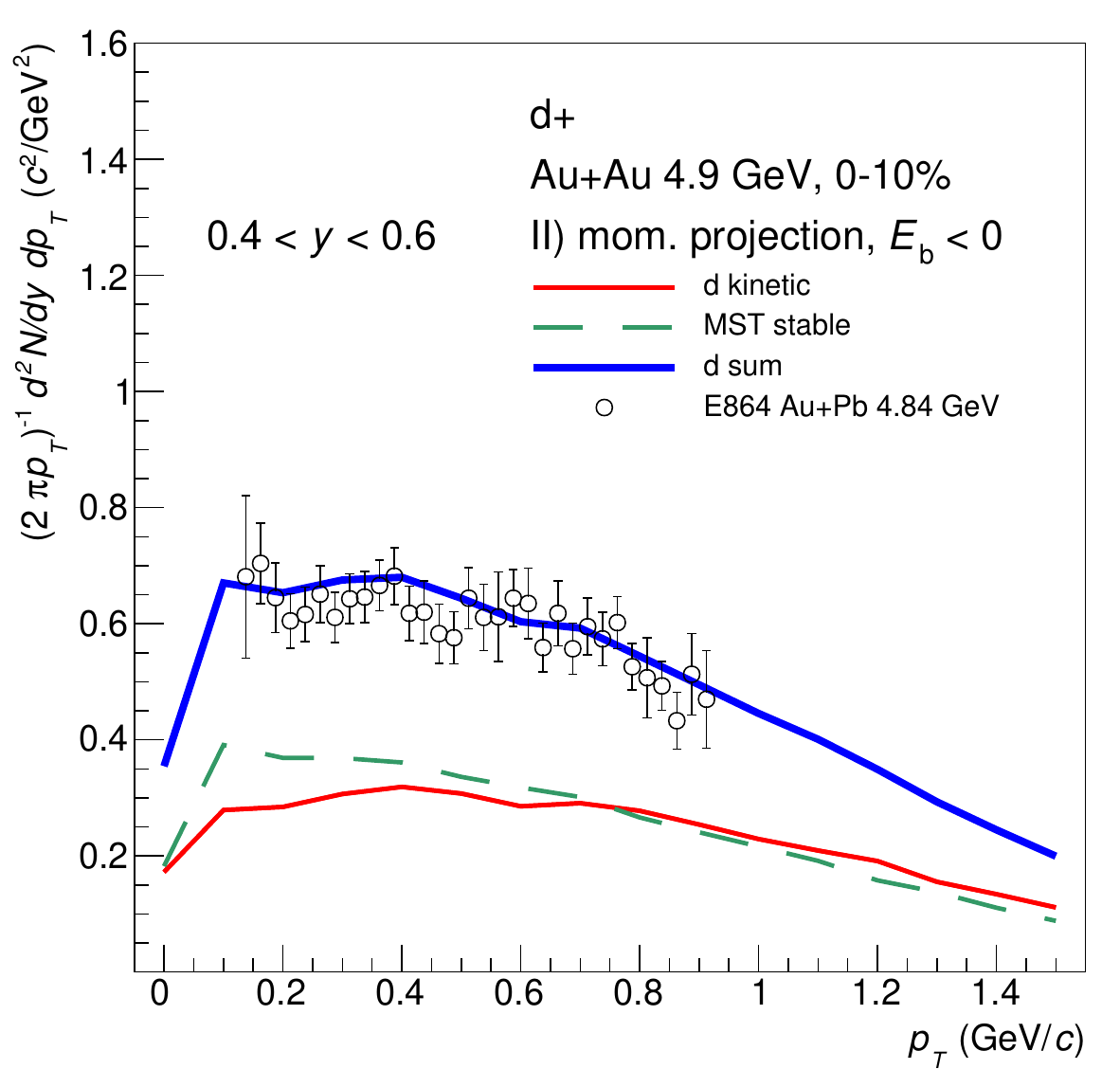}
    \end{minipage}
    \begin{minipage}{0.32\textwidth}
        \centering
        \includegraphics[width=\textwidth]{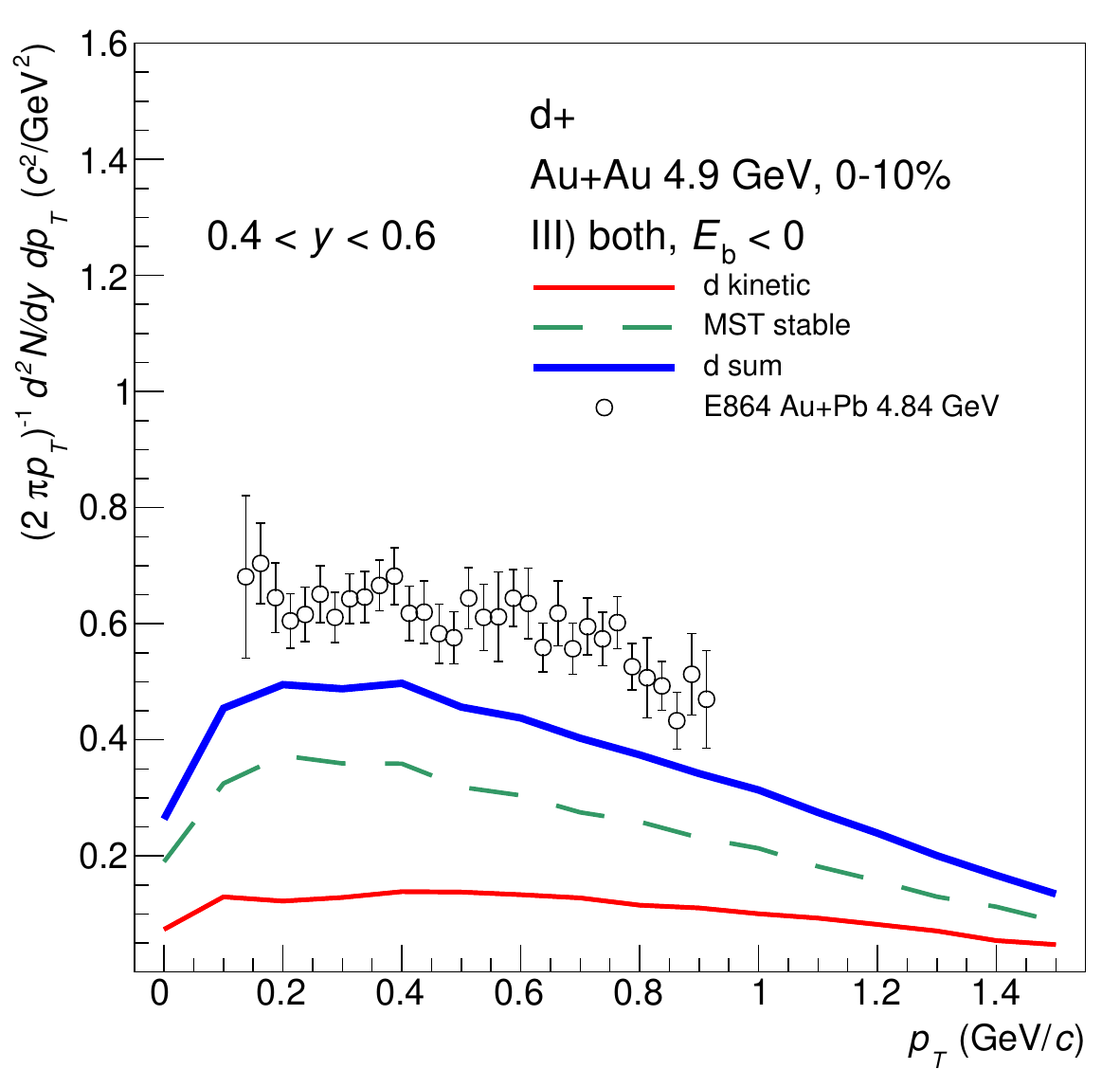}
    \end{minipage} \\
     \begin{minipage}{0.32\textwidth}
        \centering
        \includegraphics[width=\textwidth]{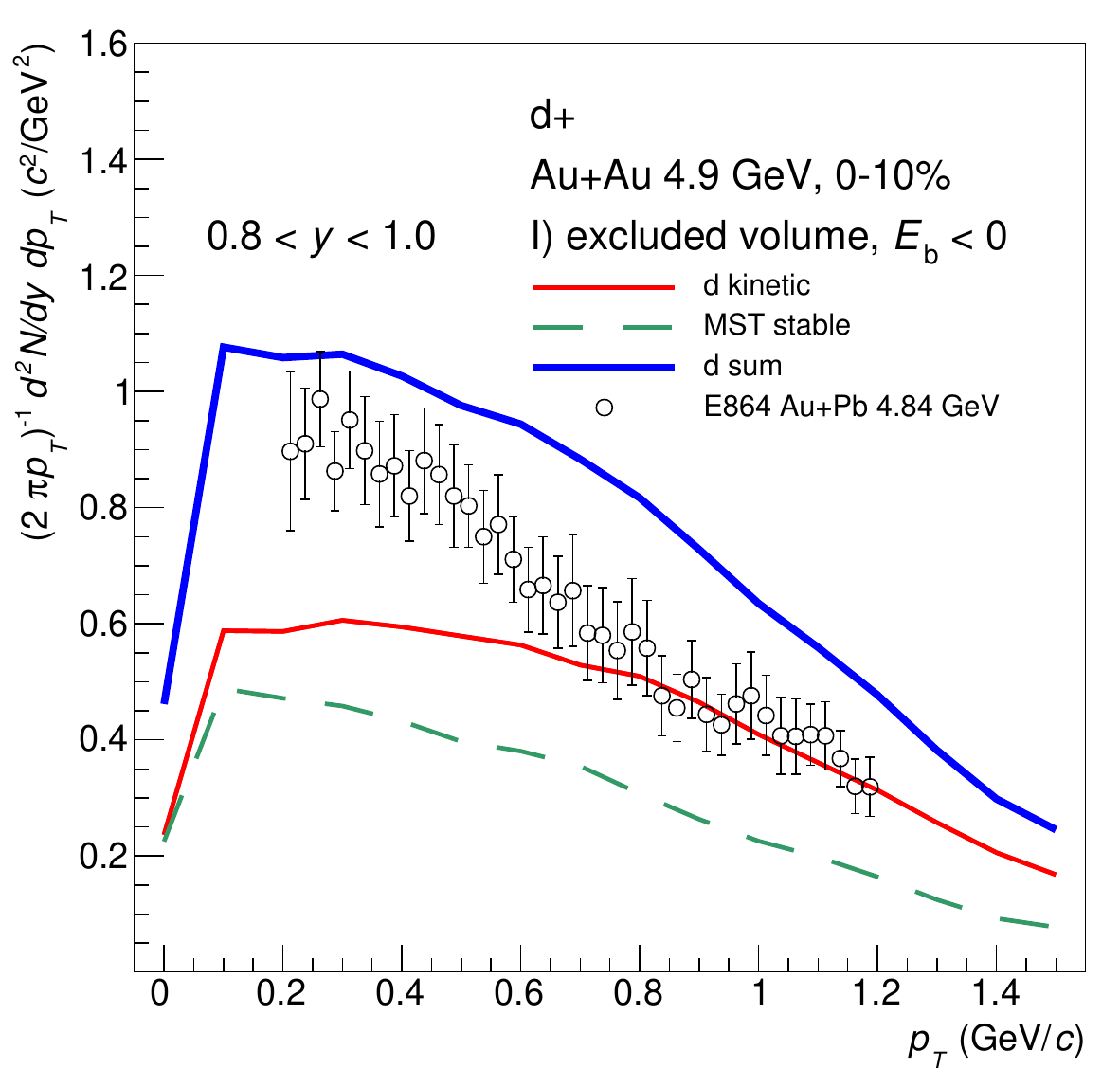}
    \end{minipage}
    \begin{minipage}{0.32\textwidth}
        \centering
        \includegraphics[width=\textwidth]{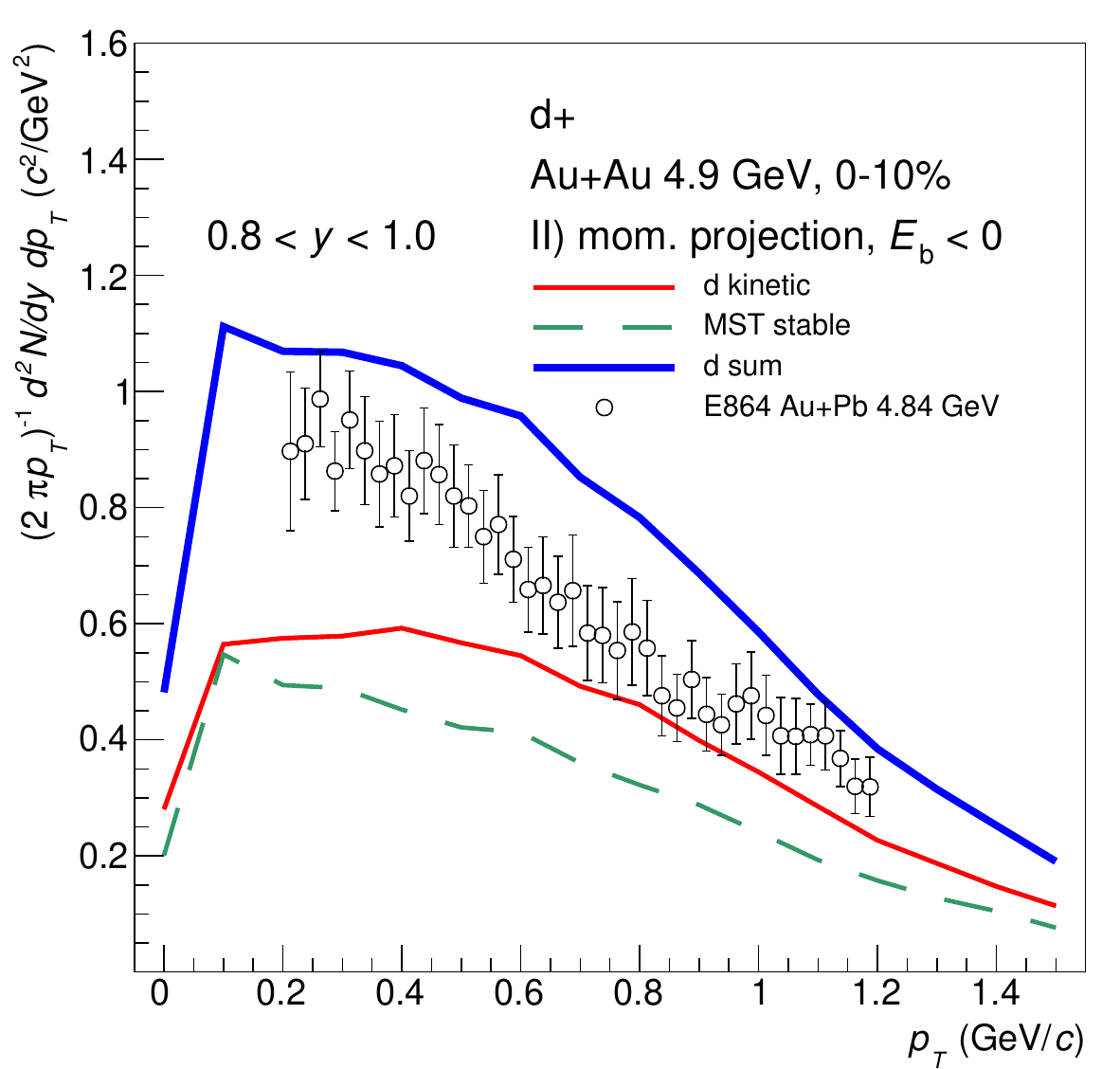}
    \end{minipage}
    \begin{minipage}{0.32\textwidth}
        \centering
        \includegraphics[width=\textwidth]{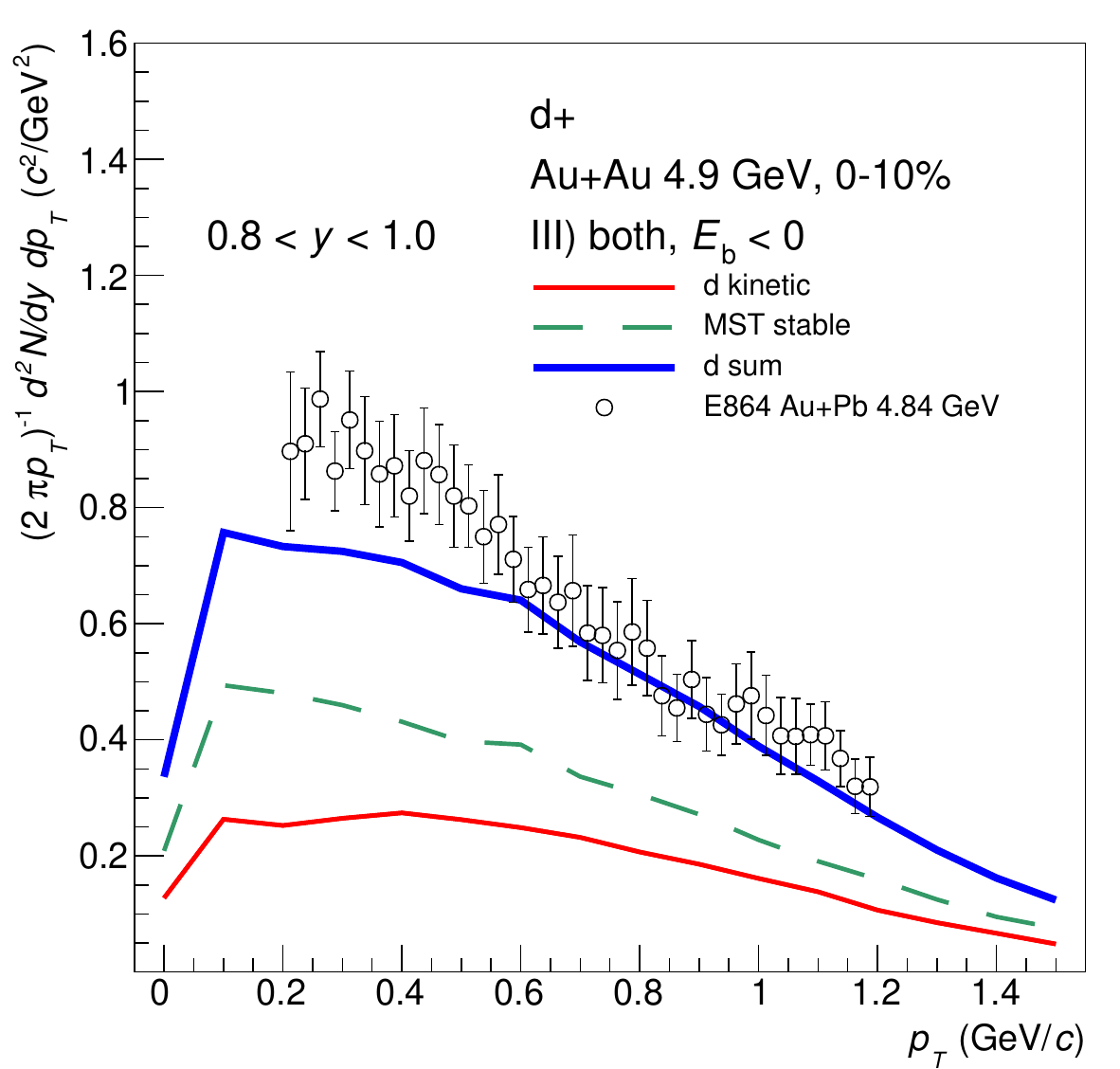}
    \end{minipage} \\
\caption{\label{fig:ptCBM} (color online) The transverse momentum distribution of deuterons in 10\% central Au+Au collisions at $E_{Lab}=11$ AGeV ($\sqrt{s}=4.9$ GeV) obtained from PHQMD calculations are compared with the experimental data from the E864 collaboration~\cite{Armstrong:2000gz}, shown as open circles. The lines denote the different deuteron contribution in PHQMD from kinetic and potential mechanisms with the same color coding as in Fig.~\ref{fig:yCBM} and described also in the legends. In each column we present the $p_T$-spectra of one of the three models of finite-size effects in kinetic production. From left to right column, respectively: I) excluded-volume condition, II) momentum projection, III) both effects are taken into account. The rows display the results for each scenario in three different rapidity intervals: $0.0<y<0.2$ (top), $0.4<y<0.6$ (middle), $0.8<y<1.0$ (bottom).}
\end{figure*}

\begin{figure}[ht!]
\centering
\includegraphics[width=0.45\textwidth]{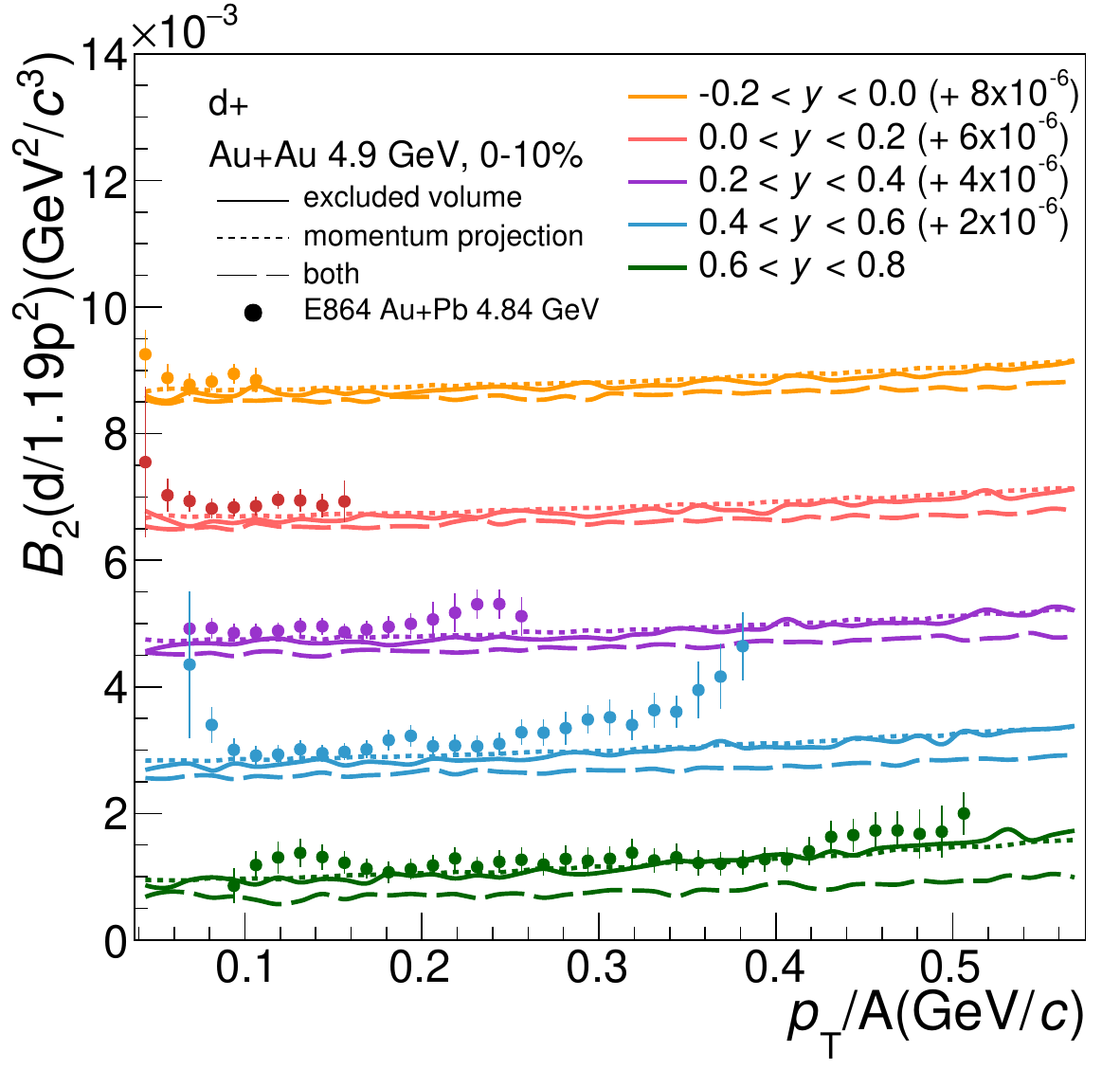}
\caption{\label{fig:B2CBM} (color online) The coalescence parameter of deuterons $B_2$ from PHQMD simulations in Au+Au central collisions at beam energy $E_{kin}=11$ AGeV ($\sqrt{s}=4.9$ GeV center-of-mass energy) is shown as colored lines as function of transverse momentum $p_T/A$, scaled by the deuteron baryon number $A=2$ in several rapidity intervals reported in the legend. The trend of PHQMD results is confronted with the experimental data from the E864 collaboration~\cite{Armstrong:2000gz} displayed with the colored full dots. In order to allow for such a comparison, a neutron to proton ratio of 1.19 is assumed, as it is explained in detail in~\cite{Glassel:2021rod}. The different lines are described in the text.}  
\end{figure}

The PHQMD results for the 10\% most central Au+Au collisions at a laboratory energy $E_{Lab}=11$ AGeV,  corresponding to  $\sqrt{s}=4.9$ GeV, are displayed in Figs.~\ref{fig:yCBM} -~\ref{fig:B2CBM}.
This system will be explored by the future Compressed Baryonic Matter (CBM) experiment at GSI FAIR. Therefore, our results can be considered as predictions until experimental data will be available. However, it is possible to compare our results with the measurements performed at the AGS accelerator for the asymmetric Au+Pb collisions at the same beam energy.

We note, that in our previous study~\cite{Glassel:2021rod} the multiplicity and $p_T$-spectra of the light nuclei $d$, $t$, $^3\!He$ were presented in Au+Pb collisions for the same beam energy and compared to the same data from the E864 experiment~\cite{Armstrong:2000gz}. As mentioned above the clusters were identified by the original MST approach described in Sec.~\ref{sec:sec2}~B.1. In that case the number of clusters was shown to decrease as a function of time due to the instabilities originating from the semi-classical nature of the QMD approach. Therefore, it was necessary to introduce a physical time for the identification of clusters which was around 50 fm/c for deuterons and 60 fm/c for tritons and $^3\!He$.\\ 
In Fig.~\ref{fig:yCBM} the rapidity distributions $(2\pi p_T)^{-1} d^2N/dp_Tdy$ of kinetic deuterons (solid red line), potential deuterons identified with aMST (dashed green line) and the sum of the two contributions (solid blue line) are compared to the data from the E864 collaboration~\cite{Armstrong:2000gz} (open circles). For the PHQMD results we apply the same selection in transverse momentum, $0.2 < p_T < 0.4$ GeV, as in the experiment. Similarly to Fig.~\ref{fig:yFOPI}, the different approaches to model the  finite-size of the deuteron for the kinetic deuterons are separated in three different panels.
The excluded-volume condition (left panel I) and momentum projection (center panel II) give about the same amount of kinetic deuterons at mid-rapidity in agreement with the experimental yield, but they start to overestimate them at larger rapidities. When both effects are applied (right panel III), the shape of the rapidity distribution agrees better with the data points, even though the total yield is slightly underestimated.  This is due to the
aMST deuterons, which have a shallower distribution.\\
In Fig.~\ref{fig:ptCBM} we show the transverse momentum distribution $(2\pi p_T)^{-1} d^2N/dp_Tdy$ of deuterons as function of $p_T$ for the same collision system and at three different rapidity intervals indicated in each plot.
The color coding  is the same as in Fig.~\ref{fig:yCBM}.  We observe that the $p_T$-slope of kinetic deuterons is insensitive to the modelling of finite-size effects and combined with the potential deuterons give a good description of the trend of the experimental data.

To conclude the analysis at this collision energy we calculate the covariant coalescence function $B_d$ which is defined by the formula
\begin{equation}
B_2=B_d = \frac{E_d \frac{dN_d}{d^3\mathbf{P}_d}}
{E_p \frac{dN_p}{d^3\mathbf{p}_p}E_n \frac{dN_n}{d^3\mathbf{p}_n}}
\end{equation}
for deuterons with momentum $P_d = 2p_p = 2p_n$, where $p_p=p_n$ is the momentum of the free nucleon. $B_2$ we present in Fig.~\ref{fig:B2CBM} as function of the transverse momentum $p_T/2$  and confront it with the experimental data from E864 collaboration~\cite{Armstrong:2000gz}. We refer to Ref.~\cite{Glassel:2021rod} for the details.

As in our previous calculations in Ref.~\cite{Glassel:2021rod} the coalescence function of deuteron $B_2$ shows a quite flat behavior as function of $p_T$ which is also observed in the experimental data, apart from the strong increase at large $p_T$ in the interval $0.4 \le y \le 0.6$. Moreover, the obtained $B_2$ from PHQMD simulations seems to be quite independent from the modelling of finite-size effects in the kinetic production. In particular, when either excluded-volume condition (solid lines) or $NN$-pair momentum projection (dotted lines) are applied separately, the results are practically the same, while when both effects are simultaneously taken into account  (dashed lines) a smaller $B_2$ is observed due to the stronger suppression of kinetic deuterons at large rapidities.
In Fig.~\ref{fig:B2CBM} the PHQMD calculations refer to the sum of the contribution of kinetic and potential deuterons. The latter are identified via the advanced MST (aMST).

\subsection{SPS energies}

\begin{figure*}[ht!]
\centering
\includegraphics[width=0.85\textwidth]{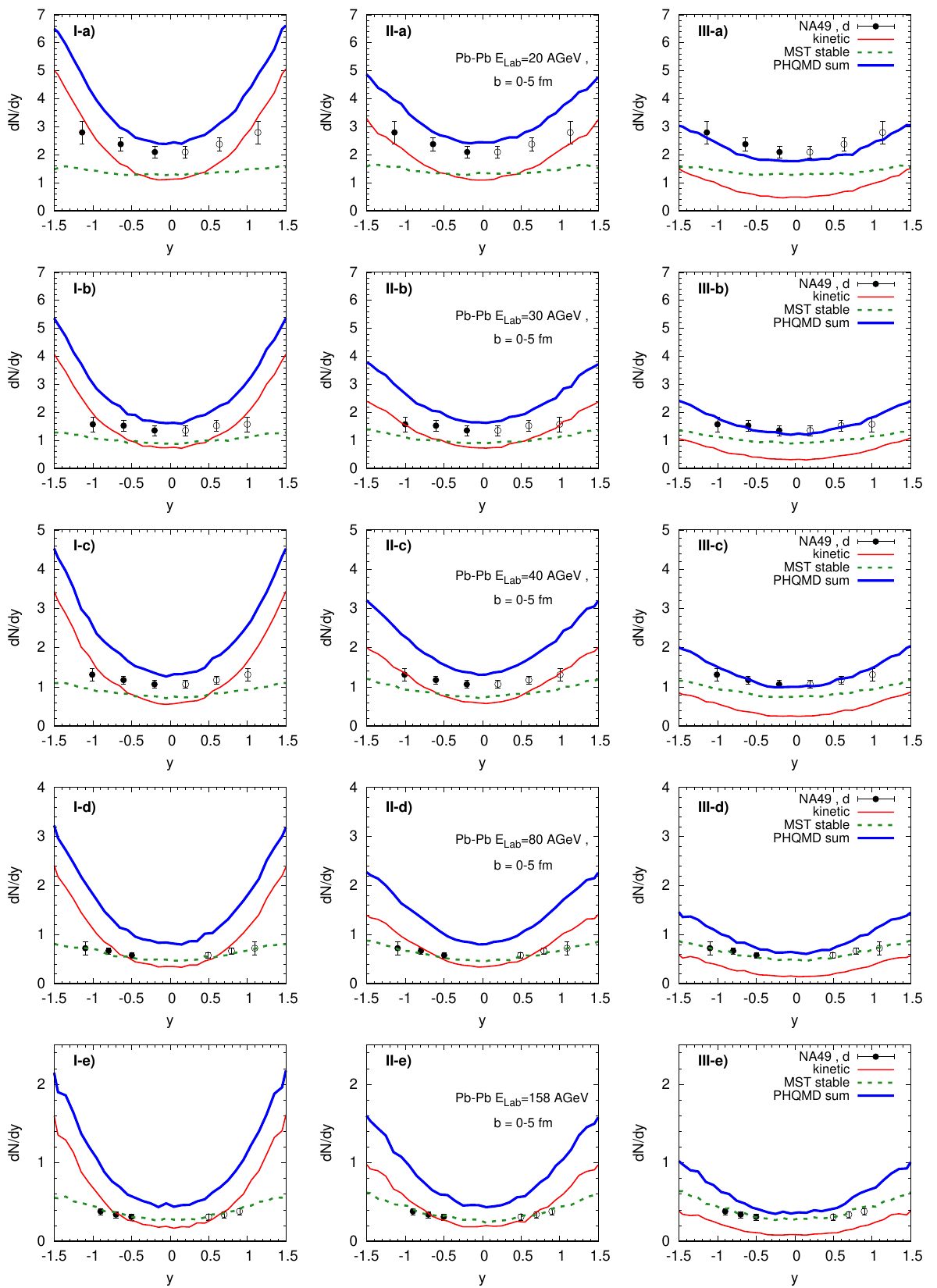}
\caption{\label{fig:ySPS} (color online) The rapidity distributions $dN/dy$ of deuterons for Pb+Pb central collisions (impact parameter interval $b=0-5 fm$) in the full beam energy range of the SPS: from top (a) to bottom (e) panels $E_{Lab}=$ 20, 30, 40, 80, 158 AGeV. The full dots are the experimental data from the NA49 collaboration~ \cite{NA49:2016qvu} (the empty dots are mirrored around mid-rapidity). The lines correspond to the PHQMD results with the same color coding as in Fig.~\ref{fig:yFOPI}: kinetic $d$ (thin red), potential $d$ from aMST, i.e. MST followed by the stabilization procedure (dashed green), total $d$ (thick solid blue). The three columns correspond to the PHQMD calculations for the three different models of finite-size; from right to left: I) only excluded-volume, II) only momentum projection, III) both effects.}  
\end{figure*}

We step up in energy and present the results of the PHQMD approach for heavy-ion collisions in the CERN SPS energy range. 
The rapidity distribution  $dN/dy$ of kinetic and potential deuterons in central  Pb+Pb collisions are shown in Fig.~\ref{fig:ySPS}. The columns represent our three different options to model  finite-size effects in the collisional deuteron  production. From left to right we display: I) excluded-volume, II) momentum projection, III) both together. The color coding is the same as in Fig.~\ref{fig:yFOPI}. The rows collect results for the full energy range of the SPS facility; from top  (a) to bottom  (e) we see: $E_{Lab}=$ 20, 30, 40, 80, 158 GeV per nucleon. The dots are the experimental data from the NA49 collaboration~\cite{NA49:2016qvu}.

By comparing the first and the second columns, we observe that even though the excluded-volume and the momentum projection give a similar suppression of deuterons at mid-rapidity (Fig.~\ref{fig:C5}), at target-projectile rapidity their effect on deuteron $dN/dy$ is quite different. There mostly spectator nucleons are localized whose density in central collisions is not very high. Their momentum distribution is close to the Fermi distribution because these baryons have not scattered or scatter with a small momentum transfer. Therefore, the excluded-volume prescription of deuteron production gives less suppression than the projection on the deuteron wave function. Combining both prescription lowers the deuteron production further (column III). The rapidity distribution of kinetic deuterons is narrower at mid-rapidity than the experimental data. If one adds the deuterons created by potential interactions (the sum is presented by the full blue line), which is wider than that of the kinetic deuterons, the experimental multiplicity as well as the rapidity distribution of deuterons, measured by the NA49 collaboration~\cite{NA49:2016qvu}, is nicely reproduced.  

\begin{figure*}[ht!]
\centering
\includegraphics[width=0.85\textwidth]{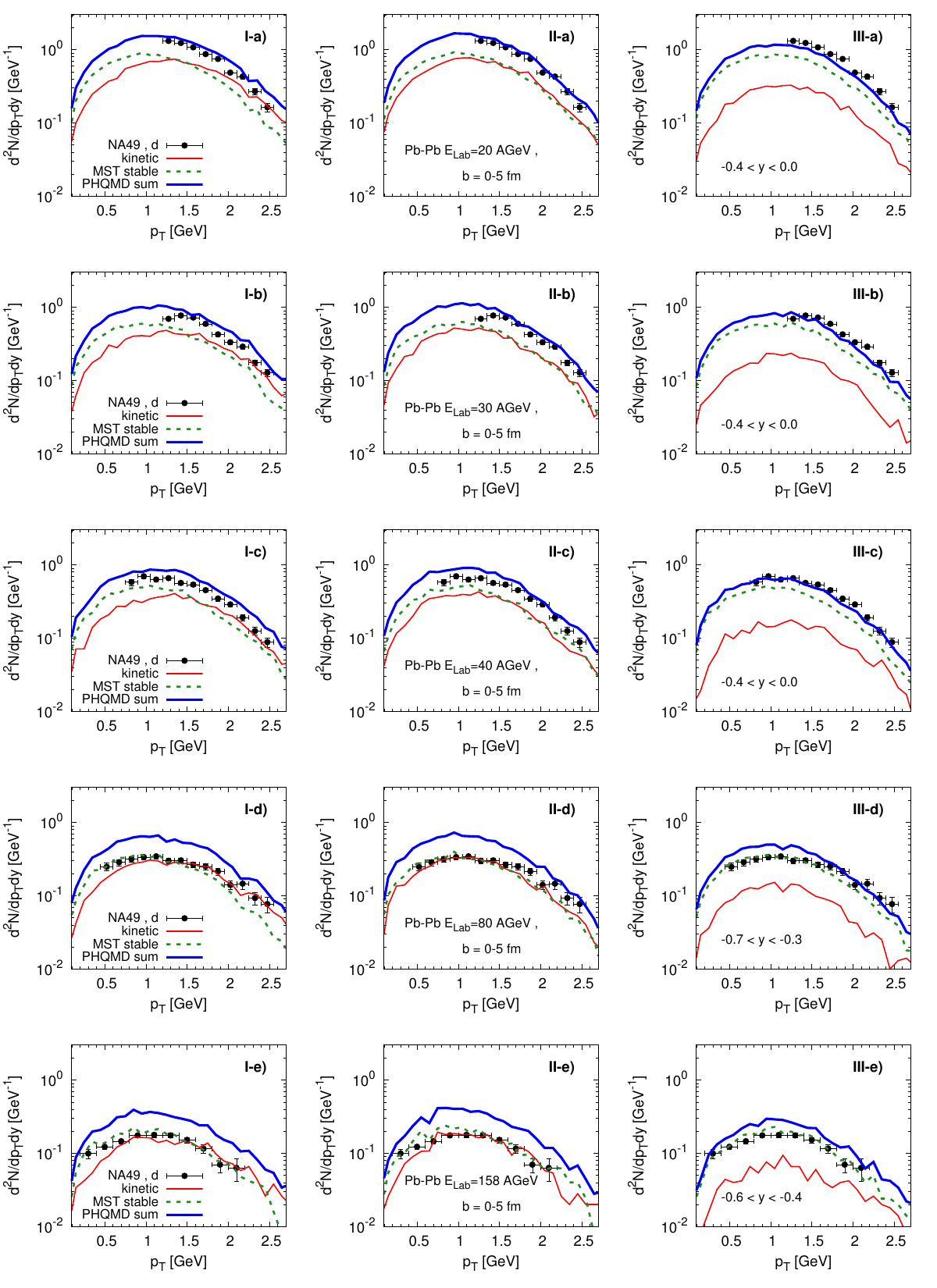}
\caption{\label{fig:pTSPS} (color online) The transverse momentum distributions $dN/dp_Tdy$ of deuterons for Pb+Pb central collisions (impact parameter interval $b=0-5 fm$) in the full beam energy range of the SPS: from top (a) to bottom (e) panels $E_{Lab}=$ 20, 30, 40, 80, 158 AGeV. The full dots are the experimental data from the NA49 collaboration~ \cite{NA49:2016qvu}. The style and color coding of the lines representing  the PHQMD results is the same as for the rapidity distributions, Fig.~\ref{fig:pTSPS}, as well as the ordering of the three columns, which denote the different finite-size models for kinetic deuterons; from right to left: I) only excluded-volume, II) only momentum projection, III) both effects. The PHQMD results for the $p_T$-spectra are calculated for the same experimental rapidity interval which is indicated in the right panels for each collision energy.}  
\end{figure*}

The transverse momentum distributions $d^2N/dp_Tdy$ of deuterons at mid-rapidity are shown in Fig.~\ref{fig:pTSPS} with the same color coding and panel structure as in Fig.~\ref{fig:ySPS}. For each energy $E_{Lab}$ the rapidity interval for the $p_T$-spectra is indicated in the right column and taken in correspondence with the NA49 experimental data~\cite{NA49:2016qvu}. While the excluded-volume (left column) and the momentum projection (middle column) give roughly the same suppression of kinetic deuterons, both effects combined yield an additional suppression factor of 2, as already seen in the rapidity distribution. The form of the $p_T$-spectra is for all three approaches to model the finite-size effects the same.

\subsection{RHIC BES energies}

\begin{figure}[ht!]
\centering
\includegraphics[width=0.45\textwidth]{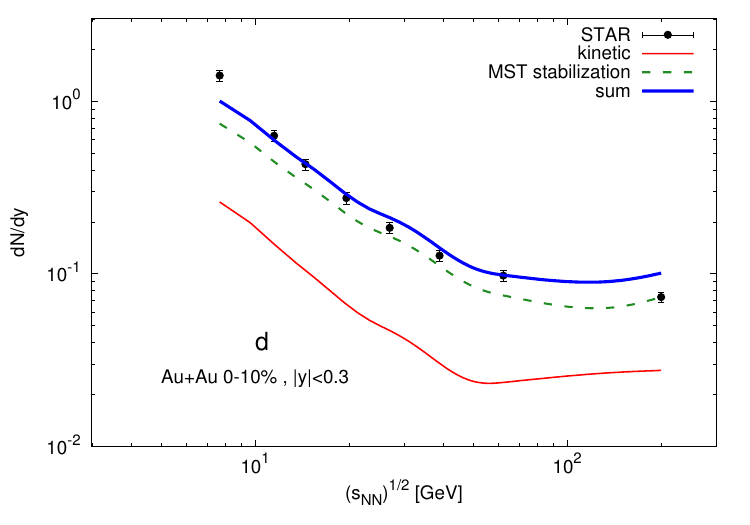}
\caption{\label{fig:BESexcid} (color online) The mid-rapidity $|y|<0.3$ excitation function for $dN/dy$ of deuterons as a function of $\sqrt{s_{NN}}$ for Au+Au $0-10\%$ central collisions in comparison with the experimental data from the STAR collaboration~\cite{STAR:2019sjh}. The different lines indicate the different deuteron contributions: kinetic production with modelling of finite-size effects in coordinate and momemtum space (solid red), potential from MST with stabilization, i.e. advanced MST (dashed green), sum (blue). The rapidity interval of the PHQMD results is the same as that measured by STAR.}  
\end{figure}

PHQMD is designed to describe clusters also at higher energies. Therefore, we can study the deuteron production in the full energy range of the RHIC Beam Energy Scan (BES). The STAR collaboration has measured in this energy region the multiplicity at mid-rapidity, as well as the mid-rapidity $p_T$ distribution~\cite{STAR:2019sjh}. Here we present the results for model \textit{III}, which accounts for both finite-size effects: in coordinate by an excluded-volume with radius $R_d=1.8$ fm and in momentum space by the projection on the DWF. The kinetic deuterons are supplemented by the potential deuterons, calculated with the advanced MST (aMST) method.
We start out by showing the excitation function of the deuteron yield $dN/dy$ at mid-rapidity, which has been already studied  in ~\cite{Glassel:2021rod}  for the standard MST deuteron recognition algorithm. It is displayed in Fig.~\ref{fig:BESexcid} for central Au+Au collisions as function of $\sqrt{s_{NN}}$. The black points represent the data at mid-rapidity, measured by the STAR experiment~\cite{STAR:2019sjh}. The lines are the PHQMD results for the same rapidity interval, $|y|<0.3$. The color coding is the same as in  Fig.~\ref{fig:yFOPI}.
The combined PHQMD results are in quite good agreement with the data, giving slightly less deuterons at the lowest RHIC BES energy $\sqrt{s_{NN}}=7.7$ GeV and slightly more deuterons at the top RHIC  energy, $\sqrt{s_{NN}}=200$ GeV, compared to experimental data points. 

\begin{figure}[ht!]
\centering
\includegraphics[width=0.45\textwidth]{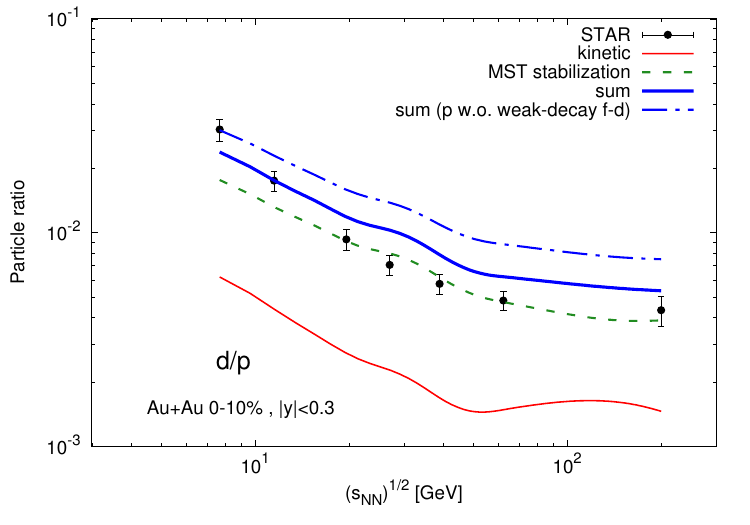}
\caption{\label{fig:BESexcidpratio} (color online) The mid-rapidity $|y|<0.3$ deuteron to proton $d/p$ ratio for Au+Au central collisions as a function of $\sqrt{s_{NN}}$. The different lines indicate the different deuteron contributions: kinetic production with finite-size effects (solid red), advanced MST identification (dashed green), sum (blue). The experimental data from the STAR collaboration~\cite{STAR:2019sjh} are indicated with the full circles. The PHQMD results are scaled in order to account for the protons from weak decay feed-down, which is included in the STAR data. The PHQMD $d/p$ result without feed-down contribution is shown with the dot-dashed blue line.}  
\end{figure}

It is visible that at these energies the kinetic contribution is small compared to that of aMST (roughly by a factor of 3 less) and, therefore, the multiplicity is dominated by potential deuterons.
The STAR collaboration has also measured the $d/p$ ratio at mid-rapidity and, hence, the fraction of mid-rapidity protons which is bound in the lightest cluster. It is shown in Fig.~\ref{fig:BESexcidpratio} as a function of the NN center-of-mass energy. Again we separate kinetic and potential deuterons. In this energy regime, as we have already seen, the kinetic deuterons contribute only around 30\% to the total yield. It is remarkable and unexpected that the form of the excitation function for kinetic and potential deuterons is very similar. If we add kinetic and potential deuterons we overpredict above $\sqrt{s}=10$ GeV this ratio by about 30\% what represents almost exactly the contribution of kinetic deuterons.  

\begin{figure*}[ht]
\centering
\includegraphics[width=\textwidth]{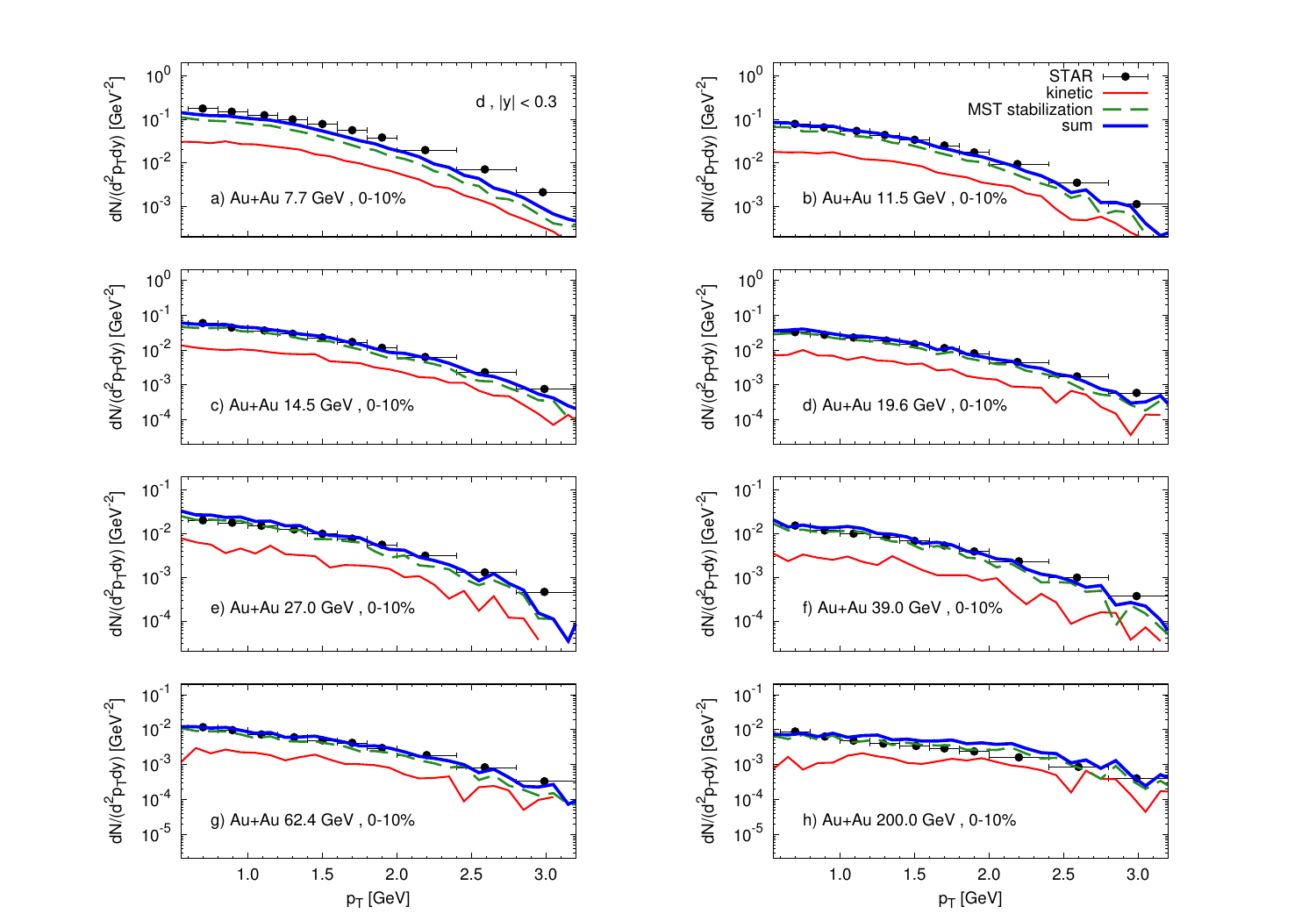}
\caption{\label{fig:BESpTd} (color online) Transverse momentum distributions of deuterons in Au+Au central collisions at all RHIC BES energies from $\sqrt{s_{NN}}= 7.7$ GeV (a) to $200$ GeV (h). The points are the experimental data from the STAR collaboration~\cite{STAR:2019sjh}. The lines correspond to the PHQMD results for the different production mechanisms: kinetic with finite-size (solid red), potential from advanced MST identification (dashed green) and sum of all contributions (thick solid blue). The PHQMD $p_T$-spectra are taken at the same mid-rapidity interval $|y|<0.3$ as the STAR measurements.}  
\end{figure*}

In Fig.~\ref{fig:BESpTd} we present the transverse momentum distribution of deuterons as function of $p_T$ for the same central Au+Au collisions and for the same mid-rapidity interval $|y|<0.3$. The color and symbol coding is identical to the one used in Fig.~\ref{fig:BESexcid}. Again the combined deuteron yield overpredicts the experimental result by roughly 30\%.
One can observe that the $p_T$-spectra of kinetic (thin solid red line) and aMST (dashed green line) deuterons have a quite similar shape and the total $p_T$-spectra (thick solid blue line) is in good agreement with the measured spectra in the wide energy range of RHIC BES.

\clearpage

\section{Conclusions}\label{sec:sec7}

In this study we have investigated the production of deuterons  in nucleus-nucleus collisions within the PHQMD microscopic transport approach. We have focused on the description of deuteron rapidity and transverse momentum distributions around mid-rapidity from  SIS $E_{Lab}=1.5$ AGeV up to top RHIC $\sqrt{s}=200$ GeV, covering essentially the whole energy range of relativistic HICs. 
In the PHQMD framework we have studied two possible mechanisms for the dynamical formation of deuterons - by collisions and by potential interaction.
The results can be summarized as follows: 
\begin{itemize}
    \item ``Kinetic" mechanism:\\
    \textit{i)} We have implemented $\pi NN \leftrightarrow \pi d$, $NNN \leftrightarrow N d$ and the sub-dominant $NN \leftrightarrow \pi d$ by means of the covariant rate formalism~\cite{Cassing:2001ds} in the PHQMD collision integral.
    The numerical implementation of the $2 \leftrightarrow 3$ and $2 \leftrightarrow 2$ reactions has been tested by the stationary ``box" calculation in comparison with the solutions of the rate equations. Moreover, it has been verified that the detailed balance condition is fulfilled for each isospin channel.
    
    \textit{ii)} Differently to the previous study by the SMASH group~ \cite{Staudenmaier:2021lrg}, we have accounted in the main $\pi NN \leftrightarrow \pi d$ and $N NN \leftrightarrow N d$ reactions for all possible reaction channels which are allowed by the conservation of total isospin. We have found that the inclusion in the $\pi-$catalysis - which at high collision energies is more dominant than $N$-catalysis due to the large pion abundance - of all $\pi$ charge exchange reactions enhances the production of ``kinetic" deuterons by about 50\% at mid-rapidity for $\sqrt{s}=7.7$ GeV STAR BES energy, while at $\sqrt{s_{NN}}=3$ GeV, the energy of the STAR FXT experiment, the $\pi$ charge exchange channels increase the deuteron yield by 20\%.

    \textit{iii)} 
    The deuteron, being an extended quantum object in coordinate space with a small relative momentum cannot be produced as a point-like hadron. We have taken this into account by two approaches: I) an excluded-volume condition which suppresses the formation of deuterons in the presence of surrounding hadrons and II) by a projection of the relative momentum of the $NN$-pairs on the deuteron wave function. We have shown that the inclusion of each of these finite-size effects leads to a significant but similar suppression of deuterons at mid-rapidity. Applying both effects together the deuteron yield is suppressed by an additional factor of two.

    \item ``Potential" mechanism:\\
     We have extended our study of deuteron formation by ``potential'' interaction between nucleons in Refs. \cite{Aichelin:2019tnk,Glassel:2021rod}, where we had to determine a time at which the cluster recognition by the Minimum-Spanning-Tree (MST) approach has been performed. 
    The newly developed ``advanced" MST (aMST) method stabilizes the clusters, which in semi-classical approaches are not stable and whose stability suffers from the Lorentz transformation between the cluster center-of-mass system (where the binding energy is determined) and the nucleus-nucleus center-of-mass, i.e. the computational frame. In aMST clusters, which are bound, cannot disintegrate after the constituents had their last collision and are outside the range of the potential interaction of other clusters and nucleons. In aMST the clusters are stable and therefore no time has to be determined at which we  analyse them. 

   \item As found in our previous studies~\cite{Glassel:2021rod,Kireyeu:2022qmv}, the clusters - produced via potential mechanisms - are created after the fast hadrons have already escaped from the reaction zone, i.e. clusters remain in transverse direction closer to the center of the heavy-ion collision than free nucleons. The ``kinetic'' deuterons analyzed here follow the same tendency. Thus, since the ``fire" is not at the same place as the ``ice", clusters can survive, what solves the ``ice in the fire" puzzle.
    
\end{itemize}

We have found that the PHQMD approach with the two mechanisms of deuteron production, consistently combined, provides a good description of the large set of available experimental data from SIS~\cite{Reisdorf:2010aa}, NA49~\cite{NA49:2016qvu} and STAR~\cite{STAR:2019sjh} collaborations. 
Finally, we mention also that our results for Au+Au collisions at AGS and RHIC BES energies are relevant for the future experiments which will be carried out at the FAIR and NICA facilities.

\section*{Acknowledgements}
The authors acknowledge inspiring discussions with M. Bleicher, W. Cassing, H. Elfner, I. Grishmanovskii, C.-M. Ko, D. Oliinychenko, L. Oliva, T. Reichert, O. Soloveva, T. Song, J. Staudenmaier, J. Steinheimer, Io. Vassiliev. 
Furthermore, we acknowledge support by the Deutsche Forschungsgemeinschaft (DFG, German Research Foundation) grant BL982-3 by the Russian Science Foundation grant 19-42-04101 and by the GSI-IN2P3 agreement under contract number 13-70.
This study is part of a project that has received funding from the European Union’s Horizon 2020 research and innovation program under grant agreement STRONG – 2020 - No 824093.
The computational resources have been provided by the Center for Scientific Computing (CSC) of the Goethe University and the "Green Cube" at GSI, Darmstadt.

\section*{Appendix A: Cross sections}
In this appendix we make a collection of scattering cross sections for the reactions of $d$ production and breakup with nucleons and pions which have been implemented in this work. 

\begin{figure}[ht]
\centering
\includegraphics[width=0.5\textwidth]{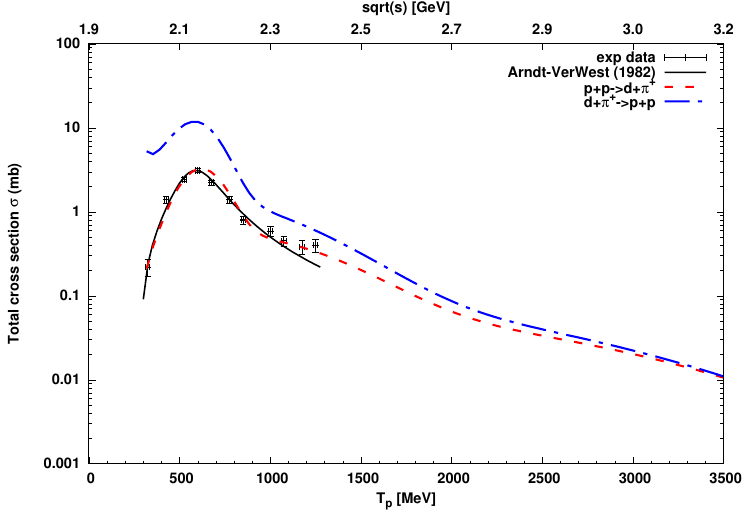}
\caption{\label{fig:A1} (color online) The total cross section for two-body inelastic $p(n)p(n) \rightarrow \pi^{+(-)} d$ for $d$ formation (red dashed line) and $\pi^{+(-)} d \rightarrow p(n)p(n)$ for d disintegration (blue solid line) as a function of the center-of-mass energy $\sqrt{s}$ (in GeV) or converted as a function of the proton laboratory kinetic energy $T_p$ (in MeV) taken from Ref.~\cite{Oh:2009gx}. The black squares and the solid black line are, respectively, experimental data and result from isospin decomposition taken from Ref.~\cite{VerWest:1981dt}.}  
\end{figure}

{\bf I.} Elastic processes $N d \rarr N d$ and $\pi d \rarr \pi d$ are characterized by cross sections of the order of $\sigma_{el} \simeq 60 \, mb$ for $\pi d$ scattering~\cite{Norem:1971nu}. 
We use the parametrization of elastic cross sections as function of invariant center-of-mass energy $\sqrt{s}$ reported in Ref.~\cite{Oh:2009gx} (see Appendix there and reference therein).
Deuterons can be inelastically produced in $p+p$ collision with projectile energy of the order of $T_p \simeq 1$  GeV and accompanied pion emission. Conversely, a projectile pion with beam energy $T_{\pi} \simeq 0.1 \, GeV$ hitting a deuteron target can be absorbed and breakup the deuteron into a final $NN$ pair without pion emission in the final state. The complete reactions $NN \lrarr \pi d$ represent a two-body inelastic process of $d$ production and disintegration. The total cross section for this reaction in both directions have been extensively analyzed within isospin decomposition formalism~\cite{VerWest:1981dt,Arndt:1997if,Oh:1997eq} 
and it has been proven that the detailed balance condition is fulfilled. This allows to relate cross section of forward and backward reactions as follows
\begin{equation}\label{eq:App1} 
\sigma(\pi d \rightarrow N N) = \frac{2}{3}\frac{(p^*_N)^2}{(p^*_\pi)^2} \sigma(NN \rightarrow \pi d) \, ,
\end{equation}  
where $p^*_N$ and $p^*_{\pi}$ are respectively the nucleon and pion momentum computed in the center-of-mass frame of the particles pair, which by simple kinematics can be written in terms of masses and $\sqrt{s}$, respectively, 
\begin{equation} 
 p^*_{N} = \frac{\sqrt{s(s-4m_{N}^2)}}{2 \sqrt{s}} \, ,
\end{equation}
for the nucleon momentum $p^*_{N}$ and 
\begin{equation} 
 p^*_{\pi} = \frac{\sqrt{(s-(m_{\pi}+m_{d})^2)(s-(m_{\pi}-m_{d})^2)}}{2 \sqrt{s}} \,
\end{equation}
for the pion momentum $p^*_{\pi}$. 
In Fig.~\ref{fig:A1} the cross section for $pp\rightarrow \pi^+ d$ as function of $\sqrt{s}$, which we also take from Ref.~\cite{Oh:2009gx}, is shown (red dashed line) in comparison with previous calculations~\cite{VerWest:1981dt} (black solid line) as a function of the kinetic energy of the projectile proton $T_p=\sqrt{P_{lab}^2+m_N^2}-m_N$, where $P_{lab}$ is the beam momentum, which is used to calculate the corresponding value of $\sqrt{s}$ according to the formula
\begin{equation}
\sqrt{s}=\left[2m_N \left(2m_N+T_p \right) \right]^{1/2} \, ,
\end{equation}
in the laboratory frame where the target proton is at rest. The black points refer to some experimental measurements of $pp\rightarrow \pi^+ d$ in the $T_p$ range of the peak \cite{Anderson:1974tp,Shimizu:1982dx}. Still in Fig.~\ref{fig:A1} the cross section for inverse process, i.e. inelastic two-body $d$ breakup by incident $\pi^+$ into $pp$ pair with no pion emission in the final state, obtained from detailed balance condition Eq.~\eqref{eq:App1} is shown with blue dash-dotted curve. We can use same arguments to implement the reaction $nn \rightarrow \pi^- d$, but for the case $pn \rightarrow \pi^0 d$ we have to account for an extra isospin factor and write
\begin{equation}
\sigma(pn \rightarrow d\pi^0) = \frac{1}{2} \sigma(pp \rightarrow d \pi^+) \, ,
\end{equation}
The factor $1/2$ is due to the fact that in terms of isospin base the $pn$ pair is an anti-symmetric superposition of isospin triplet ($T=1$) and singlet ($T=0$) with three-component projection $T_3=0$, while on the other hand the final $\pi^0 d$ depends only on the pion isospin, hence it is a pure triplet state.

\begin{figure}[ht]
\centering
\includegraphics[width=0.5\textwidth]{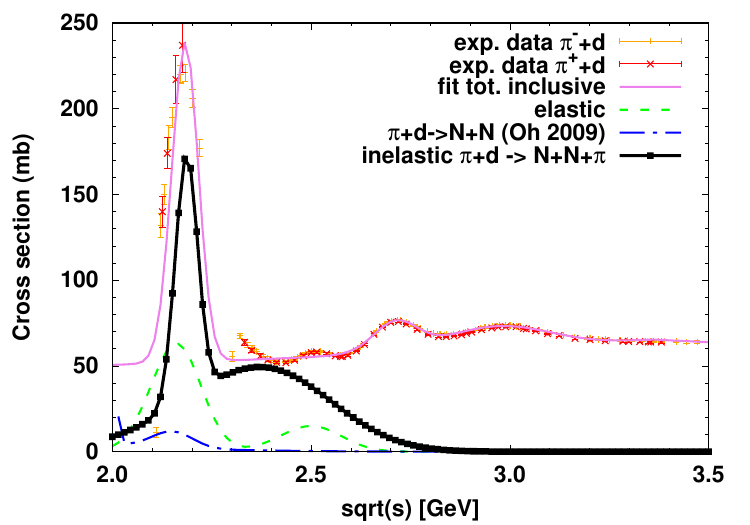}
\caption{\label{fig:A2} (color online) The inelastic cross section for the dominant $\pi d \rightarrow NN\pi$ inelastic scattering (squared-thick black line) as a function of $\sqrt{s}$ is compared to experimental data of total inclusive $\pi d$ cross section from the Particle Data Group (PDG)~\cite{Tanabashi:2018oca}. The cross sections, respectively, for the elastic $\pi d\rightarrow \pi d$ (green dashed) and the two-body inelastic $ \pi d \rightarrow NN$ scattering (blue dash-dotted) with parametrization from Ref.~\cite{Oh:2009gx} are also shown.}  
\end{figure}

As can be seen from Fig.~\ref{fig:A1} the experimental two-body cross section for $d$ breakup into $NN$-pair without pion emission in the final state is quite small and of the order of $\sigma \simeq 10$ mb at the peak.\\ 
In Fig.~\ref{fig:A2} the measured inclusive total cross section for $\pi^\pm d$ scattering from the Particle Data Group~\cite{Tanabashi:2018oca} are shown, respectively, with orange and red marks with as a function of $\sqrt{s}$. We focus on the peak region where the cross section reaches a value of the order of $\sigma_{peak} \simeq 200$ mb. Therefore, the inelastic two-body channel $\pi d \rightarrow pp$ shown in Fig.~\ref{fig:A1} exhausts less than 5\% of the total inclusive cross section. This comparison demonstrates the necessity to implement inelastic processes for $d$ breakup by energetic $\pi$ involving more than 2 particles in the final state and consequently the inverse process where a deuteron can be formed by the interaction of two nucleons catalyzed by colliding pions. A theoretical study to understand why the $\pi d$ cross section is larger than the typical inelastic hadronic cross section has been conducted in Ref.~\cite{Ikeno:2021frl}. 

{\bf II.} In this work we consider the reaction $NN\pi \leftrightarrow d\pi$ within the PHQMD transport approach where two-body and, more importantly, three-body processes are treated using the convariant rate formalism~\cite{Cassing:2001ds}.
In this formalism the only requested input is a parametrization of the total cross section as function of $\sqrt{s}$ since the transition rate should depend only on the invariant energy which means to assume an isotropic differential cross section.
Here we describe the procedure to extract such phenomenological cross section. 

 The inclusive $\pi^\pm d$ cross section is strongly peaked at $\sqrt{s} \simeq 2.2$ GeV due to excitation of underline $T=3/2$ isospin channel which, as we have just said, is possible both for $\pi^-$ and $\pi^+$ due to the presence of a proton and a neutron in the deuteron target. Referring to Fig.~\ref{fig:A2}, we firstly perform a fit (violet curve) of this total inclusive cross section using the following piecewise expression

\begin{equation}
\!\! \sigma \!=\!\!  
\begin{cases}
\sum_{i=1}^{3} a_i  e^{\left[ - (s-b_i)^2/c_i \right]} + (d + f s) \quad \quad \, , \, \sqrt{s} \le 3.35   \\
\\
\left(d_0 + d_1\sqrt{s} + d_2s\right)e^{\left[ -hs^{g/2} \right]} \quad , \, 3.35 < \sqrt{s} \le 3.7 
\end{cases}
\end{equation}

The center-of-mass energy $\sqrt{s}$ is in GeV and the $\sigma(\pi d)$ is given in mb. For $\sqrt{s}>3.7$ GeV we simply take constant $\sigma=47$ mb. The values of the fit parameters are reported in the upper part of Table~\eqref{Tab:Incltotcrossd}. 

\begin{table}[h]
\centering
\begin{tabular}{c|c|c|c}
$i$ & $a_i$ & $b_i$ &  $c_i$   \\
\hline \hline
1 & 186.690 & 4.767 &  0.042   \\
2 & 16.765 &  7.356  & 0.174  \\
3 & 12.907 &  8.808  & 1.282 \\
\hline
$d =$ & 42.586  & $f =$ &  2.009   \\
\hline
 & $d_0$ & $d_1$ &  $d_2$   \\
& 15543.600 & 1145.460 & 504.896 \\
\hline
$h =$ & 4.651  & $g =$ &  0.203   \\
\hline
\hline
$i (inel.) $ & $a_i  $ & $b_i$ &  $c_i$   \\
1 & 143.415 & 4.779  & 0.030 \\ 
2 & 49.652 & 5.587 & 1.603  \\
\hline
\end{tabular}
\caption{Upper part: fit parameters for the total inclusive $\pi^{\pm}d$ scattering cross section (in mb) as a function of center-of-energy $\sqrt{s}$. Lower part: the same fit parameters for the phenomenological cross section for $d\pi \rightarrow NN\pi$ estimated by subtracting the elastic and two-body $d \pi \rightarrow NN$ contributions to the total inclusive cross section.}
  \label{Tab:Incltotcrossd}
\end{table}

In Fig.~\ref{fig:A2}, subtracting the elastic $\pi d$ (green dashed line) and two-body inelastic $\pi d\rightarrow NN$ (blue dash-dotted line) contributions, we can infer the inelastic cross section for deuteron breakup into 2 baryons + $n$ mesons which we assume can be only pions with $n \ge 1$  
\begin{equation}
\sigma( \pi d \rightarrow BB+n*\pi) =  \sigma - \sigma_{el}(\pi d)- \sigma(\pi d \rightarrow NN) \, .
\end{equation}
In particular, the two baryons can be regarded as only nucleons $B=N$ plus excitation of $\Delta$ resonances which further decay into $N+\pi$, hence feeding the number $n$ of final pions. Inelastic processes with increasing $n-$body add up subsequently with increasing value of $\sqrt{s} \ge \sqrt{s}_{th} = \sum_{f} m_f$ where the sum runs over the masses of final produced particles. On the other hand, this causes the closure of the phase-space of few-body production at larger values of $\sqrt{s}$. 
Keeping this in mind, we can estimate the behavior of the leading inelastic process $\pi d \leftrightarrow NN \pi$ where deuteron breaks up by the incident pion into a pair of nucleons plus a single emitted pion. The cross section for such leading inelastic process is parametrized by the 2-Gaussian expression below 
\begin{equation}
\sigma( \pi d \rightarrow NN\pi) =  \sum_{i=1,2} a_i e^{\left[-(s - b_i)^2/c_i \right]} \, .
\label{eq:CrossDpi23}
 \end{equation}
 
The values of the fit parameters are reported in the lower part of Table~\eqref{Tab:Incltotcrossd} (\emph{inel.}) with the cross section in Eq.~\eqref{eq:CrossDpi23} given in mb. Finally, the resulting inelastic $\pi d \rightarrow NN \pi$ cross section is shown in Fig.~\ref{fig:A2} by the squared-thick black line.

\begin{figure}[ht]
\centering
\includegraphics[width=0.5\textwidth]{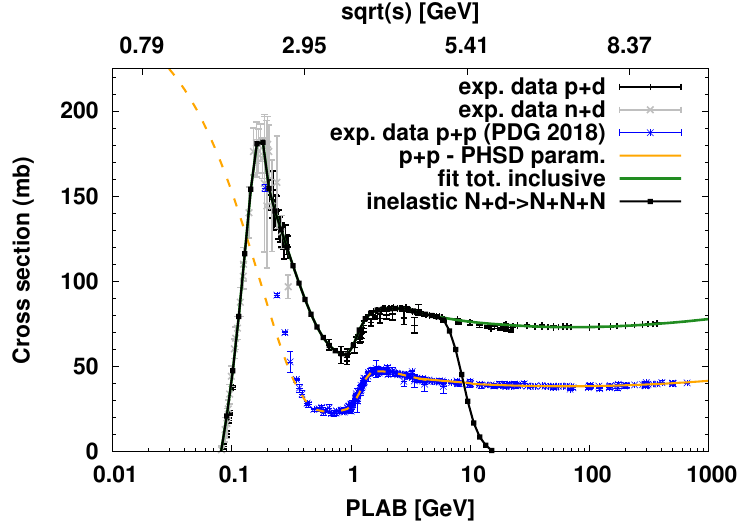}
\caption{\label{fig:A3} (color online) The PHQMD parametrization of the three-body inelastic cross section for $Nd \rightarrow NNN$ reactions (squared-thick black line) as a function of $\sqrt{s}$ is compared to experimental data of total inclusive $p(n)d$ scattering from the PDG database~\cite{Tanabashi:2018oca} shown with the black (grey) points. The solid green line corresponds to the fit of the total inclusive $Nd$ scattering cross section which is constructed using Eq.~\eqref{eq:sNdlow} for $\sqrt{s} \le 8.9$ GeV and Eq.~\eqref{eq:sNdhigh} $\sqrt{s} > 8.9$ GeV. The orange line shows the PHQMD parametrization of the $pp$ scattering cross section taken from the original (P)HSD framework.}  
\end{figure}

{\bf III.} Similarly to the $\pi d \rightarrow N N\pi$ reaction which dominates at relativistic energies due to the large pion abundance, we implement also the three-body inelastic process for $Nd \leftrightarrow NNN$ which is more important at low energy heavy-ion collisions~\cite{Kapusta:1980zz,Siemens:1979dz}. 
We derive a parametrization of the total inclusive cross section through a fit on experimental data which are shown in Fig.~\ref{fig:A3} for $pd$ experimental data (black points) and for some $nd$ data (grey points) taken from the PDG database ~\cite{Tanabashi:2018oca} and from other references~\cite{Bugg:1966zz}. The lower horizontal axis is the range of the proton beam momentum $P_{lab}$, while the upper one is the corresponding $\sqrt{s}$ calculated by relativistic kinematics. 
The total inclusive cross section is composed by an elastic and an inelastic part
\begin{equation}
\sigma(Nd) = \sigma_{el}(Nd) + \sigma_{inel}(Nd \rightarrow NNN+n*\pi) \, .
\label{eq:CrossDN}
\end{equation}
In this case the elastic and inelastic contributions are distinctly separated in kinematics due to the existing energy threshold for the $dN \rightarrow NNN + n*M$ processes which is always an endothermic process for any final number of pions $n \ge 0$. In particular, for the $Nd\rightarrow NNN$ reaction with $n=0$, this threshold corresponds to the difference $E_{th}=3m_N-(m_N+m_d)$. Replacing the deuteron mass $m_d=2m_N+E_B$, this threshold energy corresponds to nothing else than the absolute value of deuteron binding energy $|E_B|=2.2$ MeV.

Our expression for the total inclusive cross section is derived applying similar procedure to what is done for the case of $pp$ cross section. In the low energy regime we employ a functional expression in terms of the projectile nucleon momentum $P_{lab}$ which is used in Ref.~\cite{Cugnon:1996kh} to parametrize the $NN$ inelastic scattering cross section. Then, our calculated expression is welded to a parametric function of $\sqrt{s}$ which we us in the high energy regime. In PHQMD this procedure is applied also for parametrizing the $pp$ scattering cross section. The resulting curve for is depicted in Fig.~\ref{fig:A3} with an orange line (dashed for the low energy regime, solid for the high one) and it provides a good fit of the experimental $pp$ data (blue points) taken from PDG database~\cite{Tanabashi:2018oca}. In Fig.~\ref{fig:A3} this function is denoted as `` $p+p$ PHSD parametrization'', because it is implemented in PHQMD directly from the original (P)HSD original code~\cite{Cassing:2008sv,Cassing:2008nn,Cassing:2009vt,Bratkovskaya:2011wp,Linnyk:2015rco} where it is used to describe nucleon-nucleon collisions.
For the case of $Nd$ scattering the complete expression for the inclusive total cross section is reported below and the resulting curve is shown in Fig.~\ref{fig:A3} with a solid green line

\begin{eqnarray}\label{eq:sNdlow}
\sigma (s) &= (-0.316+P_{lab}^{0.46})/(6.2\cdot10^{-3}+(P_{lab}^2-0.021)^2) \nonumber \\
                               & \quad  \, P_{lab}<0.208 \, GeV \nonumber \\[5pt]
       							&= 56.6413 +117.547 \left| 1.1588 - P_{lab} \right|^{4.348} \nonumber \\
                               & \quad \, 0.208 \le P_{lab}<0.977 \, GeV \nonumber \\[5pt]
       							&= 28.0475 + 56.07/(1.0+exp(-\frac{P_{lab}\!-\!0.971}{0.1665})) \nonumber\\
                               & \quad \, 0.977 \le P_{lab}<2.96 \, GeV \nonumber \\[5pt]
                               &= 78.736 + 15.31(P_{lab}+2.932)exp(-0.952P_{lab}) \nonumber \\
                               & \quad \, 2.96 \le P_{lab}<3.8 \, GeV \nonumber \\[5pt]
                               &= 93.66 + 1.6473\log(P_{lab})^2 -11.301\log(P_{lab}) \nonumber \\
                               & \quad \, 3.8 \le P_{lab}<19.9 \, GeV \nonumber \\
\end{eqnarray}

The low-energy parametrization of the total inclusive $Nd$ cross section (in mb) as a function of $P_{lab}$ is valid within the range of $P_{lab} \le 19.92$ GeV which corresponds to a value of $\sqrt{s}\le 8.9$ GeV. Instead, the high-energy parametrization which is valid for $\sqrt{s} > 8.9 \, GeV$ and it is given by the following formula 
\begin{eqnarray}\label{eq:sNdhigh}
\sigma (s) &= 94.01*s^{-0.555} -30.7318*s^{-0.4986} \nonumber \\
								&+ 69.2663+0.42055*\log^2(\frac{s}{46.2745}) \, .
\end{eqnarray}

Finally, in order to pass from the total inclusive $\sigma(Nd)$ to the total inelastic $\sigma(Nd\rightarrow NNN)$ cross section, we perform a smooth cut of the expression Eq.~\eqref{eq:sNdhigh}. This is motivated by the fact that at high $\sqrt{s}$ values inelastic channels with final particles multiplicity larger than 3 (i.e. $NNN+\pi \, , \, NNN+2\pi \, , \, \dots$) start to contribute and the three-body $NNN$ phase-space is suppressed. Therefore, similarly to what we have done for the $\pi d\rightarrow NN\pi$ reaction, we consider a gaussian function $A\exp(-(s-B)^2/C)$ with parameters $A=37.985 \, , \, B=28.343 \, , \, C=137.733$ which we attach to the inclusive cross section formula Eq.~\eqref{eq:sNdhigh} at $\sqrt{s} \ge 5.0$ GeV.
The resulting curve is depicted in Fig.~\ref{fig:A3} (thick black solid line). At lower values of $\sqrt{s}$ the cross section for inelastic $\sigma(Nd\rightarrow NNN)$ reaction equals the total inclusive $\sigma(Nd)$. as it is clearly visible in Fig.~\ref{fig:A3} where the two lines are superimposed.

 \section*{Appendix B: Box simulations}

Additionally to the pion catalysis $\pi d \leftrightarrow pn \pi$ process considered in Sec.~\ref{sec:sec4}, in this Appendix we present the box study for other reactions  for deuteron production:\\
 \textit{i)} $2\leftrightarrow 3$ reaction of nucleon catalysis - $Nd \leftrightarrow pnN$, which plays a dominant role at low collision energies where the nucleon density is high (see Fig.~\eqref{fig:C0});
 \textit{ii)} $2\leftrightarrow 2$ reaction - $\pi d \leftrightarrow NN$, which are a subdominant channel due to the small cross section.
 
\begin{figure}[ht]
\centering
\includegraphics[width=0.45\textwidth]{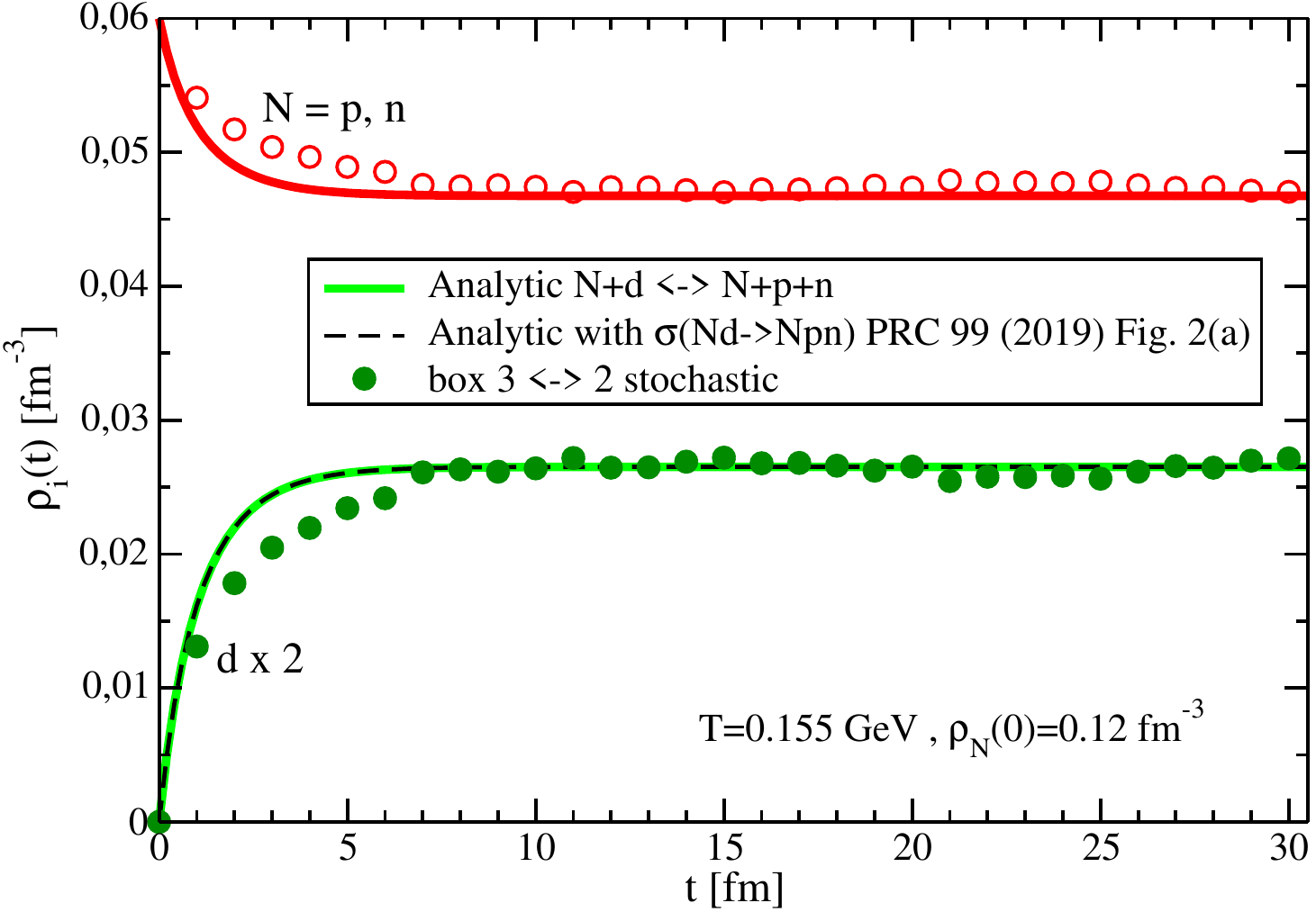}
\caption{\label{fig:B1} (color online) The time evolution of particle densities for $Nd \leftrightarrow pnN$ reactions in static hadronic box. The lines represent the solutions obtained from the rate equations, while the symbols are the results from box simulations. The dashed black line is the analytic solution for the density of deuterons obtained using the cross section $\sigma(Nd \rightarrow Npn)$ taken from Ref.~\cite{Oliinychenko:2018ugs} Fig.~2(a). This result is equal to the same analytic expectation (solid green line) employing the PHQMD parametrization of the cross section for $Nd \rightarrow NNN$ inelastic scattering which is reported in Appendix A.}  
\end{figure}

\begin{figure}[ht]
\centering
\includegraphics[width=0.45\textwidth]{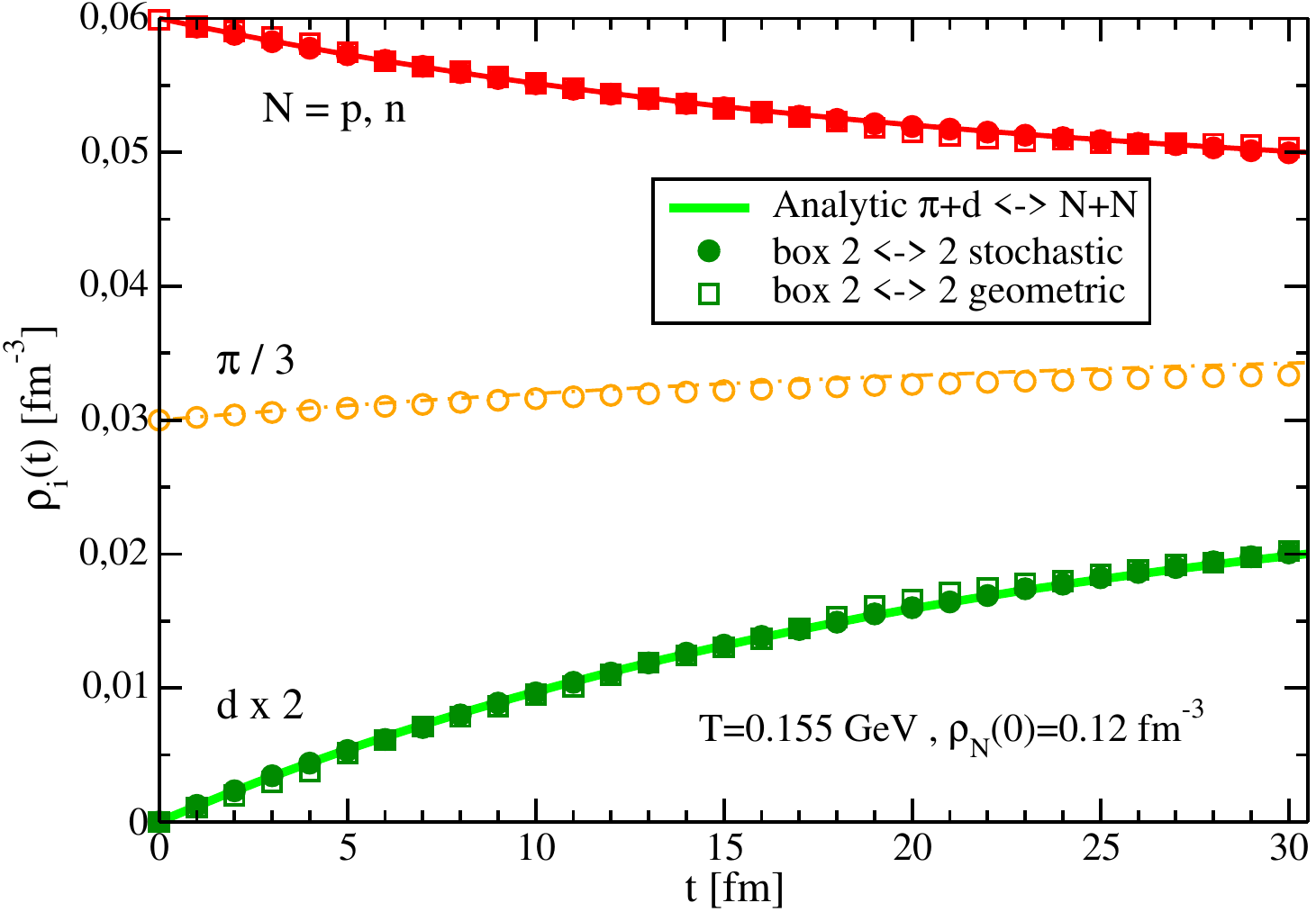}
\caption{\label{fig:B2} (color online) The time evolution of particle densities for $\pi d \leftrightarrow NN$ reactions in static hadronic box. The lines represent the solutions obtained from the rate equations, while the symbols refer to box calculations. For nucleons and deuterons, the full circles and the open squares are the numerical results when the collision integral is solved, respectively, by means of stochastic method or adopting the geometric criterium.}  
\end{figure}

In Figs.~\ref{fig:B1} and \ref{fig:B2} we show the time evolution for the density of nucleons $N=p,n$ (red), deuterons (green) and pions (orange) for simulations performed in a static box at initial temperature $T=0.155$ GeV and initial nuclear density $\rho_N(0)=2\rho_p(0)=2\rho_n(0)=0.12 \, fm^{-3}$ and pion density $\rho_\pi(0)=0.09 \, fm^{-3}$ in comparison with analytic results obtained as solution of the rate equations. These two plots are similar to Fig.~\ref{fig:B0} of Sec.~\ref{sec:sec4}, where we tested the correct numerical implementation of $\pi d \leftrightarrow pn \pi$ process. \\
In particular, in Fig.~\ref{fig:B1} we test the formation/breakup of deuterons by nucleon catalysis $Nd \leftrightarrow pnN$ by switching off all other reactions and verify that the collisions inside the box performed by stochastic method (circles) are in agreement with analytic expectations (solid lines). Here the result for pions is not shown, because they are not involved in the main reaction channel, but they can scatter elastically with deuterons in order to drive them to faster equilibration. 

In Fig.~\ref{fig:B2} we investigate  the correct implementation of the two-body inelastic $\pi d \leftrightarrow NN$ process. The analytic expectations of densities as function of evolution time are represented with the lines (red solid for $N=p,n$, green solid for $d$ and orange dash-dotted for $\pi$). For nucleons and deuterons we show box results obtained adopting either the geometric criterium $d_T \le \sqrt{\sigma^{2,2}/\pi}$ (open squares) or the stochastic collision method $P_{2,2} = \sigma^{2,2}v_{rel}\Delta t/\Delta V_{cell}$ (full squares) with cross section $\sigma^{2,2}(\sqrt{s})$ taken from Ref.~\cite{Oh:1997eq}. For pions we show only the results employing the latter collision method.
In this case the particle densities show a longer equilibration time compared to the box results where either $\pi$-catalysis or $N-$catalysis reactions are switched on. This is expected because the cross section of $\pi d\rightarrow NN$ disintegration is much smaller than those of other reactions. Consequently, the inverse process of $NN$ fusion into $\pi d$ final state, whose cross section is experimentally proved to fulfill detailed balance (see Eq.~\eqref{eq:App1}), slowly proceeds with an equilibration time about 10 times larger than that of $\pi-$catalysis.       

As follows from Figs.~\ref{fig:B1} and \ref{fig:B2} the simulated density distributions in the box are in a good agreement with the analytic thermodynamic results.

We remind here that the ``box" framework is just a toy-model to test in controlled conditions that the detailed balance as well as the agreement with the analytic solutions of the rate equations (see Sec.~\ref{sec:sec4}) is fulfilled for all the implemented ``kinetic" deuteron reactions, in particular $\pi NN \leftrightarrow \pi d$, $NNN \leftrightarrow Nd$ and $NN \leftrightarrow \pi d$.  In realistic HIC simulations with PHQMD we describe the phase-space evolution of the fireball on an expanding 3D-grid where the cells and the time-step parameters are described at the end of Sec.~\ref{sec:sec4}.

\section*{Appendix C: Isospin factors in deuteron reactions}

In detail, we derive the isospin coefficients for the $\pi$-catalysis reactions $d \pi^k \leftrightarrow N N \pi^l$ listed in Tab.~\eqref{Tab:deut3to2}. We look first at the reaction in the forward direction, i.e. to the deuteron disintegration by incident $\pi$.
Since the deuteron has zero isospin and rhe pion has isospin 1, the 2-particle state $\ket{d , \pi^k}$ is already an eigenstate of total initial isospin $T_i=1$ and the projection along the quantized axis are related to the pion charge indicated by the index $k=-1,0,+1$. In short notation we can write
\begin{equation}\label{eq:Iso1}
\ket{d , \pi^k} = \ket{T_i,M_i} = \ket{1,sgn(k) \delta_{|k|1}}
\end{equation}
 
To calculate the possible eigenstates of total final isospin $T_f$ from the 3-particle state $\ket{N,N , \pi^l}$ we use the rules for summation of angular momentum in quantum mechanics~\cite{Sakurai:2011zz}.\\
Firstly we perform the summation of the $1/2$-isospin of the two nucleons, which - as is well known - generates the singlet state
\begin{equation}\label{eq:Iso2}
\ket{T_N=0,M_N=0} = \frac{\ket{p,n}-\ket{n,p}}{\sqrt{2}}
\end{equation}
and the triplet state
\begin{align}\label{eq:Iso3}
&\ket{T_N=1,M_N=1} = \ket{p,p} \nonumber \\
&\ket{T_N=1,M_N=0} = \frac{\ket{p,n}+\ket{n,p}}{\sqrt{2}}  \\
&\ket{T_N=1,M_N=-1} = \ket{n,n} \nonumber
\end{align}
where $T_N$ indicates the eigenvalue of total isospin of the intermediate two-nucleons system and $M_N$ its projection along the quantized axis.
Then we add the contribution of the isospin quantum numbers of the pion which we indicate with small letters as $t_{\pi}=1$ and $m_{\pi} \le |t_{\pi}|$.

Therefore, the eigenstates of total isospin $T_f$ are given by the following expansion
\begin{align}\label{eq:Iso4}
\ket{T_N,t_{\pi};T_f,M_f} &= \sum_{M_N=-T_N}^{T_N} \sum_{m_{\pi}=-t_{\pi}}^{t_{\pi}} \ket{T_N,t_{\pi};M_N,m_{\pi}} \nonumber \\ 
\times & \bra{T_N,t_{\pi};M_N,m_{\pi}}\ket{T_N,t_{\pi};T_f,M_f}
\end{align}
On the right hand side, the ket of the basis $ \ket{T_N,t_{\pi};M_N,m_{\pi}}$ represent the tensor product of the two-nucleons state with defined isospin quantum numbers $(T_N,M_N)$ with the pion state $(t_{\pi},m_{\pi})$. On the left hand side, the new basis is formed by the ket with defined quantum numbers $(T_N,t_{\pi},T_f,M_f)$. The angular momentum rules require that $M_f=M_N+m_{\pi}$ which is encoded already in the Clebsch-Gordan coefficients of the expansion Eq.~\eqref{eq:Iso4}.

By construction, the new states of total isospin $T_f$ will remain eigenstates of the two-nucleons isospin $T_N$. Hence, formula Eq.~\eqref{eq:Iso4} must be separated into two orthogonal contributions according to the two different values of $T_N$ from the singlet Eq.~\eqref{eq:Iso2} and from the triplet state Eq.~\eqref{eq:Iso3}.
\begin{itemize}
\item[I)] When the two nucleons are in the singlet case $T_N=0$, the condition is similar to the 2-particle state $\ket{d , \pi}$, i.e. the total final isospin has eigenvalue $T_f=1$ from the pion contribution. In the same notation of Eq.~\eqref{eq:Iso1} we can write
\begin{align}\label{eq:Iso5}
&\ket{T_N=0,t_{\pi}=1;T_f=1,M_f} = \ket{0,1;1,M_f}  \nonumber \\
&=\frac{\ket{p,n,\pi^l} - \ket{n,p,\pi^l} }{\sqrt{2}} = \ket{0,1;1,sgn(l)\delta_{|l|1}} 
\end{align}
\item[II)] In case the two nucleons couple two a triplet state $T_N=1$ the addition of the pion isopsin $t_{\pi}$ generate independent spaces of total final isospin with eigenvalues given by the triangle rule
\begin{equation}\label{eq:Isot}
|T_N - t_{\pi}| \le T_f \le T_N+t_{\pi} \rightarrow T_f=0,1,2
\end{equation}
Due to the conservation of total isospin in strong interaction, we are interested in the eigenstates of final isospin $T_f=T_i=1$. Using Eq.~\eqref{eq:Iso4} with the correct quantum numbers and taking the corresponding Clebsch-Gordan coefficients from available tables, we obtain 
\begin{align}\label{eq:Iso6}
&\ket{T_N=1,t_{\pi}=1;T_f=1,M_f=1} = \ket{1,1;1,1} \\
&= \left( \frac{\ket{p,n,\pi^+} + \ket{n,p,\pi^+}}{\sqrt{2}} \right) \frac{-1}{\sqrt{2}}  +  \ket{p,p,\pi^0}\frac{1}{\sqrt{2}}  \nonumber
\end{align}
\begin{align}\label{eq:Iso7}
&\ket{T_N=1,t_{\pi}=1;T_f=1,M_f=0} = \ket{1,1;1,0} \\
&= \ket{n,n,\pi^+}\frac{-1}{\sqrt{2}} + \ket{p,p,\pi^-} \frac{1}{\sqrt{2}} \nonumber
\end{align}
\begin{align}\label{eq:Iso8}
&\ket{T_N=1,t_{\pi}=1;T_f=1,M_f=-1} = \ket{1,1;1,-1} \\
&= \left( \frac{\ket{p,n,\pi^-}+\ket{n,p,\pi^-}}{\sqrt{2}} \right) \frac{1}{\sqrt{2}} + \ket{n,n,\pi^0} \frac{-1}{\sqrt{2}} \nonumber
\end{align}
\end{itemize}
Eq.~\eqref{eq:Iso5} and \eqref{eq:Iso6}-\eqref{eq:Iso8} express the $T_f=1$ eigenstates in terms of the 3-particles states $\ket{N,N,\pi^l}$. Combining them the associated isospin factors for the transition $\ket{d,\pi^k} \rightarrow \ket{N,N,\pi^l}$ are nothing else than the Fourier coefficients of the following expansion
\begin{align}\label{eq:Iso9}
&\ket{d,\pi^+} \rightarrow \frac{1}{2} \left[ \frac{}{} \ket{p,p,\pi^0} + \right. \nonumber \\
 &\left. \left(-1+\frac{1}{\sqrt{2}}\right)\ket{p,n,\pi^+} - \left(1+\frac{1}{\sqrt{2}}\right)\ket{n,p,\pi^+} \right]
\end{align}
\begin{align}\label{eq:Iso10}
\ket{d,\pi^0} \rightarrow &\frac{1}{2} \left[ \frac{}{} \ket{p,n,\pi^0} - \ket{n,p,\pi^0} + \right. \nonumber \\ & \left. \ket{n,n,\pi^+} - \ket{p,p,\pi^-} \frac{}{} \right]
\end{align}
\begin{align}\label{eq:Iso11}
&\ket{d,\pi^-} \rightarrow \frac{1}{2} \left[ \frac{}{} \ket{n,n,\pi^0} + \right. \nonumber \\
 &\left. \left(1+\frac{1}{\sqrt{2}}\right)\ket{p,n,\pi^-} + \left(-1+\frac{1}{\sqrt{2}}\right)\ket{n,p,\pi^-} \right]
\end{align}

An overall factor $1/2$ is introduced in order to guarantee the normalization of the corresponding final state $\ket{T_f=1,M_f}$ to unity. Differently, in calculating the probability of each transition allowed by isospin conservation one should divide by the sum of the squares of all the Fourier coefficients.
Finally, the associated isospin probability $P_{iso}(d\pi^k \rightarrow NN\pi^l)$ for each channel are given by the square of such coefficients. Moreover, since in PHQMD the isospin number is not stored during the dynamical evolution of particles, we can make a further simplification by adding the probabilities for those 3-particle states on the right hand side of Eq.~\eqref{eq:Iso9}-\eqref{eq:Iso11} which differ only by the order of nucleons. Therefore, we obtain
\begin{align}\label{eq:Iso12}
&P_{iso}(d\pi^+ \rightarrow pn\pi^+)=\frac{3}{4} \, ; \, P_{iso}(d\pi^+ \rightarrow pp\pi^0)= \frac{1}{4} \nonumber \\
&P_{iso}(d\pi^0 \rightarrow pn\pi^0)=\frac{1}{2} \, ; \, P_{iso}(d\pi^0 \rightarrow p(n)p(n)\pi^{-(+)})=\frac{1}{4} \nonumber \\
&P_{iso}(d\pi^- \rightarrow pn\pi^-)=\frac{3}{4} \, ; \, P_{iso}(d\pi^- \rightarrow nn\pi^0)= \frac{1}{4} \nonumber \\
\end{align}
which, as discussed in Sec.~\ref{sec:sec3}, correspond to the factors employed to select the final state of each collision where a deuteron disintegrates by inelastic $2 \rightarrow 3$ pion reaction according either to the geometric criterium or to the stochastic method, both depending on the total cross section $\sigma^{2,3}(\sqrt{s})$ described in Appendix A. It is important that the sum of all probabilities over the possible final state $NN\pi^l$, to which the initial state $d\pi^k$, can decay according to total isospin conservation equals to one. This condition is clearly fulfilled by summing the terms in each row of Eq.~\eqref{eq:Iso12}.

In the backward direction, i.e. when an incident $\pi$ catalyzes the fusion of two nucleons to form a deuteron plus an emitted pion, the transition from the 3-particle state $NN\pi^l$ into the 2-particle state $d\pi^k$ can happen only if total isospin is conserved. This means, that only when the initial state $NN\pi^l$ finds itself in an eigenstate of total isospin 1, it can make the transition to the final state $d\pi^k$.
Formally, we need to invert the basis expansion Eq.~\eqref{eq:Iso4} and obtain the following 
\begin{align}\label{eq:Iso13}
\ket{N,N,\pi} = \sum_{T} \sum_{M=-T}^{T} \ket{T,M} \bra{T,M} \ket{N,N,\pi}
\end{align}
Physically, each 3-particle state $\ket{N,N,\pi}$ is written as a superposition over the eigenstates of total isospin $\ket{T,M}$ with eigenvalues $T$ again provided by the same triangle rule Eq.~\eqref{eq:Isot} and eigenvalue of the isospin 3-component $ M \le |T|$. In Eq.~\eqref{eq:Iso13} the same Clebsch-Gordan coefficients of Eq.~\eqref{eq:Iso4} appear because
\begin{align}
\bra{T,M} \ket{N,N,\pi} &= \bra{T_N,t_{\pi};T_f,M_f} \ket{T_M,t_{\pi};M_N,m_{\pi}} \nonumber \\&= \bra{T_N,t_{\pi};M_N,m_{\pi}} \ket{T_N,t_{\pi};T_f,M_f} \nonumber
\end{align}

Once we calculate all the eigenstates of total isospin, we can look at the Fourier coefficient of each state $\ket{N,N,\pi}$ associated to the eigenavalue  $T=1$ and obtain the right probability. We point out, that the use of Eq.~\eqref{eq:Iso13} requires also that the intermediate total isospin $T_N$ of the two-nucleons system is a quantum number of the basis. Therefore, the contributions which come from the singlet and the triplet state must be squared and independently summed without mixing term. 

Then, for example, the probability that the state $\ket{p,n,\pi^+}$ has total isospin $T=1$ is given by
\begin{align}\label{eq:Iso15}
&|\bra{1,0 ;1,1} \ket{p,n,\pi^+}|^2 + |\bra{1,1;1,1} \ket{p,n,\pi^+}|^2 \nonumber \\
&= \frac{1}{2} + \frac{1}{2} \frac{1}{4} = \frac{3}{4}
\end{align}
where the first term comes from the singlet $T_{N}=0$ and the second one from the triplet $T_{N}=1$ state.
Analogously we can calculate for the other 3-particle $NN\pi^l$ states and obtain the probabilities
\begin{align}\label{eq:Iso16}
&P_{iso}( pn\pi^{\pm} \rightarrow d\pi^{\pm}) =\frac{3}{4} \quad ; \quad P_{iso}( np\pi^{\pm} \rightarrow d\pi^{\pm}) =\frac{3}{4} \nonumber \\
&P_{iso}( pp\pi^0 \rightarrow d\pi^+) = P_{iso}( nn\pi^0 \rightarrow d\pi^-) = \frac{1}{2} \nonumber\\
&P_{iso}( pp\pi^- \rightarrow d\pi^0) = P_{iso}( nn\pi^+ \rightarrow d\pi^0) = \frac{1}{2} \nonumber\\
&P_{iso}( pn\pi^0 \rightarrow d\pi^0) = P_{iso}( np\pi^0 \rightarrow d\pi^0) =\frac{1}{2} \nonumber \\
\end{align}
which named shortly as $F_{iso}$ in the formula Eq.~\eqref{Prob32} for the full covariant probability of $3 \rightarrow 2$ for the reaction of deuteron formation by $\pi$-catalysis . 
 
The isospin coefficients for the $N$-catalysis reactions $d N \leftrightarrow N N N$ can be calculated in similar way. We emphasize only two differences. 
On the one hand, the final state of the forward reaction, i.e. the deuteron disintegration, contains only nucleons. As mentioned previously, in PHQMD the particle isospins are not propagated dynamically during the system evolution. This means, that we cannot distinguish between those states of the reactions which differ by the order of protons and neutrons. Therefore, the transition from $dN$ to $NNN$ reduces to one channel and consequently  
\begin{equation}\label{eq:Iso17}
P_{iso} (dN \rightarrow NNN) = P_{iso}(dN \rightarrow (pn)N) = 1
\end{equation}
On the other hand, the isospin factors for the backward reaction, i.e. the formation of deuteron by two-nucleons plus third nucleon as catalyzator, demands that the initial state $\ket{N,N,N}$ finds itself in an eigenstate of total isospin $T=1/2$. We apply the addition rules of angular momentum and construct the intermediate singlet and triplet states by summing isospins of two nucleons and then add the $1/2$-isospin of the third one. The transition probabilities can be derived from the square of the Fourier coefficients taking the contribution of the singlet and the triplet without mixing term. We obtain
\begin{equation}\label{eq:Iso18}
P_{iso} (pnp \rightarrow dp) = P_{iso}(pnn \rightarrow dn) = \frac{1}{3}
\end{equation} 
which corresponds to the coefficient $F_{iso}^{3,2}$ in Eq.~\eqref{Prob32} for the deuteron production by $N$-catalysis.

\section*{Appendix D: Phase-Space integrals}

In Eq.~\eqref{Prob32} the Lorentz invariant two-body and three-body phase-spaces appear. For $R_2(\sqrt{s},m_1,m_2)$ we simply adopt its analytic expression 
\begin{equation}\label{eq:2bodyfinal2}
R_2 \left( \sqrt{s}, \, m_{1}, \, m_{2} \right) = \frac{\sqrt{\lambda(s, \, m_1^2, \, m_2^2)}}{8 \pi s} \, ,
\end{equation}
where on the right hand side the kinematical function is $\lambda(s, \, m_1^2, \, m_2^2) = \left(s-m_1^2-m_2^2 \right)^2 - 4m_1^2m_2^2$ and $\sqrt{s}$ is the center-of-mass energy. For $R_3(\sqrt{s},m_3,m_4,m_5)$ we use the well known recursion relations~\cite{Kajantie:1971rj,Byckling:1971vca} to write it in terms of a factorized product of two-body phase-spaces
\begin{eqnarray}
R_3 \left( \sqrt{s}, \, m_3, \, m_4, \, m_5 \right) = && \nonumber \\ \int_{(m_3+m_4)^2}^{(\sqrt{s}-m_5)^2} \frac{d M_2^2}{2 \pi} R_2 \!\left( \sqrt{s}, \, m_5, \, M_2 \right) R_2 \!\left( M_2, \, m_3, \, m_4 \right) = \nonumber \\
\int_{(m_3+m_4)^2}^{(\sqrt{s}-m_5)^2} \frac{d M_2^2}{2 \pi} \frac{\sqrt{(s-m_5^2 -M_2^2)^2-4M_2^2m_5^2}}{8\pi s} \times \nonumber \\
\frac{\sqrt{(M_2^2 - m_3^2 - m_4^2)^2 - 4 m_3^2m_4^2}}{8 \pi M_2^2} \nonumber \\
\end{eqnarray}
The integration is run over the invariant mass variable $M_2$. It has been shown in Ref.~\cite{Seifert:2018sbg} that such expression can be fitted by the following function
\begin{equation}
f_3\left( t \right) = a_1*t^{a_2}*\left( 1 - \frac{1}{a_3*t + 1 +a_4}  \right) \, ,
\label{eq:3bodyfinal3}
\end{equation}  
where $t = \sqrt{s} - m_3 - m_4 - m_5$ and the parameter values depend on the physical masses $(m_3,m_4,m_5)$ involved.
For the $\pi NN$ and the $NNN$ three-body phase-spaces these parameters are reported in Table~\eqref{tab:3BPS}.

\begin{table}[ht]
\begin{tabular}{|c|c|c|c|c|c|c}
\hline
 $m_3 \, m_4 \, m_5$ & $a_1 $ &  $a_2$ & $x=2-a_2$ & $a_3 $ &  $a_4$ \\
\hline
$\pi \, N \, N$ &  0.000249  &  1.847779 &  0.152221 &  0.071509 &  9.973413 \\
\hline
$N  \, N \, N$ &  0.000350  &  1.781741 &  0.218259 &  0.052836 &  4.221995 \\
\hline
\end{tabular}
\caption{\label{tab:3BPS} The values of the fit parameters for the three-body $\pi NN$ and $NNN$ phase-spaces using formula Eq.~\eqref{eq:3bodyfinal3}.}
\end{table}

\bibliography{PHQMD_Deuteron}

\end{document}